\documentclass[fleqn,usenatbib]{mnras}

\usepackage{newtxtext,newtxmath}

\usepackage[T1]{fontenc}

\DeclareRobustCommand{\VAN}[3]{#2}
\let\VANthebibliography\thebibliography
\def\thebibliography{\DeclareRobustCommand{\VAN}[3]{##3}\VANthebibliography}

\usepackage{graphicx}	
\usepackage{amsmath}	
\usepackage{subfigure}
\usepackage{multirow}
\usepackage[normalem]{ulem}

\title[Building the Galilean moons system]{Building the Galilean moons system via pebble accretion and migration: A primordial resonant chain}

\author[Madeira et al.]{
Gustavo Madeira,$^{1}$\thanks{E-mail: gustavo.o.madeira@unesp.br}
Andr\'e Izidoro,$^{2,1}$
and Silvia M. Giuliatti Winter$^{1}$
\\
$^{1}$Grupo de Din\^amica Orbital \& Planetologia, The University of S\~ao Paulo State-UNESP , Av. Ariberto Pereira da Cunha, 333, Guaratinguet\'a SP, 12516-410, Brazil\\
$^{2}$Department of Earth, Environmental and Planetary Sciences, MS 126, Rice
University, Houston, TX 77005, USA
}

\date{Accepted 2021 April 05. Received 2021 March 24; in original form 2020 November 23}

\pubyear{2015}

\begin{document}
\label{firstpage}
\pagerange{\pageref{firstpage}--\pageref{lastpage}}
\maketitle

\begin{abstract}
The origins of the Galilean satellites  -- namely Io, Europa, Ganymede, and Callisto -- is not fully understood yet. Here  we use N-body numerical simulations to study the formation of Galilean satellites in a gaseous circumplanetary disk around Jupiter. Our model includes the effects of pebble accretion, gas-driven migration, and gas tidal damping and  drag. Satellitesimals in our simulations first grow via pebble accretion and start to migrate inwards. When they reach the trap at the disk inner edge, scattering events and  collisions take place promoting additional growth. Growing satellites eventually reach a multi-resonant configuration anchored at the disk inner edge.  Our results show that an integrated pebble flux of $\geq2\times10^{-3}M_{J}$ results in the formation of satellites with masses typically larger than those of the Galilean satellites. Our best match to the masses of the Galilean satellites is produced in simulations where the integrated pebble flux is $\sim10^{-3}M_{J}$. These simulations typically produce between 3 and 5 satellites. In our best analogues, adjacent satellite pairs are all locked in 2:1 mean motion resonances. However, they have also moderately eccentric orbits ($\sim0.1$), unlike the current real satellites. We propose that the Galilean satellites system is a primordial resonant chain, similar to exoplanet systems as TRAPPIST-1, Kepler-223, and TOI-178. Callisto was probably in resonance with Ganymede in the past but left this configuration -- without breaking the Laplacian resonance -- via divergent migration due to tidal planet-satellite interactions. These same effects further damped the orbital eccentricities of these satellites down to their current values ($\sim$0.001). Our results support the hypothesis that Io and Europa were born with water-ice rich compositions and lost all/most of their water afterwards. Firmer constraints on the primordial compositions of the Galilean satellites are crucial to distinguish formation models. 
\end{abstract}




\section{Introduction} \label{sec:intro}

The Galilean satellites of Jupiter -- namely, Io, Europa, Ganymede, and Callisto -- were discovered by Galileo Galilei in the 17th century. They were the first objects to be observed orbiting another body than the Earth and Sun. Historically, their importance lies in the fact that it was one of the first observational evidence supporting the Copernican view of the Solar System. Early mathematical studies of the Galilean satellites motion around Jupiter were also crucial to promote the development of celestial mechanics and our understanding of resonances in the Solar System and beyond it \citep{Jo78}. Today -- 400 years after their discovery --  the origins of the Galilean satellites remain an intense topic of debate. 

Space-mission explorations, ground and space-based observations have provided a  series of important constraints on Galilean satellites' formation models. The low orbital eccentricities and inclinations of these satellites and their common direction of rotation around Jupiter suggest formation in a thin common disk, in a process similar to the formation of planets around a star \citep{Lu82}. Indeed, observations \citep{Pi19,Ch19} and numerical simulations \citep{Lu99,Kl99}  suggest that the gas and dust not yet accreted to the envelope of a growing gas giant planet may give rise to a circumplanetary disc (hereafter referred as CPD). The formation of a circumplanetary disk is particularly possible if the temperature at the planet's envelope surface is not above a threshold ($\sim$ 2000~K) to prevent the planet's envelope from contracting and material in-falling into a fairly thin disk around its equator \citep{Wa10,Sz16}. 

A growing gas giant planet eventually opens a gap in the gaseous circumstellar disk \citep[hereafter referred as CSD;][]{Li86a,Li86b}. Numerical simulations show that  the CPD is fed by a fraction of the gas from the CSD that enters the planet's Hill radius~\citep{Lu99,Kl99} due to the  meridional circulation of gas in the gap’s vicinity~\citep{Ta12,Mo14,Sz14,Sc20}. This theoretical result has been recently supported by  observations of CPDs~\citep{isellaetal19,Te19}. In this work, we use N-body numerical simulations to model the formation of the Galilean satellites in a gaseous circumplanetary disk around Jupiter. Our model includes the effects of pebble accretion, gas-driven migration, tidal damping of eccentricity and inclination, and gas drag. Before presenting the very details of our model and results, we briefly discuss the physical and orbital properties of the Galilean satellites and also review existing models accounting for the origins of the Galilean satellites.

\subsection{Physical and orbital properties of the Galilean system}
\begin{table*}
\centering
\caption{Physical and orbital parameters of the Galilean satellites. From left-to-right the columns are semi-major axis ($a$), eccentricity ($e$), inclination ($I$), mass ($M$), radius ($R$), the water-ice mass fraction (ice), and bulk density ($d$) of the Galilean satellites \citep{Sc04,Og12} \label{tab:intro}}
\begin{tabular}{lcccccccc} \hline \hline
          & $a$ (${\rm R_J}$) & $e$ ($10^{-3})$ & $I$ (deg) & $M$ ($10^{-5}~{\rm M_J}$) & $R$ (km) & \% ice & $d$ (g/cm$^3$) \\ \hline
Io        & 5.9   &  4.1  & 0.04 & 4.70      & 1822           & 0       & 3.53                \\
Europa    & 9.4 &  10.0  & 0.47  & 2.53      & 1565           & 8       & 2.99               \\
Ganymede & 15.0 & 1.5 & 0.19 &  7.80 & 2631          & 45      & 1.94                \\
Callisto   & 26.4  & 7.0 &  0.28  & 5.69 &  2410         & 56      & 1.83            \\ \hline 
\end{tabular}
\end{table*}

The Galilean satellites form a dynamically compact system. The innermost satellite -- Io --  orbits Jupiter at about $\sim$6~${\rm R_J}$ while the outermost one -- Callisto -- is at $\sim$26~${\rm R_J}$, where ${\rm R_J}$ is the physical radius of Jupiter. All these satellites have almost circular and coplanar orbits. Table~\ref{tab:intro} summarizes the physical and orbital parameters of the Galilean satellites.

The Galilean satellites system forms an intricate chain of orbital resonances.  Io and Europa evolve in a 2:1 mean-motion resonance (MMR), where Io completes two orbits around Jupiter while Europa completes one. Europa and Ganymede are also in a 2:1 MMR. These resonances are associated with the characteristic resonant angles $2\lambda_E-\lambda_I-\varpi_I$, $2\lambda_E-\lambda_I-\varpi_E$, and $2\lambda_G-\lambda_E-\varpi_E$, where $\varpi$ followed by a subscript label denotes a specific satellite \citep{Pe99}. These 2-body MMRs have low-amplitude libration, and they can be combined into an associated resonant angle $\phi_{I,E,G}$ that librates around $180^{\circ}$ with a small amplitude of $0.03^{\circ}$. $\phi_{I,E,G}$  defines the so-called Laplace resonance and it is given as \citep{Gr77}
\begin{equation}
\phi_{I,E,G}=2\lambda_G-3\lambda_E+\lambda_I, \label{laplacian}
\end{equation}
where $\lambda$ is the mean anomaly, and the subscripts $_I$, $_E$, and $_G$ refer to the satellites Io, Europa, and Ganymede, respectively.

Unlike the three innermost Galilean satellites, Callisto is not locked in a first-order mean motion resonant configuration with any other satellite. This is an important constraint to formation and also dynamical evolution models.

Formation models are also constrained by the satellites' composition. Table \ref{tab:intro} shows that the water-ice content (density) of these satellites increases (decreases) with their orbital distance to Jupiter. Io is virtually dry, Europa carries about 8\% of its mass as water-ice, Ganymede and Callisto are water-ice rich bodies with $\sim$40-50\% water-ice mass content.

The low water-ice contents of Io and Europa have been originally interpreted as evidence of formation in hot regions of the CPD disk, probably mostly inside the disk snowline -- the location in the Jovian circumplanetary disk where water condensates as ice \citep{Lu82}. Their compositions have been used as strong constraints on several formation models \citep{Mo03a,Mo03b,Ca09,Ro17,Sh19}. However, it has been more recently proposed that Io and Europa may have formed mostly from water-ice rich material (similar to Ganymede and Callisto) and lost (most of) their water. The energy deposited by the accretion of solids combined with warm temperatures in the inner parts of the CPD can lead to the formation of surface water oceans and rich water-vapor atmospheres. These water reservoirs are then rapidly lost via hydrodynamic escape, partially (or fully) drying the two innermost satellites \citep{Bi20}. Additional water loss may be driven by giant impacts during their formation and enhanced tidal heating effects caused by the Laplacian resonance \citep{Dw13,Ha20}. Different from Io and Europa, the high concentrations of water-ice in Ganymede and Callisto suggest formation in cold and volatile-rich environments of the CPD \citep{Sc04}, very likely outside the disk snowline where very limited water loss via hydrodynamic escape took place, if any at all.

Formation models have also attempted to account for expected differences in the internal structures of the Galilean satellites. Different levels of core-mantle segregation are typically associated to different accretion timescales \citep{Mo03a,Mo03b,Ca02,Ca06,Ca09,Sa10,Mi16,Ro17,Sh19}.  Io, Europa, and Ganymede are most likely fully differentiated bodies, with well distinct metallic cores and silicate mantles \citep{Sc04}. Unlikely, Callisto has been thought to be at most only partially differentiated \citep{Sc04}. The weak evidence of endogenic activity on Callisto's surface found by the Galileo mission in combination with frustrated detection of a core magnetic field by magnetic observations suggest limited differentiation \citep{Jo20}. This implies that Callisto had a relatively late and protracted accretion phase, potentially completed after the extinction of short-lived radioactive nuclei  \citep[e.g. $^{\rm 26}$Al,][]{Ba08,Ba10} -- the most likely source of heating to cause large scale ice melting and core segregation. In this school of thought, it has been proposed that Callisto formed no earlier than $\sim$3~Myr after the calcium–aluminum-rich inclusions \citep{Mc06}. Io, Europa, and Ganymede should have formed earlier than that to account for their differentiated state. However, the interior structure of Callisto is still debated. It has been more recently suggested that even in the scenario of late/protracted formation, it could be very difficult to suppress  differentiation of Callisto due to  density gradients trapping heat generated by the decay of long-lived radioisotopes \citep{OR14} during the Solar System history. The presence of non-hydrostatic pressure gradients in Callisto could also allow a complete differentiated state~\citep{Ga13}.  So, if Callisto is (fully) differentiated or not remains unclear \citep{Ro20}.

Finally, it is also important to note in Table \ref{tab:intro} that the masses of the Galilean satellites do not show any clear correlation with their orbital distance to Jupiter (e.g., no radial mass ranking), which is an important constraint to formation models \citep{Cr12}. In the next section, we review Galilean satellite formation models to motivate this work.

\subsection{The minimum mass sub-nebula model (MMSN model)}

Based on the minimum mass solar nebula model for the Solar System \citep{We77,Ha81}, \cite{Lu82} and \cite{Mo03a,Mo03b} proposed a modified version of this scenario applied to the Galilean satellites system. In these models, solids in the CPD are assumed to be in form of km-sized satellitesimals (following previous studies \citep{Mo03a,Mo03b} we use the term ``satellitesimals'' to refer to km-sized objects, precursors of the satellites in the CPD) that grow by mutual collisions up to the Galilean satellites masses. In most of these simulations, satellites are formed in very short timescales, typically of about $10^2$-$10^3$~years, which would suggest that they all should have differentiated interiors (or similar internal structure). This has been one of the issues raised against the  MMSN model because several models considered that Callisto is at most only partially differentiated \citep[e.g.][]{Ca02}. Regardless if Callisto is or not differentiated in reality, other inconsistencies between the model results and the real satellite system exist.
In addition, the masses and orbits of the simulated satellites poorly match the Galilean ones (e.g. some simulations show satellite systems with radial mass ranking).

\cite{Mi16} analyzed the formation of the Galilean satellites considering different MMSN model scenarios. They invoked a semi-analytical model to simulate migration and growth including also the effects of an inner cavity in the gas disk. This disk feature was mostly neglected in previous studies but it is crucial to avoid the dramatic loss of solids via gas drag and gas driven migration \citep{Mi16} in the disk. When a body reaches the inner disk cavity its drift and migration stop which allows other bodies migrating/drifting inwards to be captured in MMRs. This process tends to repeat and leads to the formation of a resonant chain anchored at the disk inner edge \citep[analogues to the formation of super-Earth systems; see][]{Iz17,Iz19}. The authors verified that the final period-ratio of adjacent satellites in their simulated systems better reproduce the Galilean system if the migration timescale is increased, relative to those used in \cite{Mo03a,Mo03b}. However, by invoking longer migration timescales to avoid the mentioned issue, the masses of their simulated satellites did not provide a reasonable match to the Galilean satellites~\citep{Mi16}. Finally, the semi-analytical treatment invoked in \cite{Mi16} did not allow them to precisely model the secular and resonant interaction of adjacent satellite pairs neither their growth via giant impacts.

\cite{Mo18} explored the MMSN model using N-body simulations starting from a population of satellites-embryos that are allowed to type-I migrate and grow via giant impacts. Some of their simulations were successful in producing four satellites starting from a population of $\sim$20 satellite-embryos and many relatively smaller satellitesimals, but their simulations fail to match the final masses and orbital configuration of the Galilean satellites. In their best Galilean system analogue, the innermost satellite is at 17~${\rm R_J}$  whereas Io is at 5.9~${\rm R_J}$ (see Table \ref{tab:intro}). Their simulations also show that when several satellites reach the disk inner edge, forming a long-resonant chain, they all get engulfed by Jupiter. It is not clear why this phenomenon takes place in their simulations. If satellites are successively pushed inside the disk inner cavity and eventually collide with Jupiter one-by-one, one would  expect that at least one satellite should  survive anchored at the disk inner edge at the end of this process.

\subsection{The classic gas-starved disk model (GSD model)}

One of the major differences between the MMSN disk models and the gas-starved disk model (GSD model) is that the latter invokes CPDs that are orders of magnitude lower-mass than  MMSN disks. The GSD model is probably one of the most successful early models for the origins of the Galilean satellites \citep{Ca02,Ca06,Ca09}. However, this model requires to be revisited because our paradigm of planet formation has evolved significantly in the last 10 years. 

The GSD model is built on the assumption that a semi-steady gas flows from the circumstellar disk to Jupiter's circumplanetary disk simultaneously delivers gas and solid material to the CPD with roughly solar dust-to-gas ratio composition.

The original GSD model invokes that all dust delivered to the CPD by gas in-fall coagulates and grows into satellitesimals of masses of about $5\times 10^{-7}~{\rm M_J}$. Then, satellitesimals grow to satellites by mutual collisions~\citep{Ca02,Ca06,Ca09}.
 
Although very appealing when initially proposed, this idea presents some conflicts with our current understanding of planet formation. Recent simulations show that a 10-20 Earth-mass planet, gravitationally interacting with the gas disk, creates a pressure bump outside its orbit \citep{La14,Bi18} that prevents sufficiently large dust grains in CSD from being delivered to the planet's circumplanetary disk. So, in fact, the dust-to-gas ratio in Jupiter's CPD incoming gas is expected to be lower than that in the sun's CSD, perhaps several orders of magnitude lower than the solar value \citep{Ro18,weber2018characterizing}. This view is also supported by mass-independent isotopic anomalies measured in carbonaceous and non-carbonaceous meteorites. The isotopic differences between these two classes of meteorites have been interpreted as evidence of very efficient separation of the inner and outer Solar System pebble reservoirs, potentially caused by Jupiter's formation \citep[see][]{Kr17,brasser2020partitioning}. The filtering of pebbles promoted by Jupiter would have also affected the abundance of pebbles in its own CPD \citep{Ro20}. This is a critical issue because it challenges the in-situ formation of satellites(imals) in the CPD. Satellitesimal formation via streaming instability \citep{Yo07,Si16,Ar16,Dr18} -- the favorite scenario to explain how mm-cm size dust grains grow to km-sized objects -- requires dust-to-gas ratio of at least a few percent \citep[e.g.][]{Ya17}.

Finally, the origins of satellitesimals in the CPD is probably more easily explained via capture of planetesimals or fragments (produced in planetesimal-planetesimal collisions) from the CSD. Planetesimals or fragments on eccentric orbits around the Sun may eventually cross the orbit of the growing Jupiter \citep[e.g.][]{Ra17} and get temporarily or even permanently captured in the CPD~\citep{Es06,Ca09}. This is possible because gas drag dissipative effects act to damp the orbits of these objects when they travel across the CPD~\citep{Ad76,Es06,Mo10,Fu13,DA15,Su16,Su17}. Planetesimals traveling across the CPD are also ablated and this mechanism is probably the main source of pebbles (mm-cm-sized dust grains) to the CPD \citep{Es06,Es09,Mo10,Fu13,DA15,Su16,Su17,Ro20}. The total mass in planetesimals/fragments captured and pebbles created via this process depends on  planetesimals/fragments sizes, the total mass in planetesimals/fragments, and gas density in the giant's planet region which are not strongly constrained \citep[e.g.][]{Ra17,Ro20}. Nevertheless, this scenario is very appealing because it invokes a single mechanism to explain the origins of pebbles and satellitesimals in the CPD.

\subsection{More Recent Models}

Galilean satellites' formation models have also invoked gas-drag assisted accretion of millimeter and centimeter size pebbles to account for their origin~\citep{Ro17,Sh19,Ro20}. This regime of growth is popularly known as pebble accretion \citep{Or10,La12,La14,Mo15,Le15,Bi18}

One of the key advantages of invoking pebble accretion to explain the formation of the Galilean system is that the growth of satellitesimals up to the masses of these satellites does not necessarily require a large (N$>$500-1000) population of satellitesimals to exist in the CPD -- as assumed in traditional models \citep{Mo03a,Mo03b,Ca02,Ca06,Ca09}. As already discussed, the in-situ formation of satellitesimals in the CPD may be problematic. However, if at least a handful of sufficiently massive satellitesimals exist in the CPD, pebble accretion may be efficient in forming a system with a few relatively massive final satellites.

\cite{Sh19} studied the growth of the Galilean satellites via pebble accretion. Their simulations include the effects of gas drag and type-I migration. They invoke integrated pebble fluxes in the CPD of about $1.5\times 10^{-3}~{\rm M_J}$. \cite{Sh19} set the CPD's  inner edge at  Io's current position (but see also \cite{Sa10,Mi16,Mo18}). \cite{Sh19} performed simulations considering initially 4 satellitesimals in the CPD. Satellitesimals are individually inserted in the CPD at different times, to mimic planetesimal capture from the CSD. They admittedly fine-tuned their simulations to produce systems that match well the masses, orbits, and water ice fractions of the Galilean satellites. Although this is an interesting approach, giving the large number of free parameters in the model, one of the key caveats of their scenario  is that it is built on semi-analytical calculations rather than in N-body numerical simulations. Their simulations do not account for the gravitational interaction between satellites as they accrete pebbles and migrate \citep[see also][]{Ci18}. The efficiency of pebble accretion is strongly dependent on the orbital parameters of the growing satellites ~\cite[e.g.][]{levisonetal15b}. Thus, they can not precisely assess the final architecture of their systems. One of the questions that remains unanswered  is whether or not an initial number of satellitesimals larger than four is also successful in reproducing the Galilean system. This is one of the questions we try to answer in this paper. We also advance to the reader that all our simulations starting initially with 4 satellitesimals produced less than 4 satellites at the end. 

In a recent study, \cite{Ba20}  proposed that the Galilean satellites' formation occurred in a vertically-fed CPD disk  that spreads viscously outwards (the vertically averaged radial gas velocity is $v_r >0$ everywhere in the CPD). In their model, satellitesimals in the CPD grow to satellites in an oligarchic growth fashion -- via satellitesimal-satellitesimal collisions -- rather than via pebble accretion. Numerical  simulations  in \cite{Ba20} start from planetesimals with masses of $4\times 10^{-7}~{\rm M_J}$ (4 times more massive than the initial masses  considered in this paper). The authors find that in their disk model pebble accretion can be simply neglected, which is not necessarily the case for other disk models \citep[e.g.][]{Ro17,Ro20}. Their model successfully explains some characteristics of the Galilean system, as the overall masses of the Galilean satellites and the Laplacian resonance, however, it remains to be demonstrated that the total dust reservoir assumed in their model can in fact settle into the CPD disk mid-plane to promote efficient satellitesimal formation via some sort of gravitational-hydrodynamic instability \citep{Ba20}.

In this work, we use N-body numerical simulations to model the formation of the Galilean satellites in a GSD-style circumplanetary disk. Motivated by previous studies,  the flux of pebbles assumed in our simulations is consistent with pebble fluxes estimated via ablation of planetesimals entering the circumplanetary disk \citep{Ro20}. The initial total number of satellitesimals in the CPD is not strongly constrained,  so in our simulations, we test 4, 30, and 50 satellitesimals. Our study represents a further step towards the understanding of the origins of the Galilean satellites because previous studies modeling their formation via pebble accretion have typically invoked simple semi-analytical models that neglect the mutual interaction of the satellites when they grow and migrate in the disk. Here, we self-consistently model the growth and mutual dynamical interaction of satellitesimals allowing also for growth via giant collisions.

The model that we propose here has the very same basic ingredients invoked in models for the formation of the so-called close-in super-Earths and giant planets around other stars \citep[e.g.][]{Iz19,Bi19,La19} -- namely pebble accretion and migration. Typical close-in super-Earths have masses of $\sim10 ^{-5}~{\rm M_{star}}$. Interestingly, the mass ratio of individual Galilean satellites and Jupiter is also $\sim10^{-5}$. The resonant dynamical architecture of the Galilean satellites also recalls that of some super-Earths systems \citep[e.g. Kepler-223 system;][]{Mib16}. So if one can explain the formation of these both types of systems via the same processes it would be reassuring. 

The structure of this paper is as follows. In Section~\ref{sec:methods} we describe the methods used in this work. In Section~\ref{sec:simulation} we describe our simulations, and in Section~\ref{sec:results} we present the main results. In Section~\ref{sec:faketides} we analyze the long-term evolution of our systems. We discuss our results and model in Section~\ref{sec:discussion}. Finally, we summarize our main findings in Section~\ref{sec:conclusions}.

\section{Methods}\label{sec:methods}

Our numerical simulations were performed using an adapted version of the MERCURY package \citep{Ch99} including artificial forces to mimic the effects of the gas disk. These forces are: 1) gas drag; 2) type I migration, and eccentricity and inclination damping (Section \ref{subsec:ge}). Our pebble accretion prescription is described in details in Section \ref{subsec:pa}. Satellitesimals are allowed to grow via pebble accretion and collisions. Collisions are modeled as perfect merging events that conserve mass and linear momentum.

\subsection{Circumplanetary disk model}\label{subsec:cdm}

We assume that as Jupiter grows via runaway gas accretion,  opens a deep gap in the circumstellar disk, and a disk mostly composed by gas forms around its equator. The circumplanetary disk is continually supplied by the in-fall of material from the CSD. Assuming a semi-steady flow of gas and balance between the in-fall of material from the CSD and the mass accretion rate onto Jupiter, \cite{Ca02} obtained that the radial surface density of gas is given by
\begin{equation}
\Sigma_g(r)=\frac{\dot{\rm M}_{\rm g}}{3\pi\nu(r)}\left\{\begin{array}{ll} 1-\frac{4}{5}\sqrt{\frac{R_c}{R_d}}-\frac{1}{5}\left(\frac{r}{R_c}\right)^2 & \textrm{for}~r\leq R_c \\ \frac{4}{5}\sqrt{\frac{R_c}{r}}-\frac{4}{5}\sqrt{\frac{R_c}{R_d}} & \textrm{for}~r>R_c,\end{array}\right.
\end{equation}
where $R_c$ and $R_d$ are the centrifugal and outer radius of the disk, respectively, $\dot{M}_g$ the mass in-fall rate, and $\nu$ the turbulent viscosity. Note that our disk model is qualitatively consistent with the delivery of gas to the CPD via meridional circulation of gas near the planet's gap \citep{Ta12,Mo14,Sz14,Sc20,Ba20}.  In our model, we assume that the gas is deposited into the CPD midplane at around the centrifugal radius ($R_c$), and then spreads viscously~\citep{Ca02}. $R_c$ corresponds to the location in the CPD where the angular momentum of the inflowing material is equal to the Keplerian angular momentum. $R_c$ is treated as a free parameter in our model~\cite[see also][]{Ba20}.

The mass in-fall rate from the CSD on the CPD in our simulations was set as \citep{Sa10,Ro17}
\begin{equation}
\dot{\rm M}_{\rm g}=10^{-7}e^{-\frac{t}{\tau_d}}~{\rm M_J}{\rm /yr},
\end{equation}
where $t$ is the time and $\tau_d= 1.0$~Myr is the disk decay timescale \citep{Ca02}. For simplicity, in all our simulations, we neglect Jupiter's growth via gas accretion and set its mass as the current one. 

We set the centrifugal radius at the fixed distance $R_c=26~{\rm R_J}$ \citep{Ro17} and the outer edge of the disk at $R_d=150~{\rm R_J}$, based on the results of hydrodynamic simulations \citep{Ta12}. The interaction of the CPD with Jupiter's magnetic field tends to slow down the planet's rotation and promotes the formation of an inner disk cavity. \cite{Ba18} found that magnetic effects dominate the gas dynamics in the inner regions of the CPD up to $4-5~{\rm R_J}$. Motivated by this result, we follow \cite{Iz17} and impose a  disk inner edge at $R_i=5~{\rm R_J}$ \citep{Ba20} in our CPD  by re-scaling the gas surface density by
\begin{equation}
\mathcal{R}=\tanh\left(\frac{r-R_i}{0.05R_i}\right)
\end{equation}

We assumed the standard $\alpha$-viscosity prescription to represent the disk viscosity \citep{Sh73}
\begin{equation}
\nu=\alpha_zc_sH_g,
\end{equation}
where $\alpha_z=10^{-3}$ \citep{Ro17} is the coefficient of turbulent viscosity, $c_s$ is the isothermal sound speed and $H_g$ the gas scale height ($H_g=c_s/\Omega_k$, where $\Omega_k$ is the keplerian orbital frequency). For a CPD in hydrostatic equilibrium, the sound speed relates to the disk temperature $T(r)$ as $c_s^2=2.56\times 10^{23}~g^{-1}~k_bT$, where $k_b$ is the Boltzmann constant \citep{Ha81}. The snowline is initially located  at $r\sim 14.5~{\rm R_J}$ and the radial temperature profile is given by \citep{Ro17}
\begin{equation}
T=225\left(\frac{r}{10~{\rm R}_{\rm J}}\right)^{-3/4}K \label{eq:Tprofile}
\end{equation}

\begin{figure}
\subfigure[]{\includegraphics[width=\columnwidth]{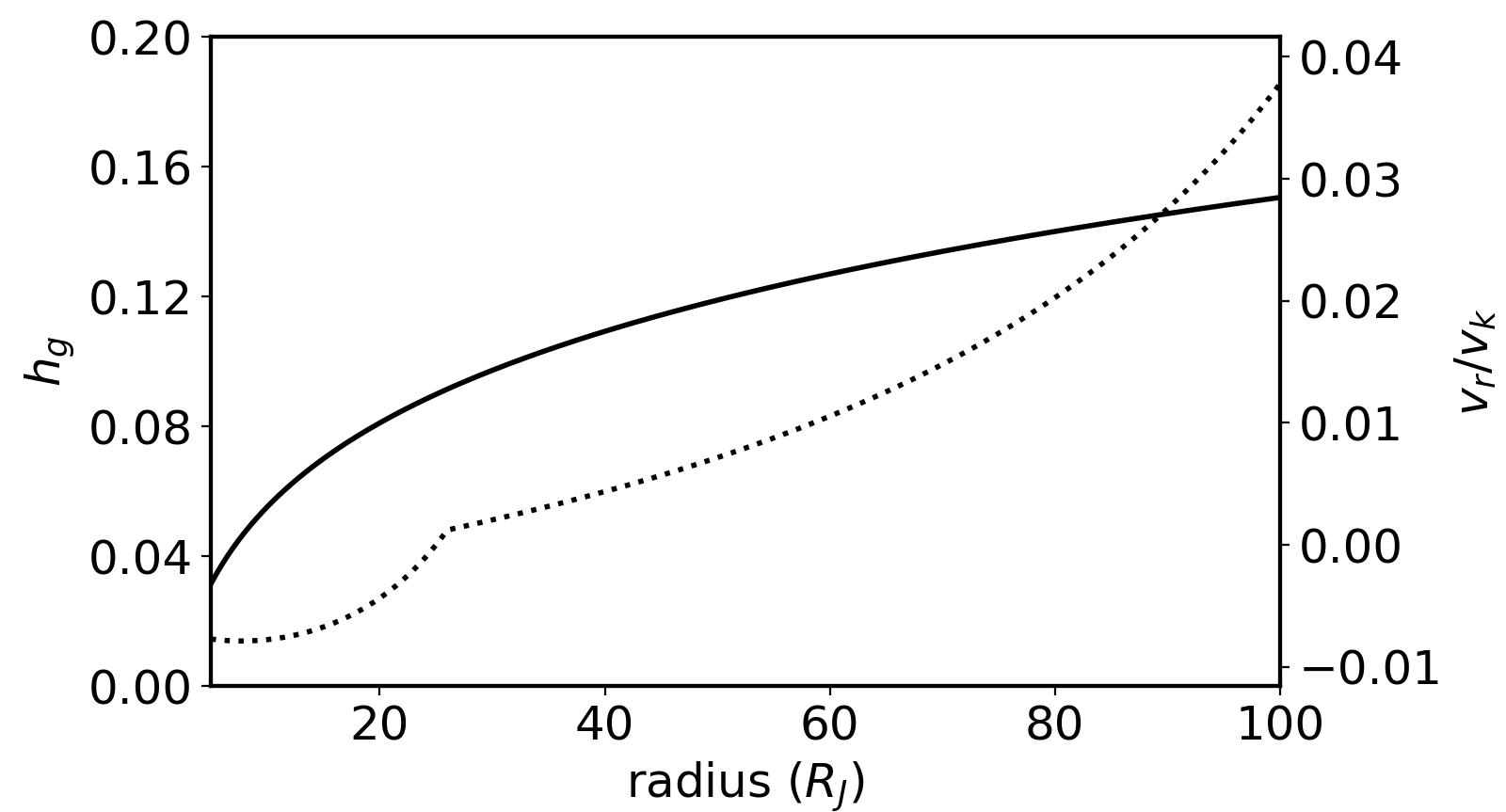} \label{fig:hvrad}}
\subfigure[]{\includegraphics[width=\columnwidth]{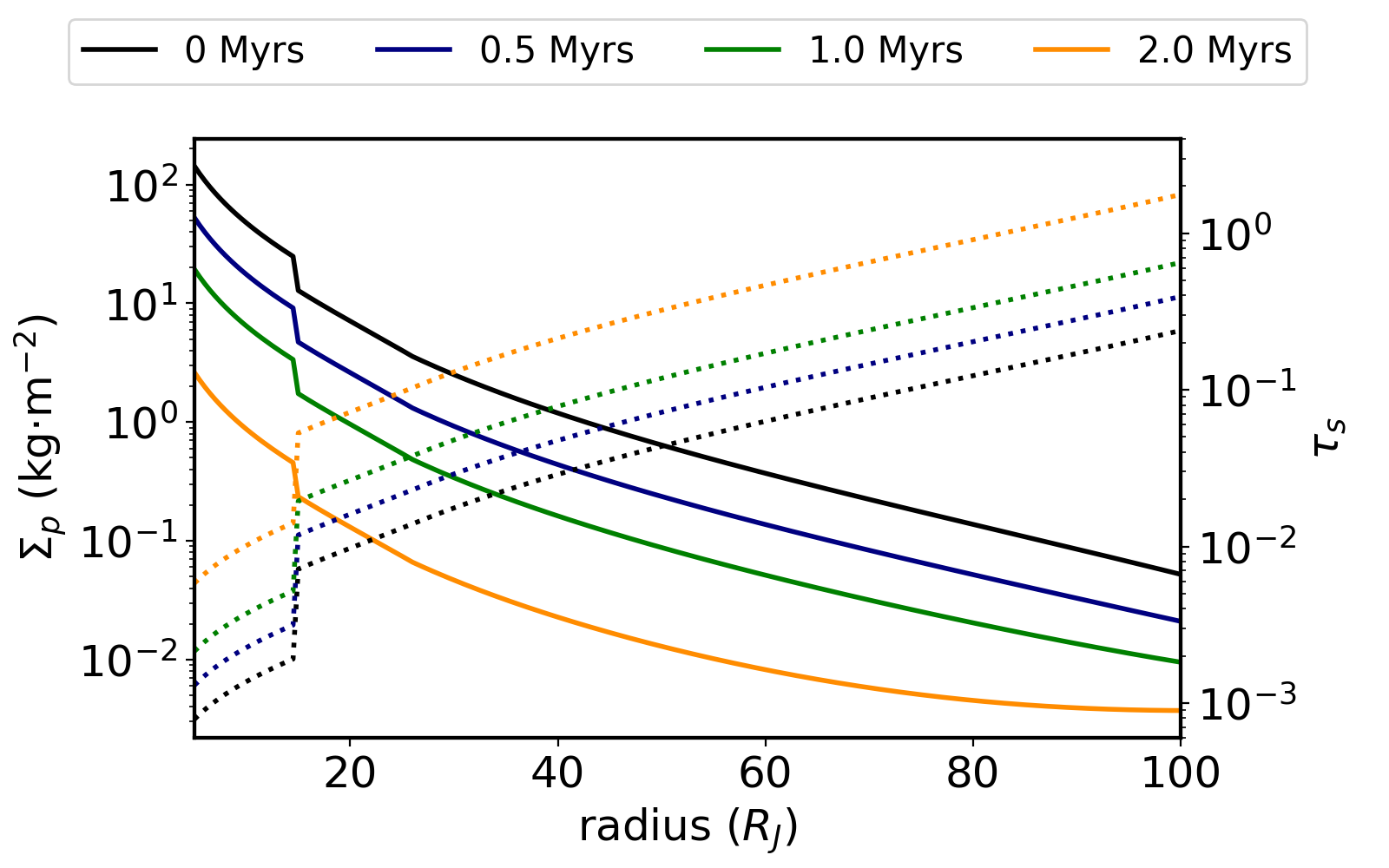} \label{fig:sigmapeb}}
\subfigure[]{\includegraphics[width=\columnwidth]{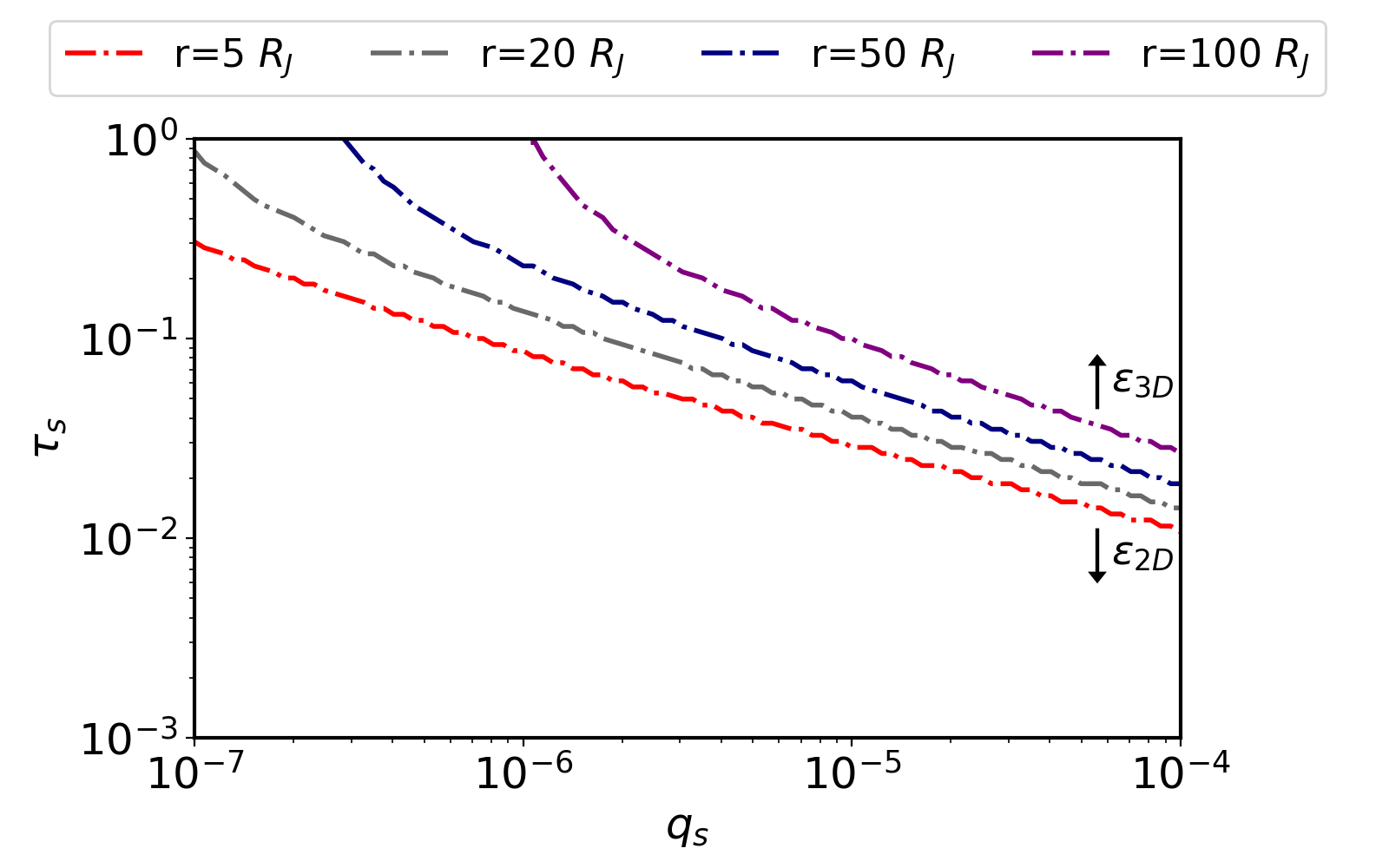} \label{fig:e2d3d}}
\caption{(a) CPD aspect ratio ($h_g=H_g/r$, solid line) and vertically averaged gas radial velocity normalized by the keplerian velocity ($v_r/v_k$, dotted line); (b) pebble surface density (solid lines) and Stokes number (dotted lines) as a  function of the distance to the planet. Each color shows different times: 0~Myr (black), 0.5~Myr (blue), 1.0~Myr (green), and 2.0~Myr (orange); (c) Threshold curves of 2D and 3D pebble accretion regimes for different satellite masses and Stokes numbers. The different colored lines correspond to different locations of the disk: 5~${\rm R_J}$ (red), 20~${\rm R_J}$ (gray), 50~${\rm R_J}$ (navy blue), and 100~${\rm R_J}$ (purple). The region above (below) of each curve corresponds to the region where the 3D (2D) accretion efficiency is higher than the 2D (3D) one (see Eq.~\ref{eftotal}). The initial pebble flux is $\dot{\rm M}_{\rm p0}=1.5\times 10^{-9}~{\rm M_J}$/yr. The integrated pebble flux over time is $10^{-3}$~M$\mathrm{_J}$. \label{fig:disk}}
\end{figure}

Figure~\ref{fig:hvrad} shows the CPD aspect ratio ($h_g=H_g/r$, solid line) and ratio between the vertically averaged gas radial and keplerian velocities ($v_r/v_k$, dotted line). In our disk model, the gas inside  $\sim R_c$ flows inwards whereas gas outside $\sim R_c$ flows outwards \citep[see also][]{Ba20}, as one can note in Figure~\ref{fig:hvrad}.

\subsection{Gas effects}\label{subsec:ge}

Our simulations start with satellitesimals of masses of  $m\sim 10^{-7}~{\rm M_J}$ and bulk density of ${\rm 2~g/cm^3}$. We have verified that in our disk setup gas-drag plays an important role in the dynamics of satellitesimals of these sizes \cite[see also][]{Mi16}. Thus,  our model includes both the effects of gas drag and satellitesimal-disk gravitational interactions. We now describe how we model these effects.

\subsubsection{Gas Drag}

The CPD around Jupiter rotates at sub-Keplerian speed because it is pressure supported. Small satellitesimals in the CPD orbiting at keplerian speed feel an strong headwind, lose energy, and tend to spiral inwards. The azimuthal gas disk velocity is given by 
\begin{align}
v_g=(1-\eta)v_k, \label{velgas}
\end{align}
where $v_k=\Omega_kr$ is the Keplerian velocity and ${\rm \eta}$ characterizes the sub-Keplerian velocity of the gas disk. It is given by 
\begin{equation}
\eta=-\frac{h_g^2}{2}\frac{\partial \ln{c_s^2\rho_g}}{\partial \ln{r}},
\end{equation}\label{eq:eta}

Setting $\rho_g=\int_{-\infty}^\infty \Sigma_g dz$, the gas volumetric density $\rho_g$ is given by \citep{We77}
\begin{equation}
\rho_g=\frac{1}{\sqrt{2\pi}}\frac{\Sigma_g}{H_g}e^{-z^2/2H_g^2}
\end{equation}
where $z$ is the reference-frame vertical component. 

The gas-drag acceleration on a body of radius $R_s$ and bulk density $\rho_s$  is given by \citep{Ad76}
\begin{equation}
\vec{a}_{gd}=-\frac{3}{8}\frac{C_d\rho_g v_{rel}}{\rho_sR_s}\vec{v}_{rel}
\end{equation}
where $C_d$ is the drag coefficient, $\vec{v}_{rel}$ is velocity of the body with respect to the gas.

The drag coefficient computes the intensity of the interaction between the gas and satellitesimal, and it is given by \citep{Br07}
\begin{equation}
C_d=\left\{ \begin{array}{ll} 2 & \mathcal{M}\geq 1\\ 0.44+1.56\mathcal{M}^2 & \mathcal{M}<1,~Re\geq 10^3\\ \frac{24(1+0.15Re^{0.687})}{Re} & \mathcal{M}<1,~Re<10^3, \end{array}\right. 
\end{equation}
where $\mathcal{M}=v_{rel}/c_s$ is the Mach number and $Re$ the Reynolds number written as
\begin{equation}
Re\approx 2.66\times 10^8\rho_g R_s\mathcal{M}.
\end{equation}

\subsubsection{Type-I migration}
Sufficiently large satellitesimals interact gravitationally with the CPD launching spiral density waves that transport angular momentum. This interaction tends to promote eccentricity and inclination damping of orbits and radial migration. Satellites with masses of the order of the Galilean satellites ($m\sim 10^{-5}~{\rm M_J}$) or lower mass are not expected to open gaps in our CPD disk and migrate in the type-I regime \citep{Ca06}. The total type-I migration torque consists of contributions from the Lindblad and Co-rotational torques \citep{Pa10,Pa11}.

The surface density  and  temperature  gradient profiles, $x$ and $\beta$, respectively,  are given by 
\begin{equation}
x=-\frac{\partial\ln{\Sigma_g}}{\partial\ln{r}}
\end{equation}
and
\begin{equation}
\beta=-\frac{\partial\ln{T}}{\partial\ln{r}}.
\end{equation}
The scaling torque at the satellite's location is given by \citep{Cr08} 
\begin{equation}
\Gamma_0=\left(\frac{q_s}{h_g}\right)^2\Sigma_gr^4\Omega_k^2,
\end{equation}
where $q_s=m/{\rm M_J}$ is the satellite-Jupiter mass ratio.

The Lindblad torque felt by a satellite on  circular and coplanar orbit is parameterized as \citep{Pa10,Pa11}
\begin{equation}
\Gamma_L=(-2.5-1.5\beta+0.1x)\frac{\Gamma_0}{\gamma_{\textrm{eff}}},
\end{equation}
where $\gamma_{\textrm{eff}}$ is the effective adiabatic index defined by \citep{Pa10,Pa11}
\begin{equation}
\gamma_{\textrm{eff}}=\frac{2\gamma Q}{\gamma Q+\frac{1}{2}\sqrt{2\psi+2\gamma^2 Q^2-2}},
\end{equation}
\begin{equation}
\psi=\sqrt{(\gamma^2 Q^2+1)^2-16Q^2(\gamma-1)},    
\end{equation}
and
\begin{equation}
Q=\frac{2\chi}{3h_g^3r^2\Omega_k}
\end{equation}
where $\gamma$ is the adiabatic index assumed as $\gamma=1.4$ in this work. The thermal diffusion coefficient depends on the opacity $\kappa$ and  is given by \citep{Pa10,Pa11}
\begin{equation}
\chi=\frac{16\gamma(\gamma-1)k_BT^4}{3\kappa\rho_g^2(h_gr)^2\Omega_k^2}.    
\end{equation}
The CPD disk opacity is calculated as in \cite{belllin94}.

A satellite on circular and coplanar orbit experiences a corotation torque that reads as \citep{Pa10,Pa11}
\begin{equation}
\begin{split}
\Gamma_C=&\left[\frac{7.9\xi}{\gamma_{\textrm{eff}}}F(p_\nu)G(p_\nu)+0.7\left(\frac{3}{2}-x\right)(1-K(p_\nu))+\right.\\ 
&+\left(2.2\xi-\frac{1.4\xi}{\gamma_{\textrm{eff}}}\right)F(p_\nu)F(p_\chi)\sqrt{G(p_\nu)G(p_\chi)}\\
&+\left.1.1\left(\frac{3}{2}-x\right)\sqrt{(1-K(p_\nu))(1-K(p_\chi))}\right]\frac{\Gamma_0}{\gamma_{\textrm{eff}}},
\end{split}
\end{equation}
where $\xi=\beta-(\gamma-1)x$, and $p_\nu$ and $p_\chi$ are parameters that measure the viscous and thermal saturations of the co-orbital torque, respectively. $F$, $G$ and $K$ are functions of these parameters.

The parameters $p_\nu$ and $p_\chi$ are given by \citep{Pa10,Pa11}
\begin{equation}
p_\nu=\frac{2}{3}\sqrt{\frac{r^2\Omega_k}{2\pi\nu}\left(\frac{1.1}{\gamma_{\textrm{eff}}^{1/4}}\sqrt{\frac{q_s}{h_g}}\right)^3}
\end{equation}
and
\begin{equation}
p_\chi=\frac{2}{3}\sqrt{\frac{r^2\Omega_k}{2\pi\chi}\left(\frac{1.1}{\gamma_{\textrm{eff}}^{1/4}}\sqrt{\frac{q_s}{h_g}}\right)^3}
\end{equation}

The functions $F$, $G$ e $K$ are \citep{Pa10,Pa11}
\begin{equation}
F(p)=\frac{1}{1+\left(\frac{p}{1.3}\right)^2},
\end{equation}
\begin{equation}
G(p)=\left\{ \begin{array}{ll} \frac{16}{25}\left(\frac{45\pi}{8}^{3/4}p^{3/2}\right) & p<\sqrt{\frac{8}{45\pi}}\\
 1-\frac{9}{25}\left(\frac{8}{45\pi}^{4/3}p^{-8/3}\right) & p\geq\sqrt{\frac{8}{45\pi}}  \end{array}\right.,
\end{equation}
and
\begin{equation}
 K(p)=\left\{ \begin{array}{ll} \frac{16}{25}\left(\frac{45\pi}{28}^{3/4}p^{3/2}\right) & p<\sqrt{\frac{28}{45\pi}}\\
 1-\frac{9}{25}\left(\frac{28}{45\pi}^{4/3}p^{-8/3}\right) & p\geq\sqrt{\frac{28}{45\pi}}  \end{array}\right. 
\end{equation}

The torques $\Gamma_L$ and $\Gamma_C$ need to be modified to account for satellitesimals/satellites on eccentric or inclined orbits. The Lindblad torque is reduced by factor $\Delta_L$ and the corotation torque by $\Delta_C$ \citep{Cr08,Co14,Fe13} given by
\begin{equation}
\begin{split}
\Delta_L=&\left\{P_e+\frac{P_e}{|P_e|}\left[0.07\left(\frac{i_s}{h_g}\right)+0.085\left(\frac{i_s}{h_g}\right)^4\right.\right.\\
&\left.\left.-0.08\left(\frac{e_s}{h_g}\right)\left(\frac{i_s}{h_g}\right)^2\right]\right\}^{-1}
\end{split}
\end{equation}
and
\begin{equation}
\Delta_C=\textrm{exp}\left(-\frac{e_s}{0.5h_g+0.01}\right)\left[1-\tanh{\left(\frac{i_s}{h_g}\right)}\right]
\end{equation}
where
\begin{equation}
P_e=\frac{1+\left(\frac{e_s}{2.25h_g}\right)^{1.2}+\left(\frac{e_s}{2.84h_g}\right)^{6}}{1-\left(\frac{e_s}{2.02h_g}\right)^{4}}
\end{equation}

Finally, we can write the total torque associated to the type I migration as 
\begin{equation}
\Gamma=\Delta_L\Gamma_L+\Delta_C\Gamma_C
\end{equation}
and the acceleration $\vec{a}_m$ of a satellite with angular momentum $L$ is \citep{Cr08} 
\begin{equation}
\vec{a}_m=\frac{\vec{v}}{L}\Gamma
\end{equation}

The gas disk also damps  eccentricity and inclination of sufficiently massive bodies on timescales given by \citep{Cr08}
\begin{equation}
\begin{split}
t_e=&\frac{t_{wv}}{0.780}\left[1-0.14\left(\frac{e_s}{h_g}\right)^2+0.06\left(\frac{e_s}{h_g}\right)^3\right.\\
&\left.+0.18\left(\frac{e_s}{h_g}\right)\left(\frac{i_s}{h_g}\right)^2\right]
\end{split}
\end{equation}
and
\begin{equation}
\begin{split}
t_i=&\frac{t_{wv}}{0.544}\left[1-0.30\left(\frac{i_s}{h_g}\right)^2+0.24\left(\frac{i_s}{h_g}\right)^3\right.\\
&\left.+0.14\left(\frac{e_s}{h_g}\right)^2\left(\frac{i_s}{h_g}\right)\right]
\end{split}
\end{equation}
where $t_{\rm wv}$ is the wave timescale \citep{Cr08}
\begin{equation}
t_{\rm wv}=\left(\frac{M*}{q_s\Sigma_ga_s^2}\right)\frac{h_g^4}{\Omega_k},
\end{equation}
$a_s$ is the satellite/satellitesimal semi-major axis.

The accelerations experienced by the bodies due to the eccentricity and inclination damping are \citep{Cr08}
\begin{equation}
\vec{a}_e=-2\frac{(\vec{v}\cdot\vec{r})\vec{r}}{r^2t_e}
\end{equation}
and
\begin{equation}
\vec{a}_i=-2\frac{(\vec{v}\cdot\hat{z})\hat{z}}{t_i}
\end{equation}
\subsection{Pebble Accretion}\label{subsec:pa}

The pebble surface density $\Sigma_p$ in our model is \citep{Sh19}
\begin{equation}
\Sigma_p=\frac{\dot{\rm M}_{\rm p}}{4\pi \tau_s\eta v_kr}(1+\tau_s^2),    
\end{equation}
where $\tau_s$ is the Stokes number and $\dot{\rm M}_{\rm p}$ is the pebble mass flux given by $\dot{\rm M}_{\rm p}=\dot{\rm M}_{\rm p0}e^{-\frac{t}{\tau_d}}$ where $\dot{\rm M}_{\rm p0}$ is a free scaling parameter that  we take to vary between $\dot{\rm M}_{\rm p0}=10^{-9}~{\rm M_J}$/yr and $5\times 10^{-9}~{\rm M_J}$/yr. Thus the integrated pebble fluxes in our simulations vary between $8\times 10^{-4}~{\rm M_J}$ and $4\times 10^{-3}~{\rm M_J}$. These values are consistent with the pebble fluxes estimated via ablation of  planetesimals \citep{Ro20}. 

The Stokes number of a pebble with a physical radius $R$ and bulk density $\rho$ is given by \citep{La12} 
\begin{equation}
\tau_s=\frac{\sqrt{2\pi}\rho R}{\Sigma_g}
\end{equation}
In this work, we assumed a bi-modal population of pebbles in the CPD. 
In our CPD the snowline is at $\sim 14.5~{\rm R_J}$. Pebbles outside the snowline are assumed to have sizes of $R=1.0$~cm and $\rho=2.0$~g/cm$^3$. As pebbles drift inwards via gas drag they eventually cross the CPD snowline. At this location, we reduce the pebble flux by a factor of 2 to account for the sublimation of the ice-pebble component. Ice pebbles sublimate at the snowline releasing small silicate dust grains. To account for this effect, pebbles inside the snowline have sizes of $R=0.1$~cm and $\rho=5.5$~g/cm$^3$. Figure~\ref{fig:sigmapeb} shows the pebble surface density evolution and the Stokes number in  function of the radial distance, in units of $R_{\rm J}$.
The pebble accretion rate is a product of the pebble flux and the total accretion efficiency $\epsilon$ \citep{ida16}. Following \cite{Li18} and \cite{Or18}, the 2D accretion efficiency in the settling regime is given by
\begin{equation}
\epsilon_{2d,set}=0.32\sqrt{\frac{q_s}{\tau_s\eta^2}\left(\frac{\Delta V}{v_k}\right)}f_{\rm set} 
\end{equation}
where $\Delta V$ is the relative velocity between pebble and satellite and $f_{\rm set}$ is a function that fits well the transition between different accretion regimes.

The relative velocity of a satellitesimal on a circular orbit and a pebble is \citep{Li18}
\begin{equation}
V_{\rm cir}=\left[1+5.7\left(\frac{q_s\tau_s}{\eta^3}\right)\right]^{-1}\eta v_k+0.52(q_s\tau_s)^{1/3}v_k
\end{equation}
while the relative velocity of satellite on a non-circular but coplanar orbits reads \citep{Li18}
\begin{equation}
V_{\rm ecc}=0.76e_sv_k
\end{equation}
One can write the relative velocity between a pebble and a satellite in both regimes as \citep{Li18,Or18}
\begin{equation}
\Delta V_y=\textrm{max}(V_{\rm cir},V_{\rm ecc})
\end{equation}

The relative velocity between a pebble and a satellite on a slightly inclined orbit can be written as $\Delta V_z=0.68i_sv_k$. Finally, the total relative velocity becomes \citep{Li18,Or18}
\begin{equation}
\Delta V=\sqrt{\Delta V_y^2+\Delta V_z^2}
\end{equation}

The transition function $f_{\rm set}$ is given by \citep{Or18}
\begin{equation}
\begin{split}
f_{\rm set}=&\exp\left[-0.5\left(\frac{\Delta V_y^2}{V^2_*}+\frac{\Delta V_z^2}{V_*^2+0.33\sigma_{pz}^2}\right)\right]\\
&\times\frac{V_*}{\sqrt{V_*^2+0.33\sigma_{pz}^2}}
\end{split}
\end{equation}
where $V_*$ is the transition velocity
\begin{equation}
V_*=\sqrt[3]{\frac{q_s}{\tau_s}}v_k
\end{equation}
and $\sigma_{pz}$ is the turbulent velocity in the z-direction \citep{Yo07}:
\begin{equation}
\sigma_{pz}=\sqrt{\frac{\alpha_z}{1+\tau_s}}\left(1+\frac{\tau_s}{1+\tau_s}\right)^{-1/2}h_gv_k
\end{equation}

The pebble disk aspect ratio is given by \citep{Yo07}
\begin{equation}
h_p=\sqrt{\frac{\alpha_z}{\alpha_z+\tau_s}}\left(1+\frac{\tau_s}{1+\tau_s}\right)^{-1/2}h_g,
\end{equation}
and the pebble volume density is \citep{Or18}
\begin{equation}
\rho_p=\frac{\Sigma_p}{\sqrt{2\pi}rh_p}
\end{equation}

According to \cite{Li18} and \cite{Or18}, the tridimensional pebble accretion efficiency in the settling regime is 
\begin{equation}
\epsilon_{\rm 3d,set}=0.39\frac{q_s}{\eta h_{\textrm{eff}}}f_{\rm set}^2    
\end{equation}
where $h_{\textrm{eff}}$ is the effective aspect ratio of pebbles in relation to a satellite in inclined orbit \citep{Or18}
\begin{equation}
h_{\textrm{eff}}=\sqrt{h_p^2+\frac{\pi i_s^2}{2}\left[1-\exp\left(-\frac{i_s}{2h_p}\right)\right]}
\end{equation}

The accretion efficiency in the 2D and 3D ballistic regimes are \citep{Li18,Or18}
\begin{equation}
\epsilon_{\rm 2d,bal}=\frac{R_s}{2\pi\tau_s\eta r}\sqrt{\frac{2q_s r}{R_s}\left(\frac{\Delta V}{v_k}\right)^2}(1-f_{\rm set})
\end{equation}
and
\begin{equation}
\epsilon_{\rm 3d,bal}=\frac{1}{4\sqrt{2\pi}\eta\tau_sh_p}\left(2q_s\frac{v_k}{\Delta V}\frac{R_s}{r}+\frac{R_s^2}{r^2}\frac{\Delta V}{v_k}\right)(1-f_{\rm set}^2)
\end{equation}

Finally, the total pebble accretion efficiency is
\begin{equation}
\epsilon=\frac{f_{\rm set}}{\sqrt{\epsilon_{\rm 2d,set}^{-2}+\epsilon_{\rm 3d,set}^{-2}}}+\frac{1-f_{\rm set}}{\sqrt{\epsilon_{\rm 2d,bal}^{-2}+\epsilon_{\rm 3d,bal}^{-2}}}  \label{eftotal}
\end{equation} 

Figure~\ref{fig:e2d3d} shows the curves $\epsilon_{2d}=\epsilon_{3d}$ for different Stokes number and satellite's masses. Each curve samples the accretion regime at one specific location of the disk. Regions above each curve corresponds to regions where the 3D accretion rate is higher than the 2D rate. Below each curve, it is the opposite.

When a satellite reaches the isolation mass M$_{\rm iso}$, its Hill radius becomes greater than the disk height creating a pressure bump that deflects gas and pebbles. If the satellite reaches M$_{\rm iso}$ the pebble accretion breaks off and the satellite grow only by impacts.
The pebble isolation mass in ${\rm M_J}$ is given by \citep{At18}
\begin{equation}
\begin{split}
{\rm M_{iso}}=&h_g^3\sqrt{37.3\alpha_z+0.01}\left\{1+0.2\left(\frac{\sqrt{\alpha_z}}{h_g}\sqrt{\frac{1}{\tau_s^2}+4}\right)^{0.7}\right\} \label{miso}
\end{split}
\end{equation}

\section{Simulations}\label{sec:simulation}

We have performed 120 simulations considering different pebble fluxes and the initial number of satellitesimals. The initial number of satellitesimals in the CPD is poorly constrained and it may also increase in time if additional planetesimals are captured from the CSD as the system evolves \citep{Mo10,Ro20}. We decided to neglect the capture of new planetesimals after our simulation starts by assuming that only sufficiently early captured planetesimals would successively grow by pebble accretion. We have performed simulations starting with 4, 30, and 50 satellitesimals in the CPD. While it would be ideal to systematically test the effects of different initial number of satellitesimals in our model, our  N-body simulations are computationally expensive what limits our approach. Our simulations starting with 4 satellitesimals are designed to test the scenario proposed by \cite{Sh19} using self-consistent N-body simulations. Satellitesimals are initially distributed between 20~${\rm R_J}$ and 120~${\rm R_J}$. Their initial masses are set $10^{-7}~{\rm M_J}$  and bulk densities $\rho=2.0$~g/cm$^3$ which are consistent with the typical masses/sizes of planetesimals formed via streaming instability \citep[$R\sim250$~km,][]{Jo14a,Ar16,Si16a}. Satellitesimals are initially separated from each other by  5 to 10  mutual Hill radii $R_H$ \citep[e.g.][]{Ko00,Kr14}
\begin{equation}
R_H=\frac{a_i+a_j}{2}\left(\frac{2\times 10^{-7}}{3}\right)^{1/3}   
\end{equation}
where $a_i$ and $a_j$ are the semi-major axes of a pair of adjacent satellitesimals. They are set initially on nearly circular and coplanar orbits ($e\leq 10^{-4}$ and $I\leq 10^{-4}$). Other angular orbital elements are randomly and uniformly selected between 0 and 360 deg. 

Our simulations are numerically integrated  for 2~Myr, considering the gas disk effects. In a few cases, to evaluate the long-term dynamical stability of our final systems, we extended our simulations up to $\sim$10~Myr, assuming that the gaseous circumplanetary disk dissipates at 2~Myr. For simplicity, we relate the pebble flux in our simulations to the gas accretion flow in the CPD \citep[e.g.][]{Sh19}. We performed simulations with $\dot{\rm M}_{\rm p0}=10^{-9}~{\rm M_J}$/yr, $1.5\times 10^{-9}~{\rm M_J}$/yr, $3\times 10^{-9}~{\rm M_J}$/yr, and $5\times 10^{-9}~{\rm M_J}$/yr. The integrated pebble flux from the lowest pebbles flux to the highest one are $8\times 10^{-4}~{\rm M_J}$,  $10^{-3}~{\rm M_J}$,  $3\times 10^{-3}~{\rm M_J}$, and $4\times 10^{-3}~{\rm M_J}$.
In these simulations, we have neglected the evolution of the CPD's temperature, but the gas surface dissipates exponentially with an e-fold timescale of 1~Myr.

\subsection{Constraining our model}\label{sec:constraints}

Our model is strongly constrained by key features of the Galilean satellites system as the number of satellites, orbital configuration, final masses, and compositions. To evaluate how our simulations match the Galilean satellites system, we define a list of relatively generous constraints. A Galilean system analogue must satisfy the following conditions: 
\begin{itemize}
    \item[{\bf i})] the final system must contain at least four satellites;
    \item[{\bf ii})] the two innermost satellite-pairs must be locked in a 2:1 MMR;
    \item[{\bf iii})] the individual masses of all satellites must be between  0.8${\rm M_E}$ and 1.2${\rm M_G}$, where ${\rm M_E}$ and ${\rm M_G}$ are the masses of Europa and Ganymede, respectively;
    \item[{\bf iv})] the two outermost satellites must have water-ice rich compositions ($>$0.3 water mass fraction).
\end{itemize}

Although Io is extremely water-ice depleted today and Europa contains only $\sim$8\% of its mass as water-ice (Table \ref{tab:intro}), we do not consider these estimates to be strong constraints to our model, because it is possible that these satellites formed with water richer compositions and lost all/most of their water \citep{Bi20}. Also, we stress that Callisto is not locked in a first-order mean motion resonance with Ganymede today. We also do not consider this observation a critical constrain to our model because Callisto may have left the resonance chain via divergent migration due to Jupiter-satellites tidal interactions \citep{Fu16,downey2020inclination}. We will further discuss these two issues later in the paper.

\section{Results} \label{sec:results}

We start by presenting the results of our sets of simulations considering initially 4 satellitesimals in the CPD, inspired by  simulations of \cite{Sh19}. In \cite{Sh19}, satellitesimals are assumed to ``appear'' in the CPD successively, at very specific times. In our simulations, they are assumed to appear simultaneously at the beginning of the simulation in arbitrarily selected positions. Note that, in our case, the four satellitesimals start about 20~${\rm R_J}$ from each other, which roughly mimics their approach. By disposing our 4 satellitesimals initially fairly apart from each other, we avoid early collisions and allow them to grow at least one order of magnitude in mass before they start to strongly interact with each other (which may affect the efficiency of pebble accretion).

\begin{figure}
\subfigure[]{\includegraphics[width=0.95\columnwidth]{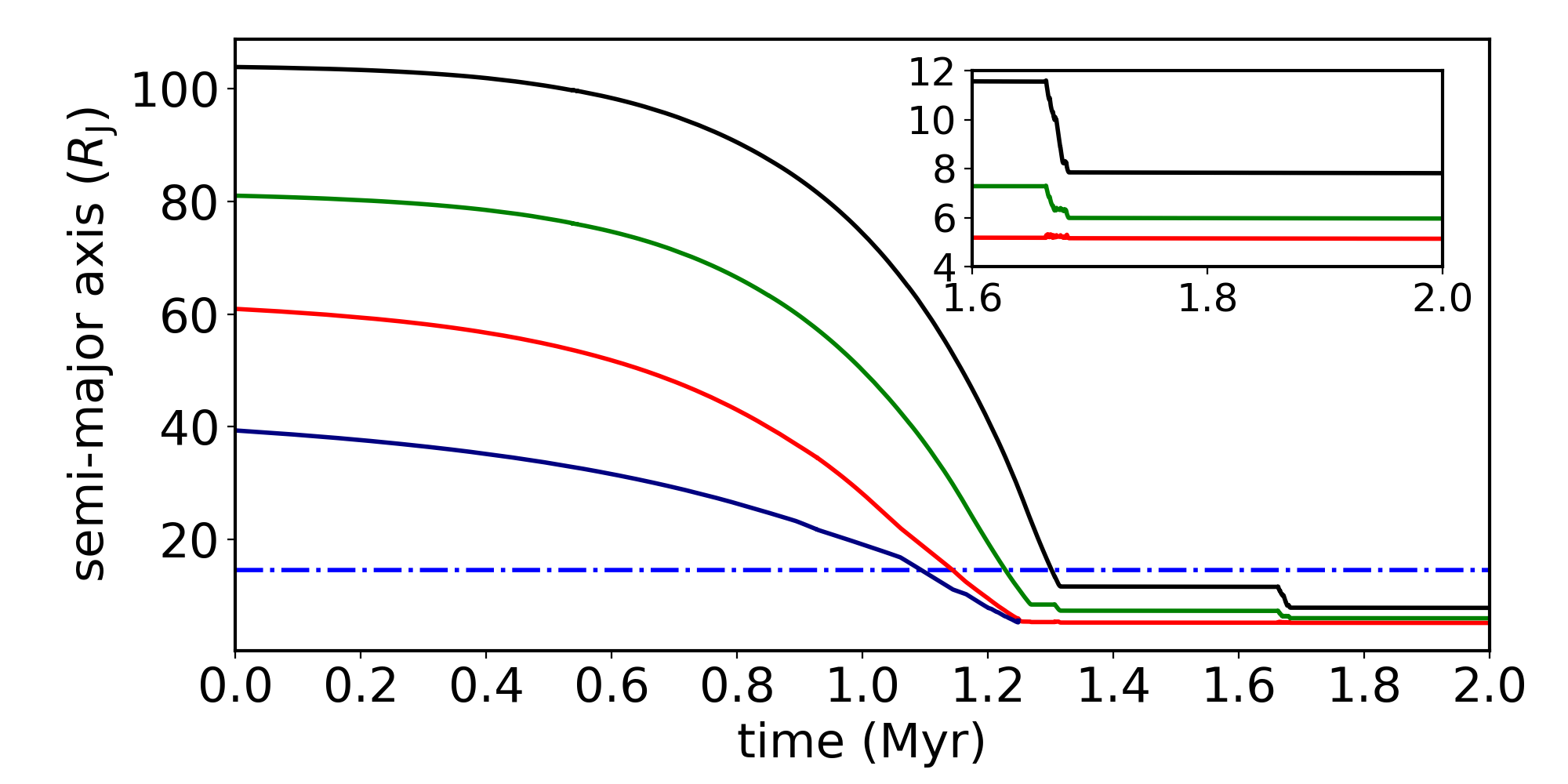}}
\subfigure[]{\includegraphics[width=0.95\columnwidth]{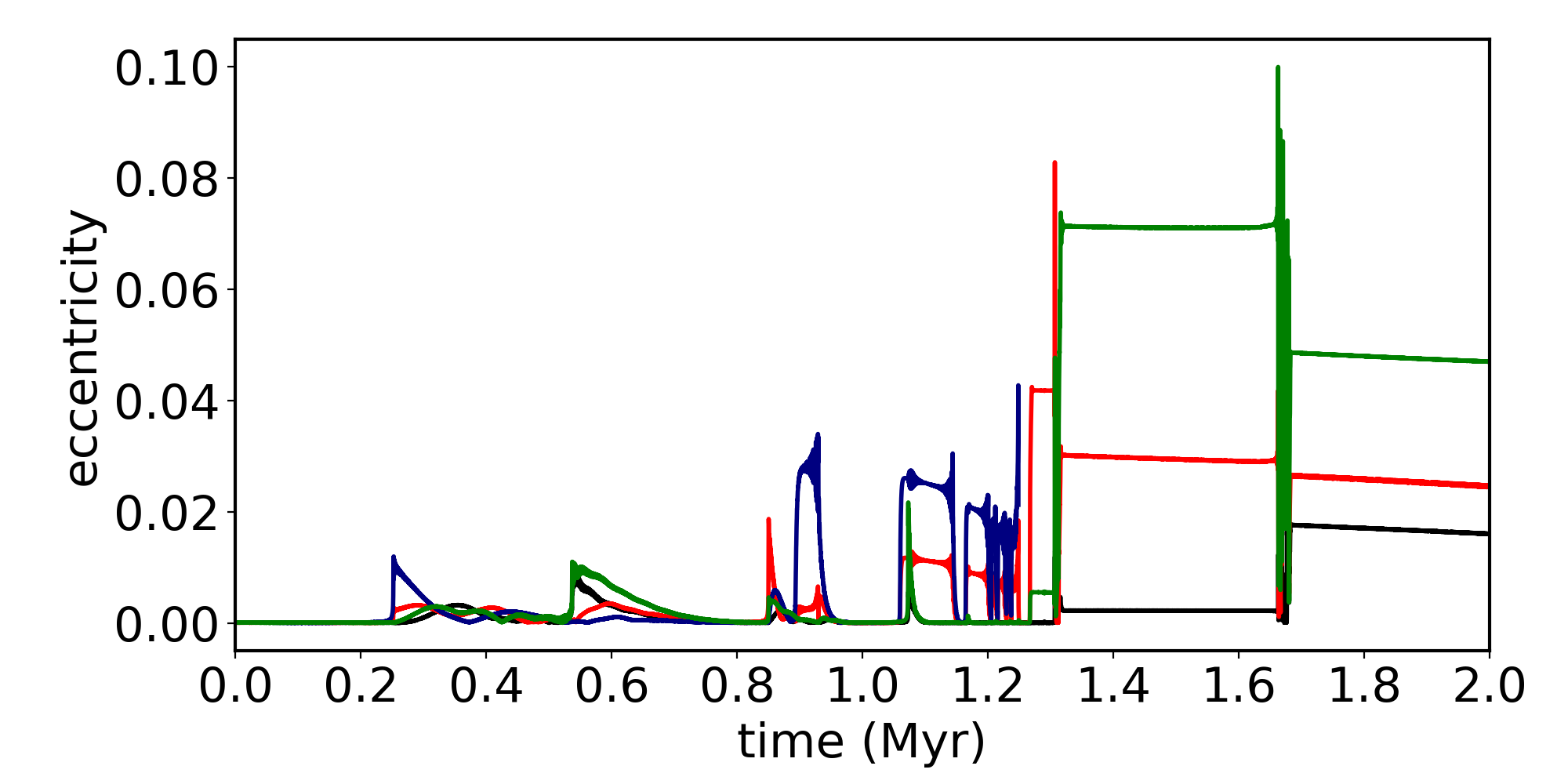}}
\subfigure[]{\includegraphics[width=0.95\columnwidth]{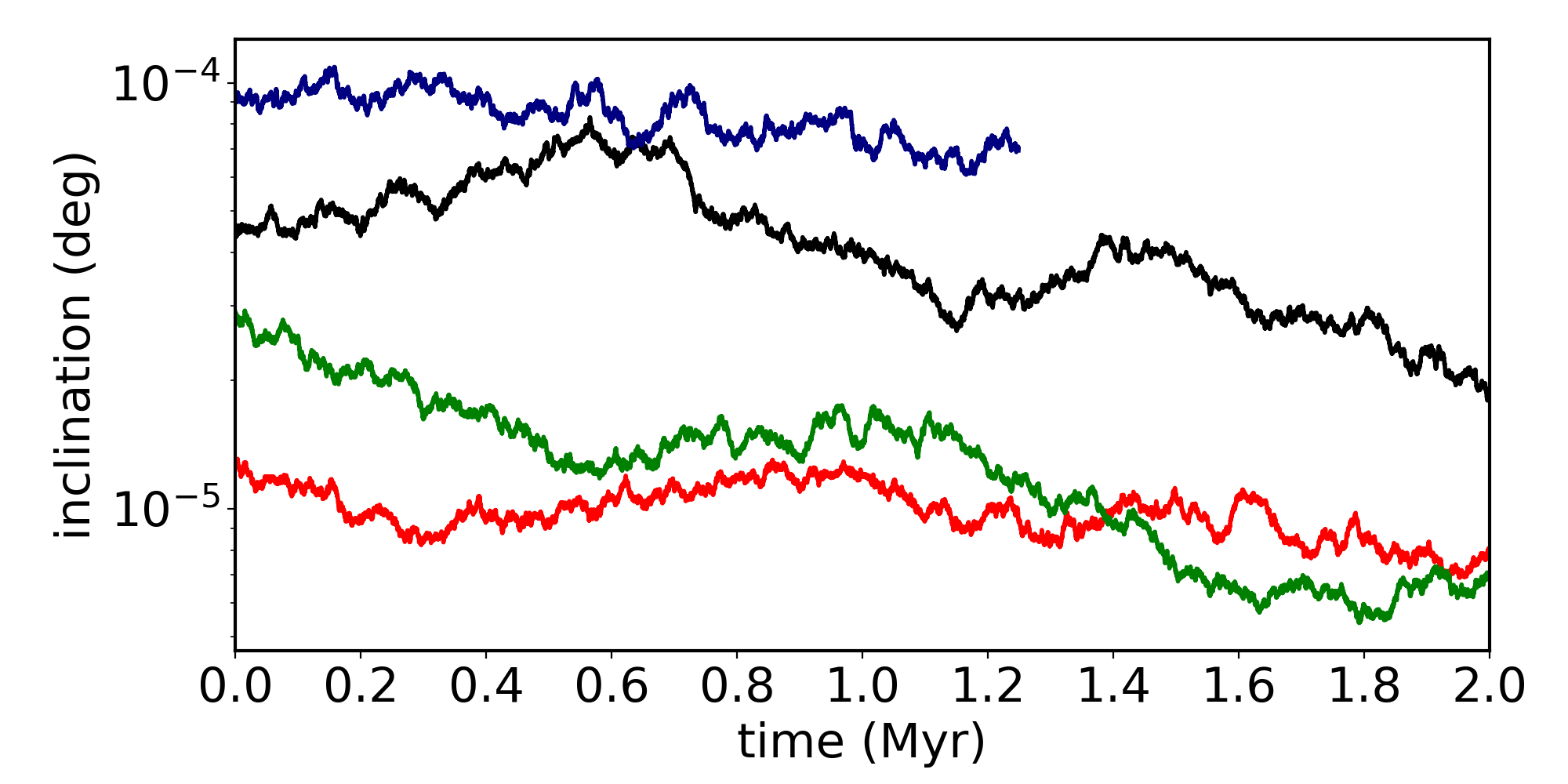}}
\subfigure[]{\includegraphics[width=0.95\columnwidth]{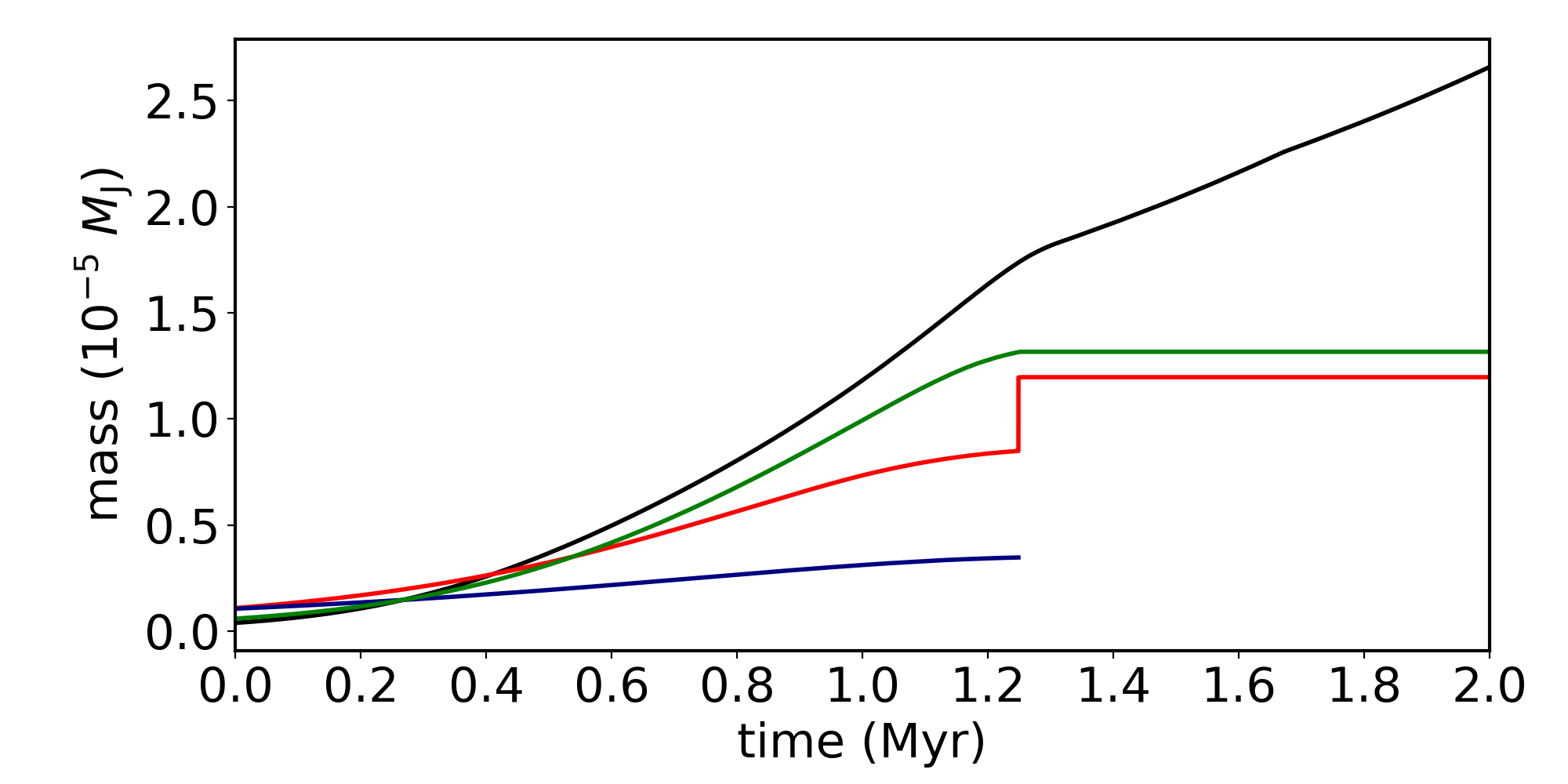}}
\caption{{\bf From top to bottom:} Evolution of the semi-major axes, orbital eccentricities, inclinations, and mass of satellitesimals in a simulation starting  with four satellitesimals and $\dot{\rm M}_{\rm p0}=10^{-9}~{\rm M_J}$/yr. Satellitesimals initially grow via pebble accretion  and migrate inwards. When they reach the inner edge of the disk locate at about $5~{\rm R_J}$, the innermost satellite pair collides at $\sim1.2$~Myr forming a system with only three final satellites. The blue dot-dashed line in the top panel shows the evolution of the snowline location in the disk.\label{fig:4sat} }
\end{figure}

Figure~\ref{fig:4sat} shows the evolution of one of our representative systems. In this simulation, we set  $\dot{\rm M}_{\rm p0}=10^{-9}~{\rm M_J}$/yr. Figure~\ref{fig:4sat} shows that satellitesimals first grow by pebble accretion and start to migrate inwards. When they reach the inner edge of the disk a dynamical instability takes place and leads to a collision. At the end of this simulation, at 2~Myr, 3 satellites survive with individual masses of the order of $10^{-5}~{\rm M_J}$. The innermost and  outermost satellite pairs are locked in a compact 5:4 and 3:2 MMR, respectively. The orbital eccentricities of satellites at the end of our simulations are between 0.02 and 0.05, which are larger than those of the Galilean satellites.
 
We verified that each one of our 40 numerical simulations starting with 4 satellitesimals shows at least one collision during the gas disk phase, typically when satellites approach the disk inner edge. This suggests that more than four satellitesimals are required to explain the Galilean system.  This is in conflict with the results of \cite{Sh19}, and these simulations violate our constraint i). In fact, even the final period ratio of the satellites in our simulations does not agree with those \cite{Sh19} have found. They have assumed that migrating satellites are successively captured in the 2:1 MMR when the migration timescale is longer than a critical timescale \citep{Og13}. The critical timescale criteria used in the semi-analytical model of \cite{Sh19} does not fully account for the eccentricity/inclination evolution of the satellites due to secular and resonant interactions. The capture in MMR also depends on the resonance order and mass-ratio of the migrating satellites \citep{2015MNRAS.451.2589B}. For instance, when the inner satellite is less massive than the outer one (e.g. Europa and Ganymede), the 2:1MMR can be skipped even when the adiabatic criteria for capture is attended \citep{2015MNRAS.451.2589B}. Thus, the criteria for capture in resonance assumed by \cite{Sh19} is in fact too simplistic. It is just a good proxy to infer (non-)capture in mean motion resonance if the eccentricities of the satellites can be neglected \citep{Og13}, which we show here to not be the case.

\begin{figure}
\includegraphics[width=\columnwidth]{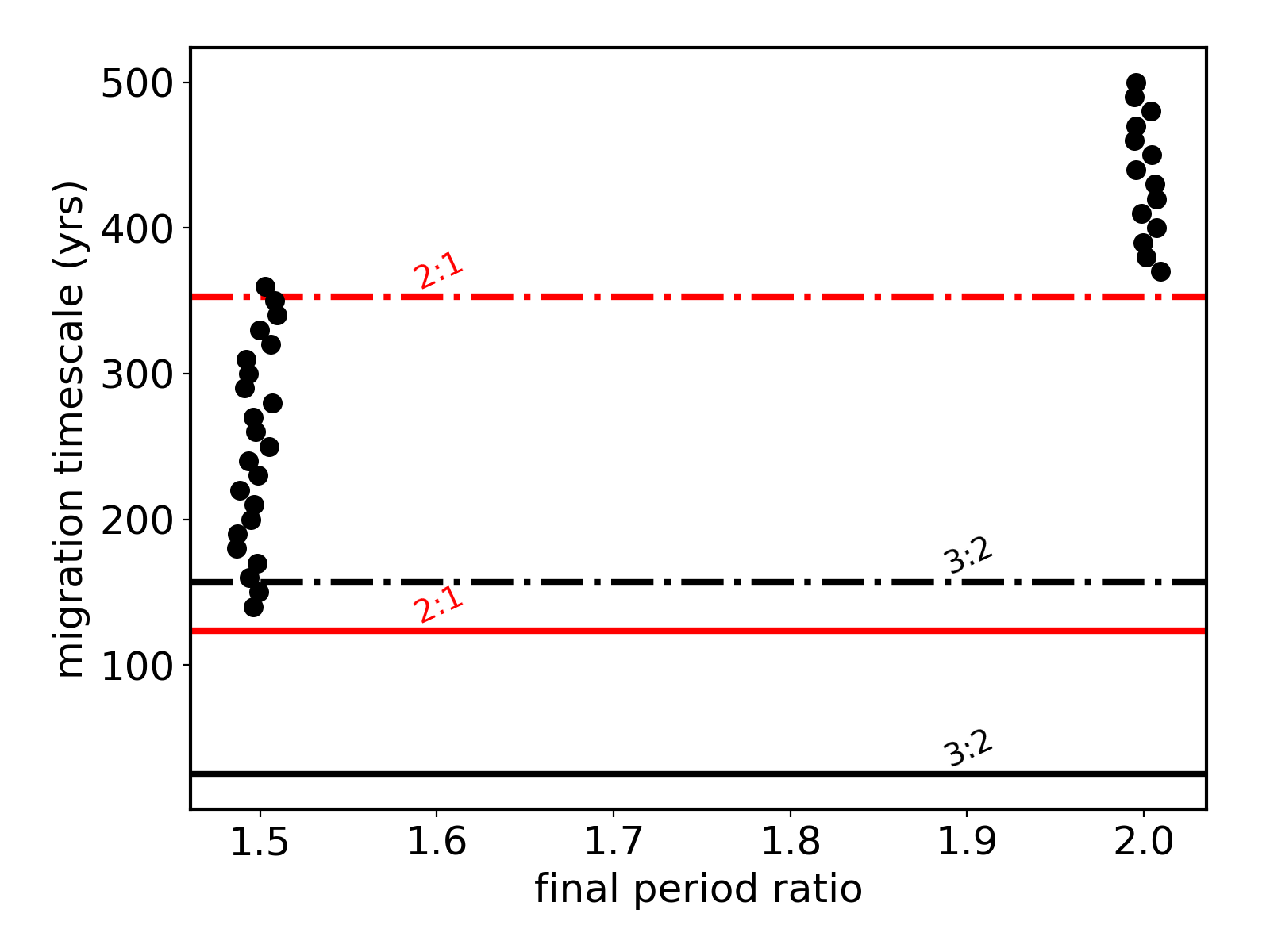}
\caption{Correlation between final period ratio and  migration timescale in simulations considering only two satellites with masses analogues of those of Io and Europa. The black dots represent the results of our numerical simulations, and horizontal lines are estimated critical timescales that lead to capture into 3:2 MMR (black line) and 2:1 MMR (red lines). The solid lines show the critical timescales inferred by Ogihara \& Kobayashi (2013) and the dashed lines show the critical timescales given by Goldreich \& Schlichting (2014a). We have found that migration timescales shorter than $\sim$130~yrs tend to lead to dynamical instabilities and collisions of the satellites. \label{fig:mig} }
\end{figure}

We performed some simple simulations considering the very disk model of \cite{Sh19}. This specific set of simulations starts with an Io-mass satellite fully formed and residing slightly outside the disk inner cavity,  and an Europa-mass satellite initially placed at 15~${\rm R_J}$. These satellites are released to migrate inwards -- but they are not allowed to accrete pebbles.  Our goal here is just compare the final resonant architecture of these satellites in our simulations and that predicted by the approach invoked in \cite{Sh19}. To conduct this set of experiments, we assume that satellites migrate with constant migration timescales, that do not vary with the distance to Jupiter. Figure~\ref{fig:mig} shows these results. In the vertical axis of Figure~\ref{fig:mig}, we show the migration timescale and in the horizontal axis the final period ratio of the satellites. Critical migration timescales that lead to capture in 2:1 and 3:2~MMR used in \cite{Sh19} are shown as solid black and red lines, respectively. The dot-dashed black and red lines show the migration timescales that lead to capture in 2:1 and 3:2 MMR when orbital eccentricities are taken into account \citep{Go14}. We use the eccentricity of the satellites in our simulations as input to calculate the latter timescales. The black dots show the results of our numerical simulations which agree very well with those of \cite{Go14}. The difference observed between our results and those of \cite{Sh19} is caused by the increase in the eccentricity of the satellites when they approach each other, which breaks down the validity of their criteria for capture in resonance. We have found that the timescale predicted to lead to capture in 2:1 MMR in simulations of \cite{Sh19} in fact tends to lead to collisions. 

\subsection{Effects of the pebble flux} \label{subsec:parameters}

\begin{figure*}
\centering
\subfigure[$\dot{\rm M}_{\rm p0}=5\times 10^{-9}~{\rm M_J}$/yr and 30 satellitesimals]{\includegraphics[width=0.8\columnwidth]{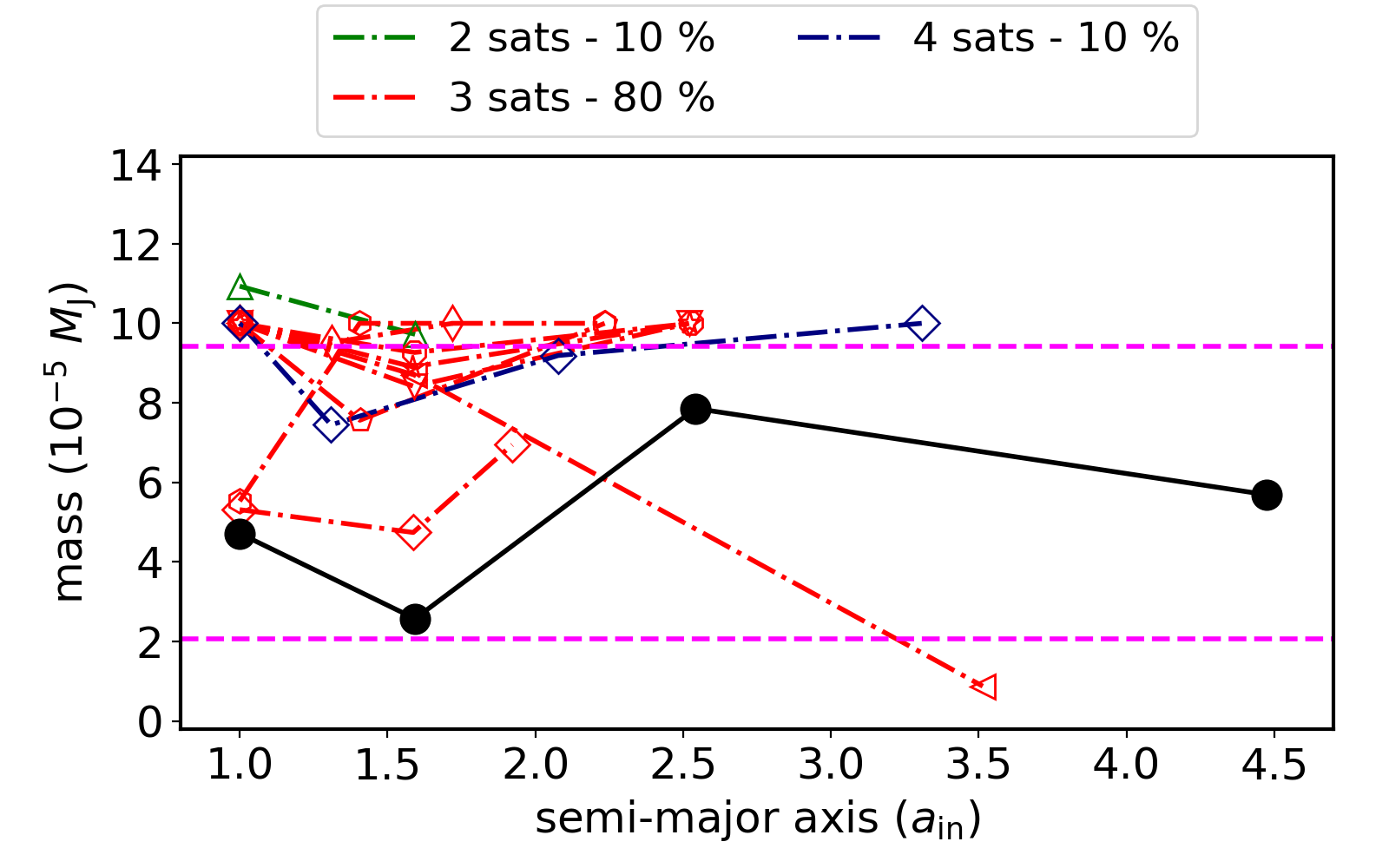} }
\quad \quad \quad
\subfigure[$\dot{\rm M}_{\rm p0}=5\times 10^{-9}~{\rm M_J}$/yr and 50 satellitesimals]{\includegraphics[width=0.8\columnwidth]{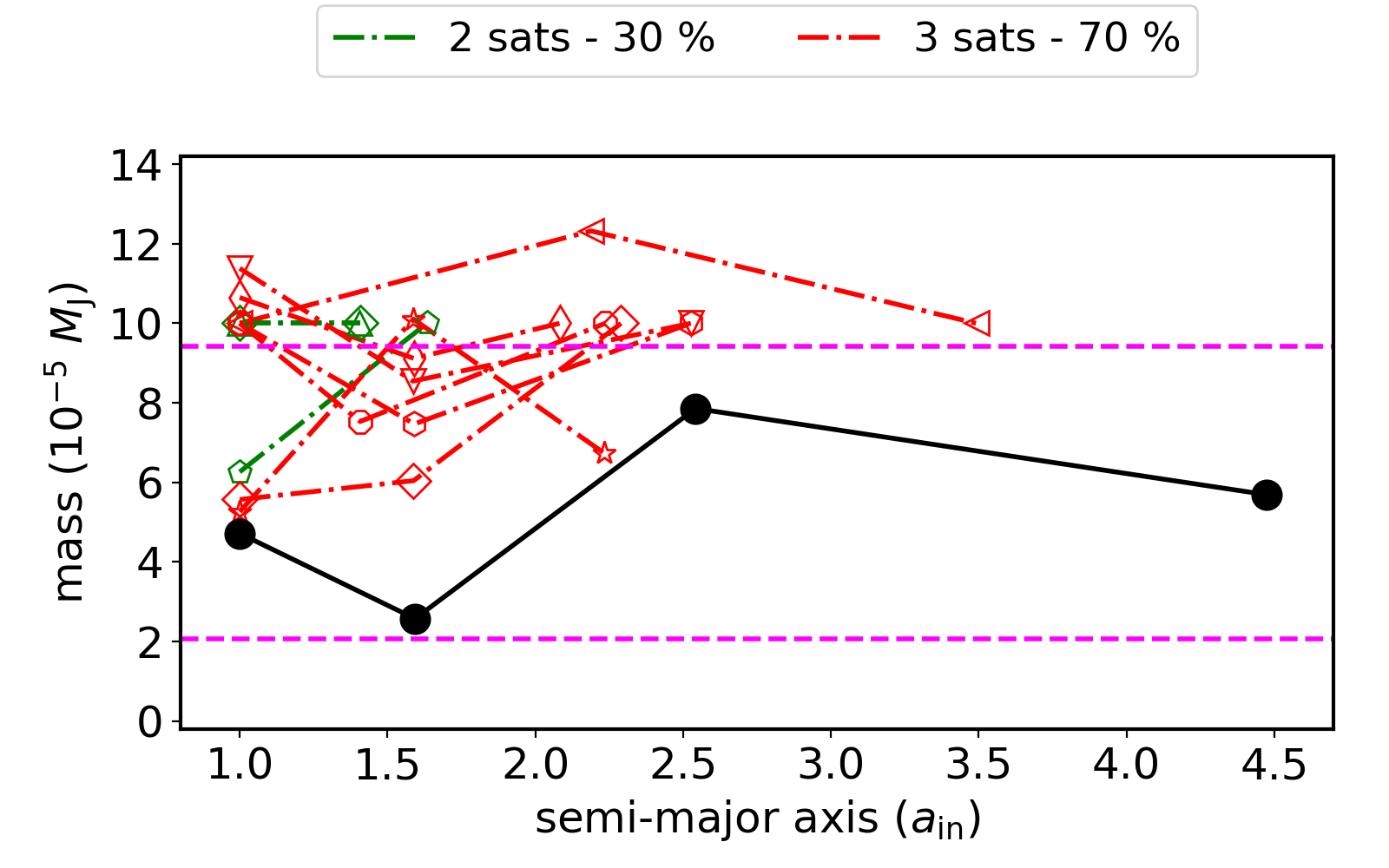} }
\centering
\subfigure[$\dot{\rm M}_{\rm p0}=3\times 10^{-9}~{\rm M_J}$/yr and 30 satellitesimals]{\includegraphics[width=0.8\columnwidth]{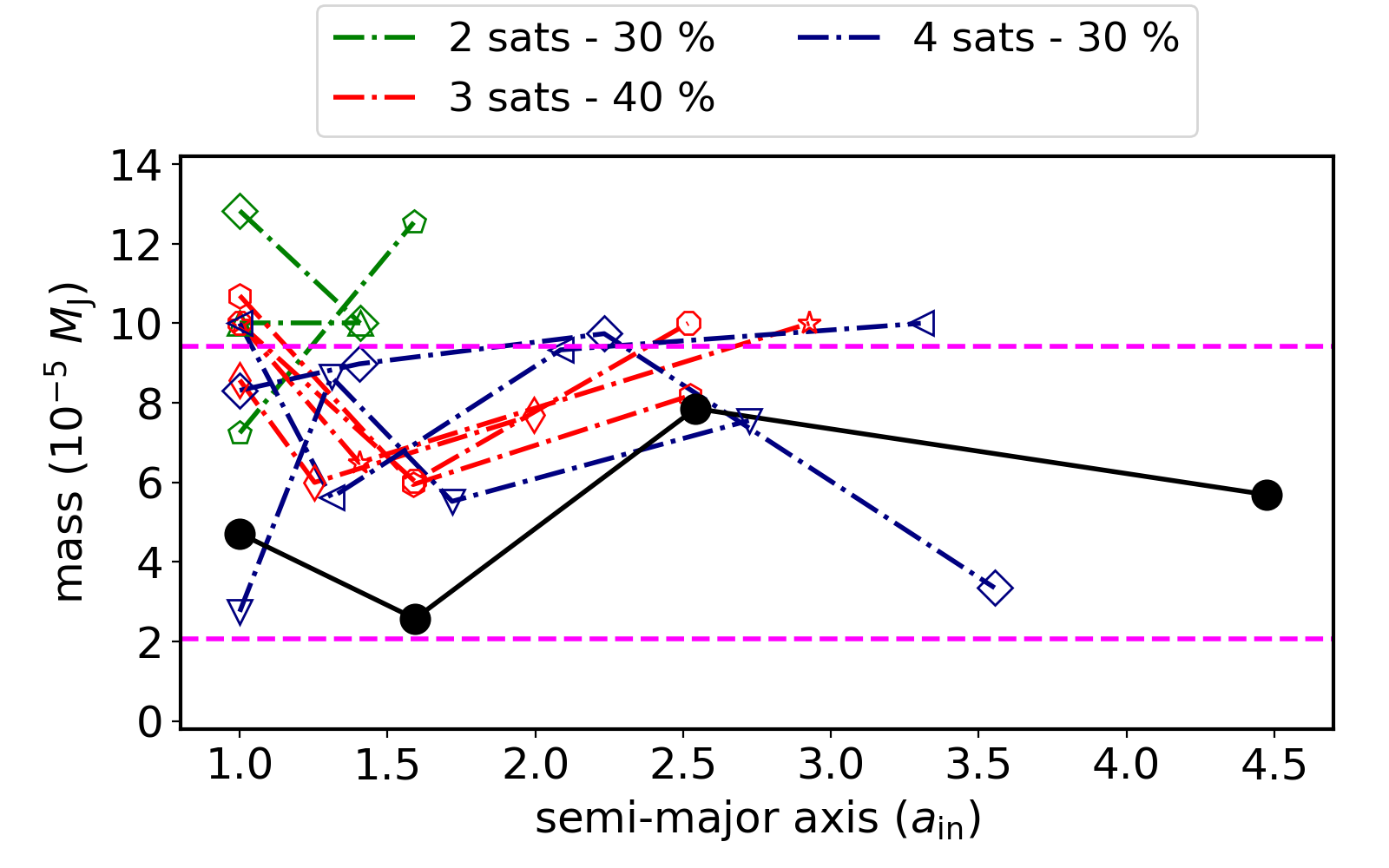} }
\quad \quad \quad
\subfigure[$\dot{\rm M}_{\rm p0}=3\times 10^{-9}~{\rm M_J}$/yr and 50 satellitesimals]{\includegraphics[width=0.8\columnwidth]{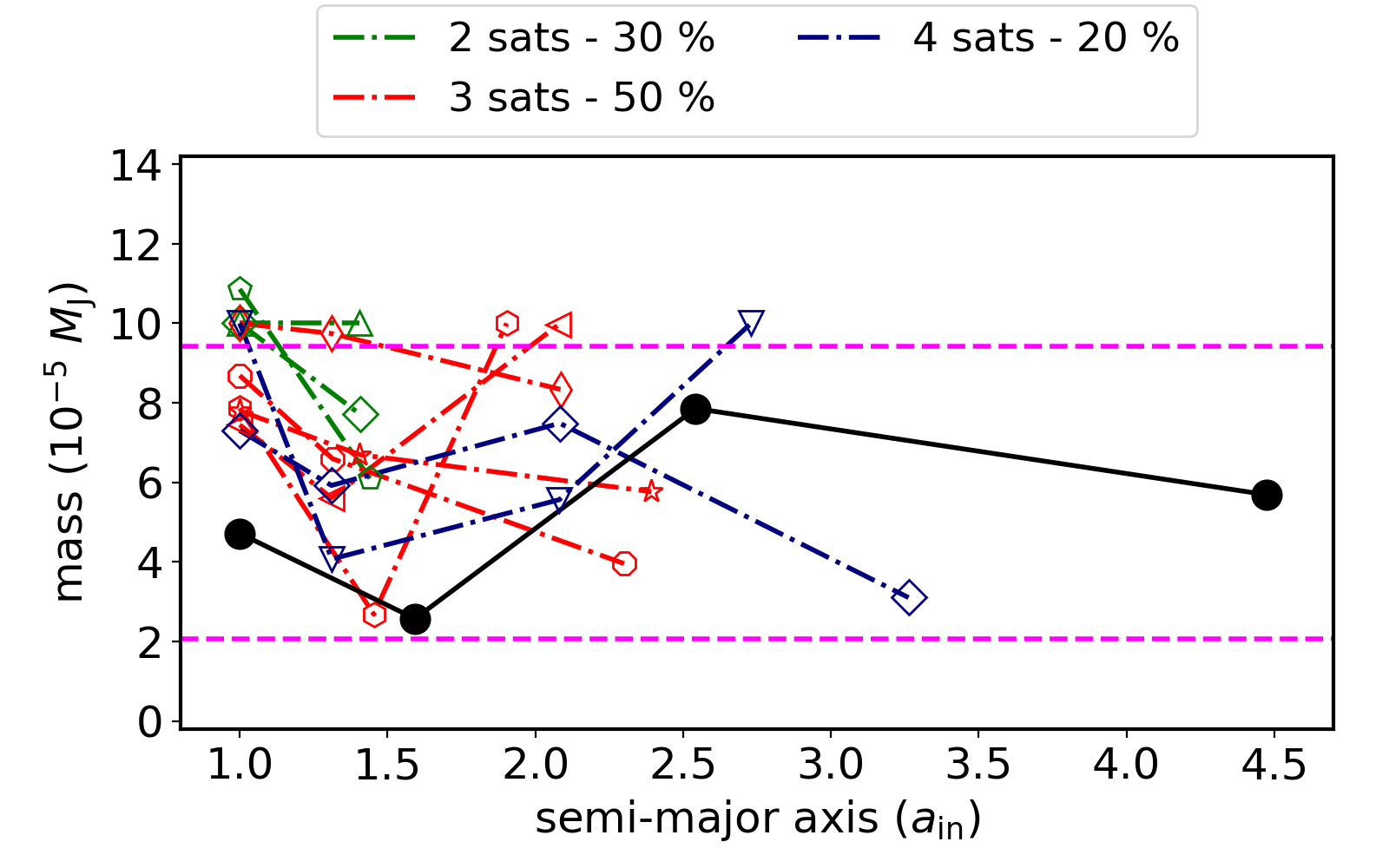} }
\centering
\subfigure[$\dot{\rm M}_{\rm p0}=1.5\times 10^{-9}~{\rm M_J}$/yr and 30 satellitesimals]{\includegraphics[width=0.8\columnwidth]{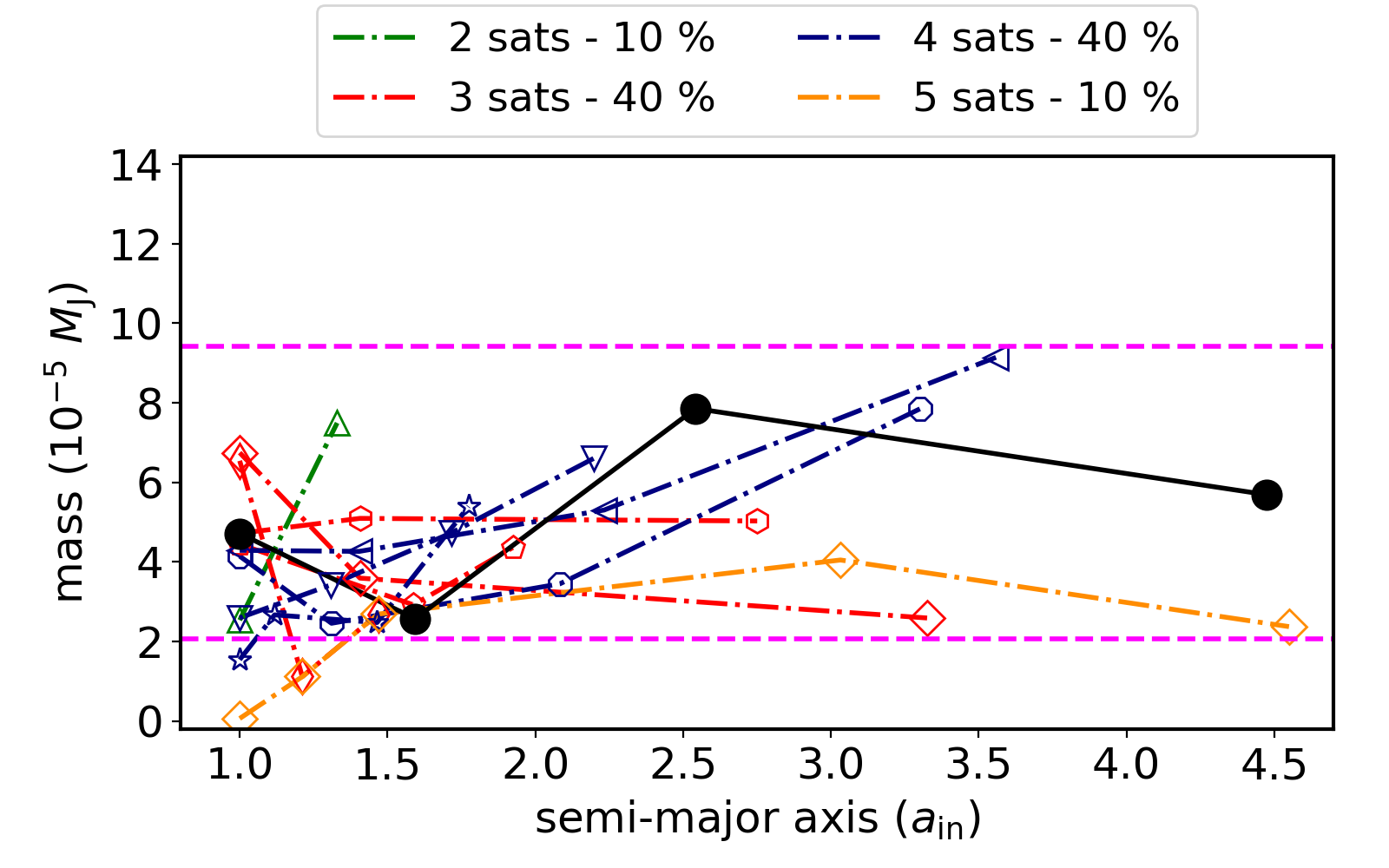} }
\quad \quad \quad
\subfigure[$\dot{\rm M}_{\rm p0}=1.5\times 10^{-9}~{\rm M_J}$/yr and 50 satellitesimals]{\includegraphics[width=0.8\columnwidth]{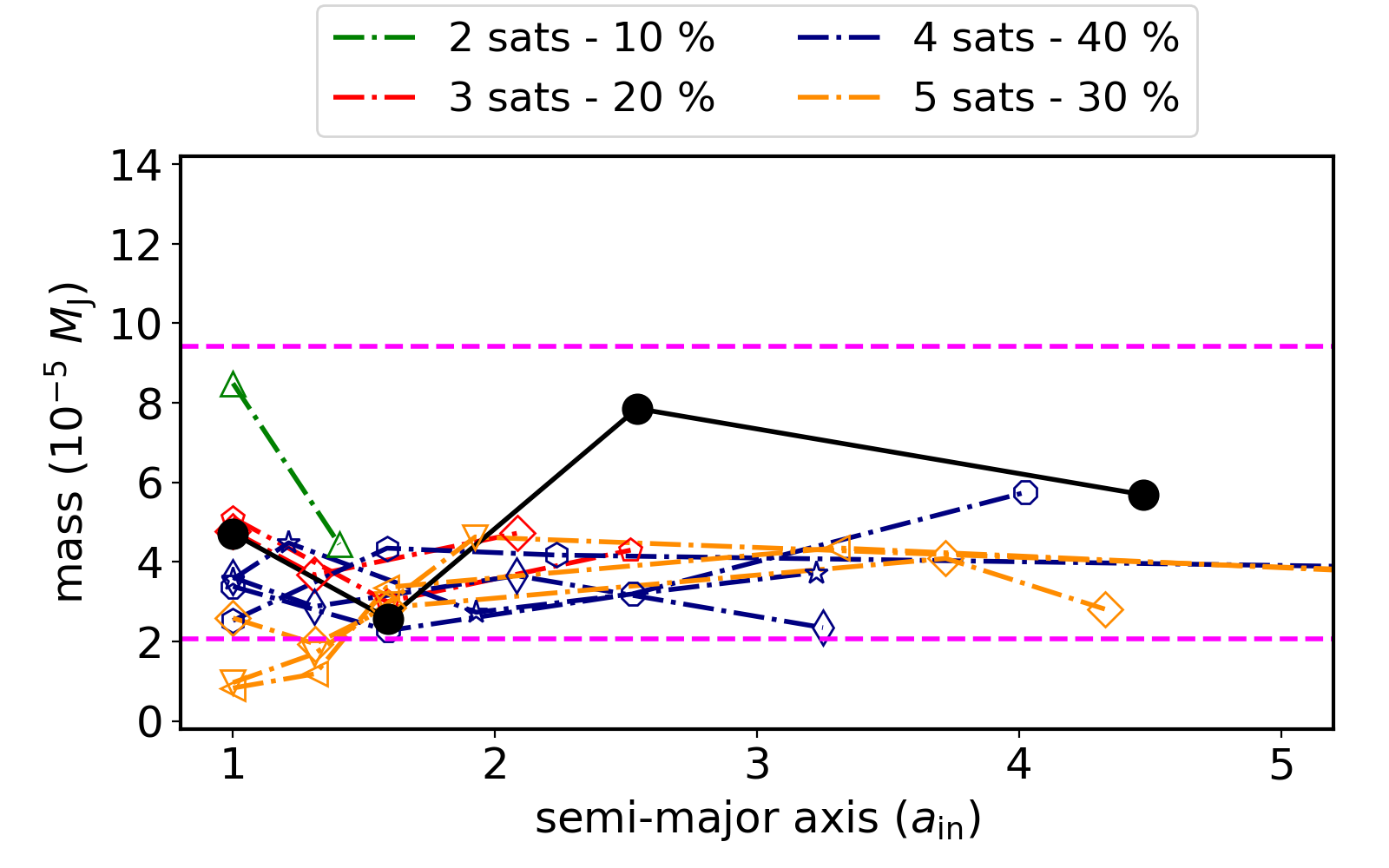} }
\centering
\subfigure[$\dot{\rm M}_{\rm p0}=10^{-9}~{\rm M_J}$/yr and 30 satellitesimals]{\includegraphics[width=0.8\columnwidth]{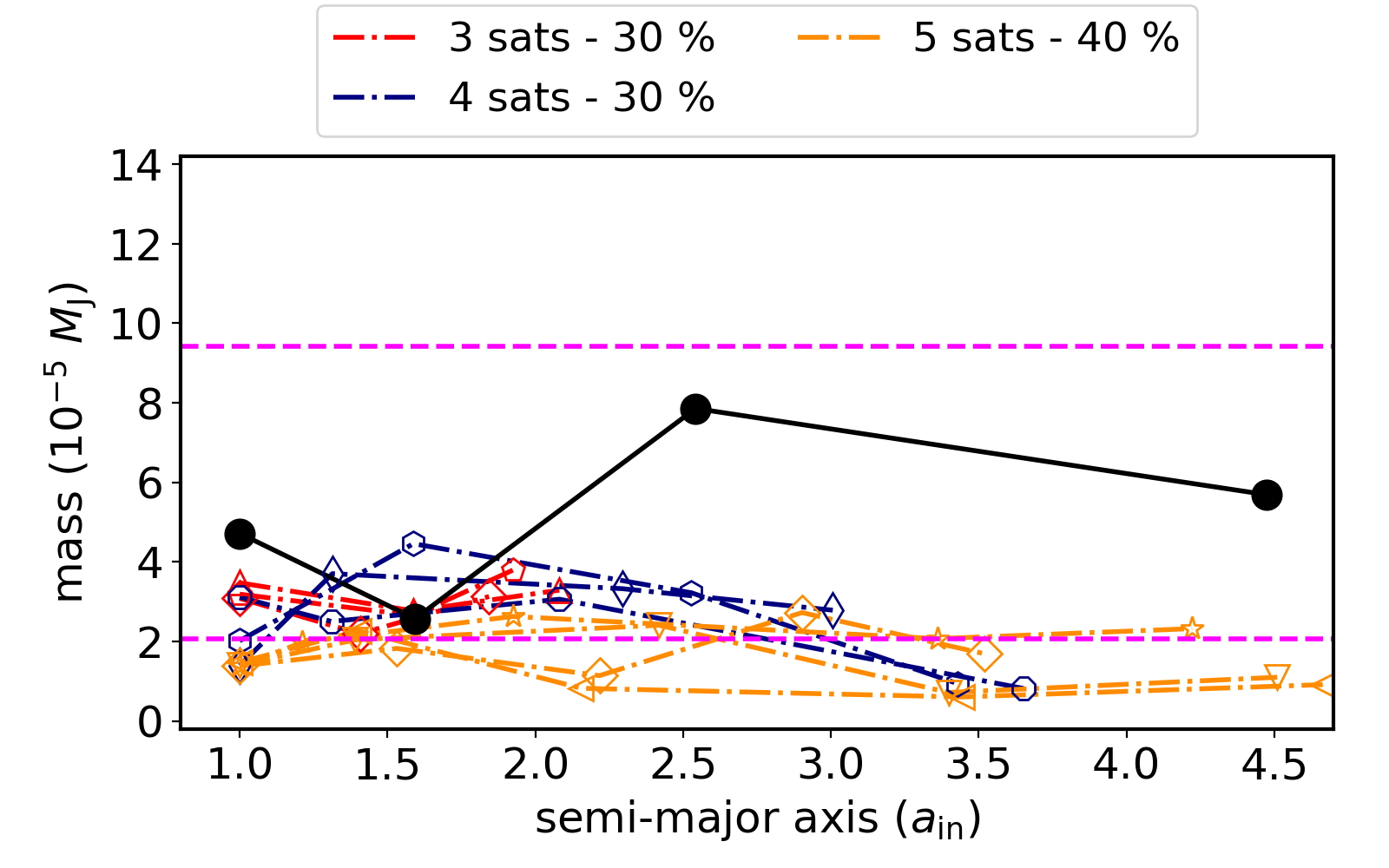} }
\quad \quad \quad
\subfigure[$\dot{\rm M}_{\rm p0}=10^{-9}~{\rm M_J}$/yr and 50 satellitesimals]{\includegraphics[width=0.8\columnwidth]{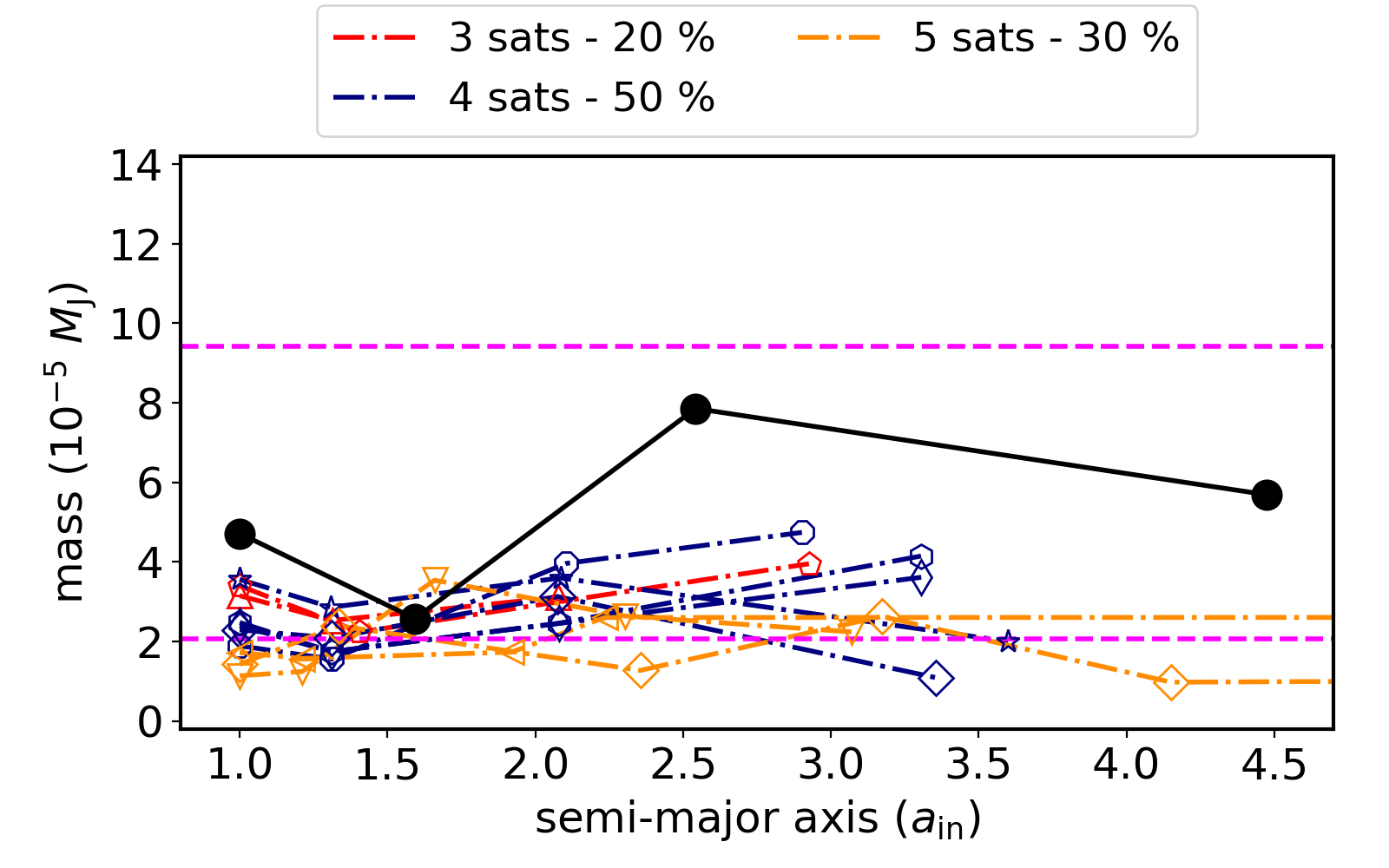} }
\caption{Final satellite systems produced in simulations starting with different initial number of satellitesimals and pebble fluxes. The left and right side panels show the results of simulations starting with 30 and 50 satellitesimals, respectively. From top-to-bottom the panels show the results of simulations  with different pebble fluxes:  a) and b) $\dot{\rm M}_{\rm p0}=5\times 10^{-9}~{\rm M_J}$/yr, c) and d) $\dot{\rm M}_{\rm p0}=3\times  10^{-9}~{\rm M_J}$/yr, e) and f) $\dot{\rm M}_{\rm p0}=1.5\times  10^{-9}~{\rm M_J}$/yr, and g) and h) $\dot{\rm M}_{\rm p0}=10^{-9}~{\rm M_J}$/yr. The lines connecting different points (symbols) show satellites in a same system. The fraction of simulations that produce two (green dot-dash lines), three (red dot-dash), four (blue dot-dash), and five satellites (orange dot-dash) are given at the top of each panel. The black solid line shows the real Galilean system. The horizontal pink lines correspond to 0.8${\rm M_E}$ and 1.2${\rm M_G}$ (see constraint iii)). Note that we have re-scaled the position of the satellites  by a factor of order of unity to make the position of the innermost satellite in each of our simulations to correspond to the distance of Io to Jupiter ($a_{\rm in}$).\label{fig:number}}.
\end{figure*}

In this section, we present the results of our simulations starting with 30 and 50 satellitesimals and compare the effects of different pebble fluxes. Figure~\ref{fig:number} shows the final satellite systems produced in our simulations at the end of the gas disk phase ($\sim$2~Myr). Simulations starting with 30 satellitesimals are shown on the left panels, and simulations starting with 50 satellitesimals are presented on the right panels.  The horizontal axis of each panel shows the final semi-major axis  and the vertical axis shows the mass of satellites. Each panel shows the results of 10 different simulations, where in each simulation, satellitesimals start with slightly different orbital parameters. Colorful dot-dashed lines connecting different points (symbols)  show different satellite systems. The black filled circles show the real Galilean satellite system for reference.  For presentation purposes, in Figure \ref{fig:number}, we have re-scaled the position of the satellites by a factor of order of unity to make the position of the innermost satellite in our simulations to correspond to the distance of Io to Jupiter ($a_{\rm in}$). At the top of each panel, we show the fraction of systems that produced a given final number of satellites.

\begin{figure*}
\centering
\subfigure[$\dot{\rm M}_{\rm p0}=5\times 10^{-9}~{\rm M_J}$/yr and 30 satellitesimals]{\includegraphics[width=0.8\columnwidth]{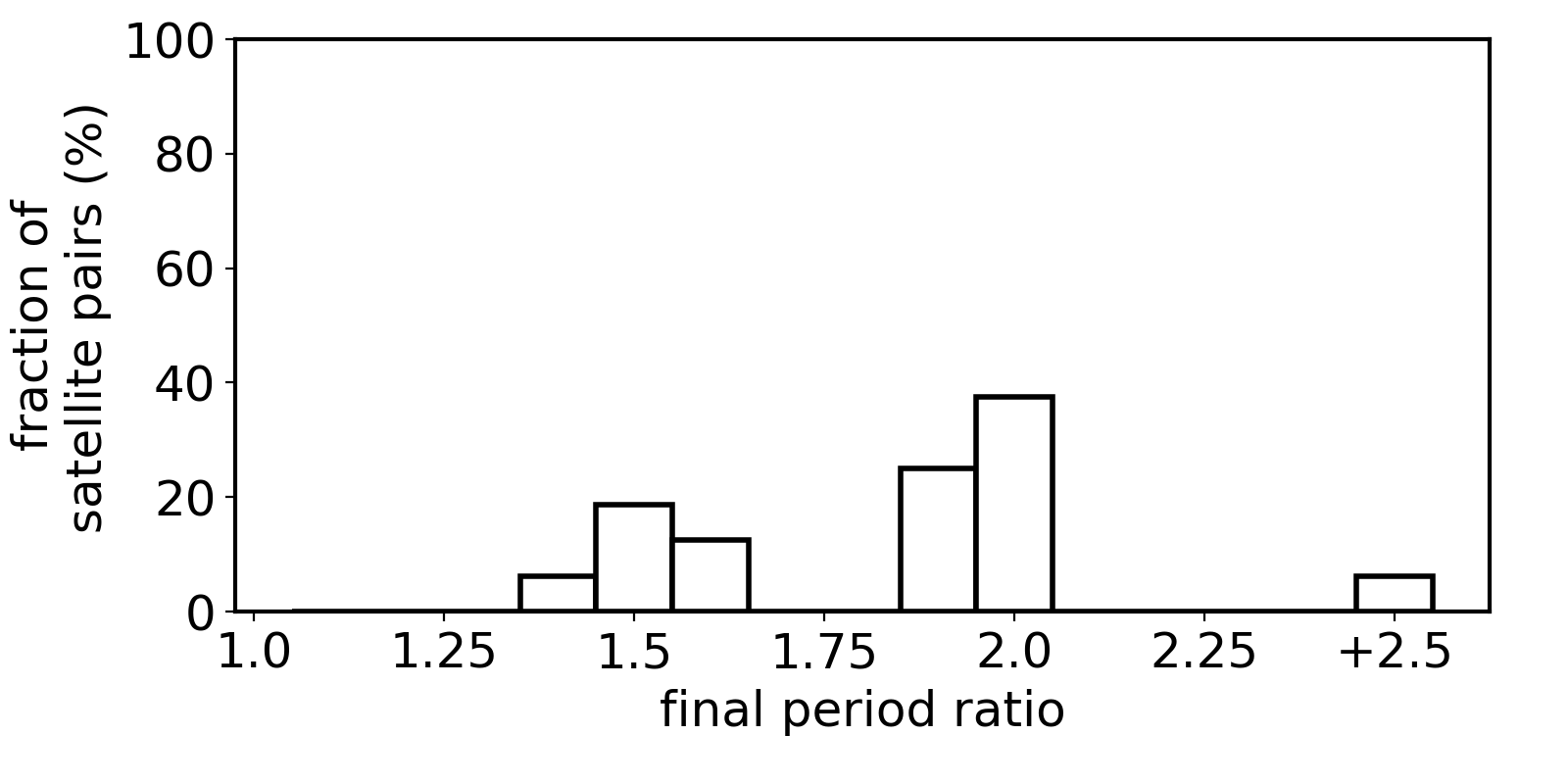} }
\quad \quad \quad
\subfigure[$\dot{\rm M}_{\rm p0}=5\times 10^{-9}~{\rm M_J}$/yr and 50 satellitesimals]{\includegraphics[width=0.8\columnwidth]{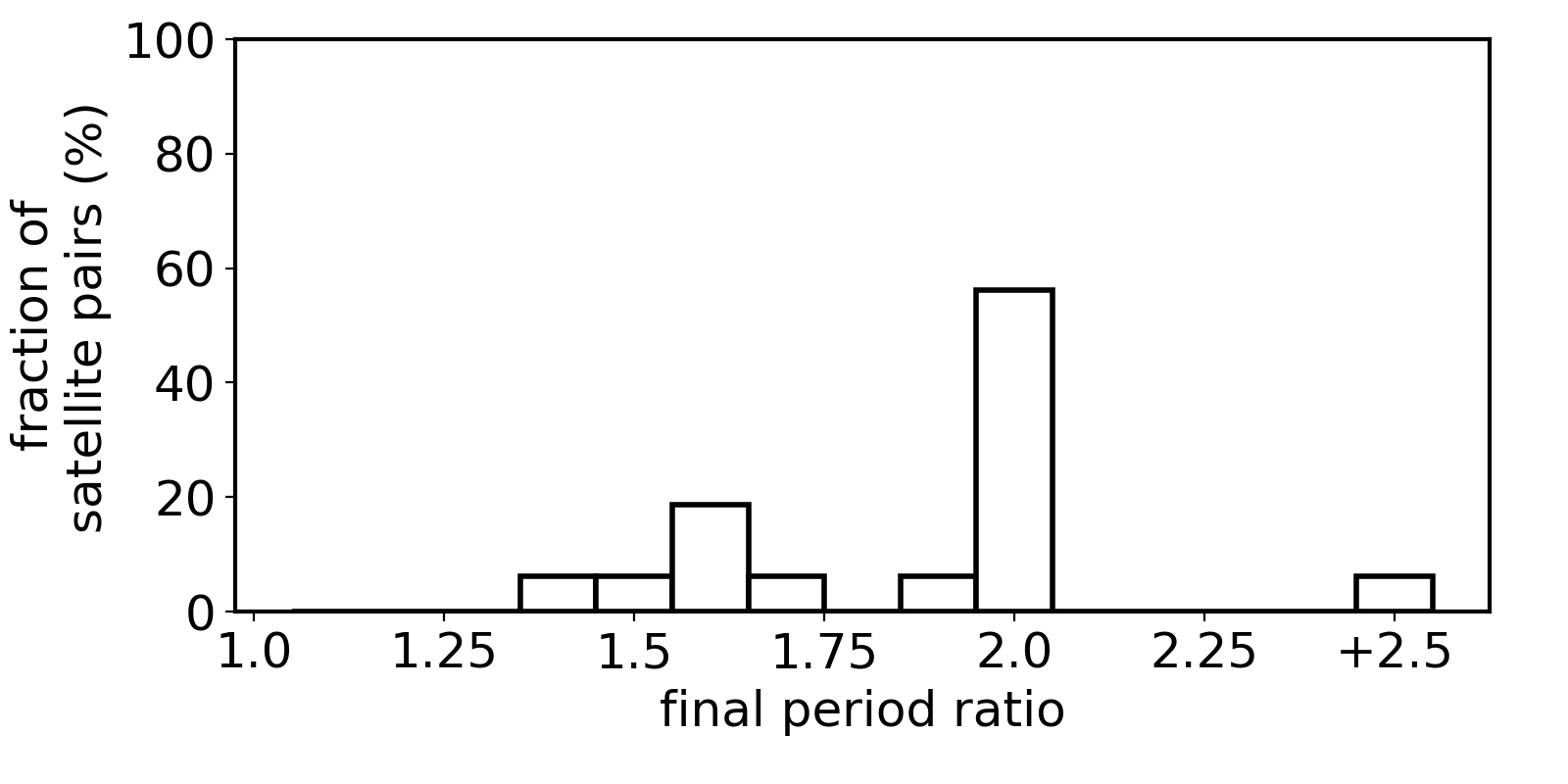} }
\centering
\subfigure[$\dot{\rm M}_{\rm p0}=3\times 10^{-9}~{\rm M_J}$/yr and 30 satellitesimals]{\includegraphics[width=0.8\columnwidth]{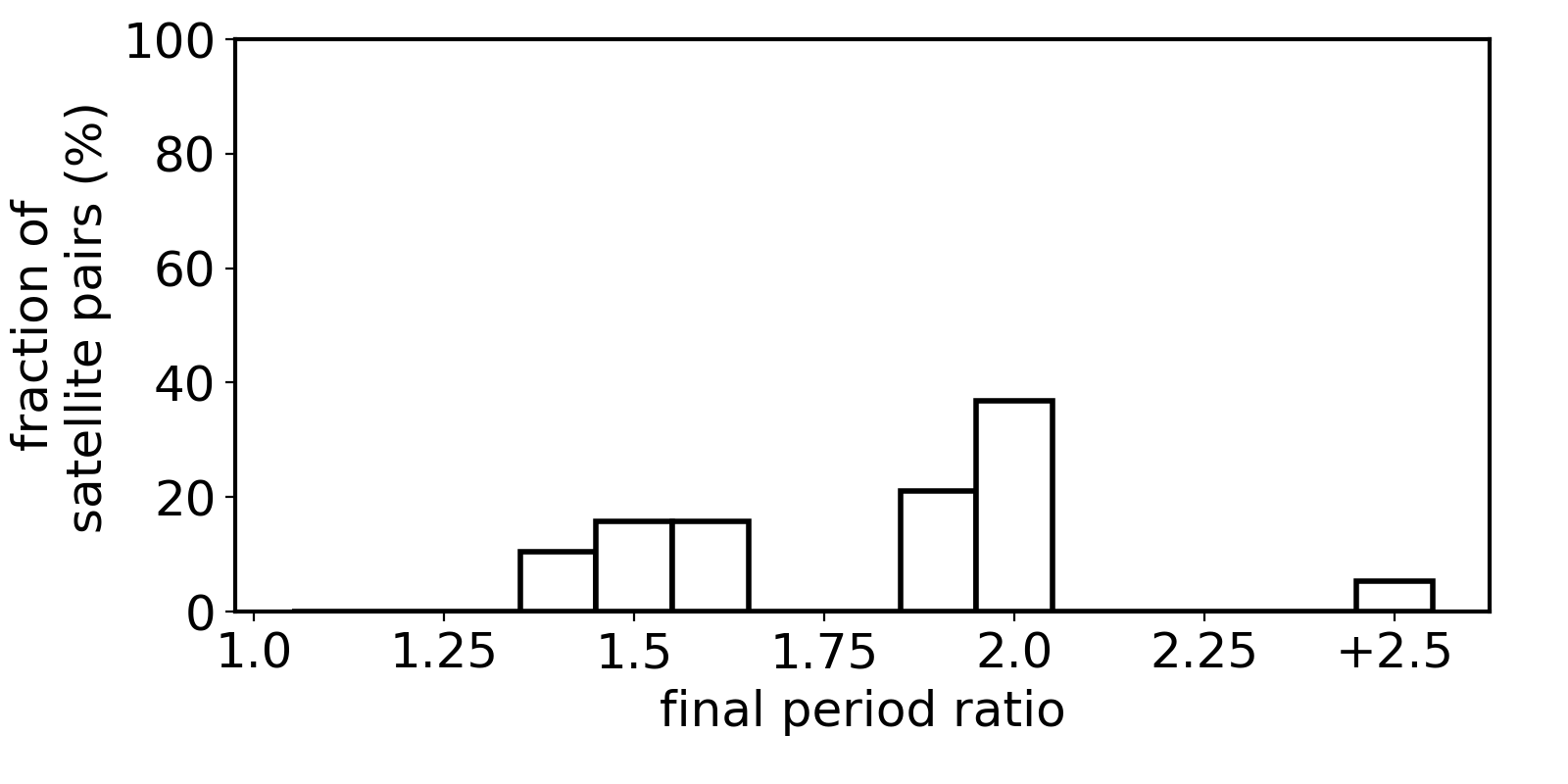} }
\quad \quad \quad
\subfigure[$\dot{\rm M}_{\rm p0}=3\times 10^{-9}~{\rm M_J}$/yr and 50 satellitesimals]{\includegraphics[width=0.8\columnwidth]{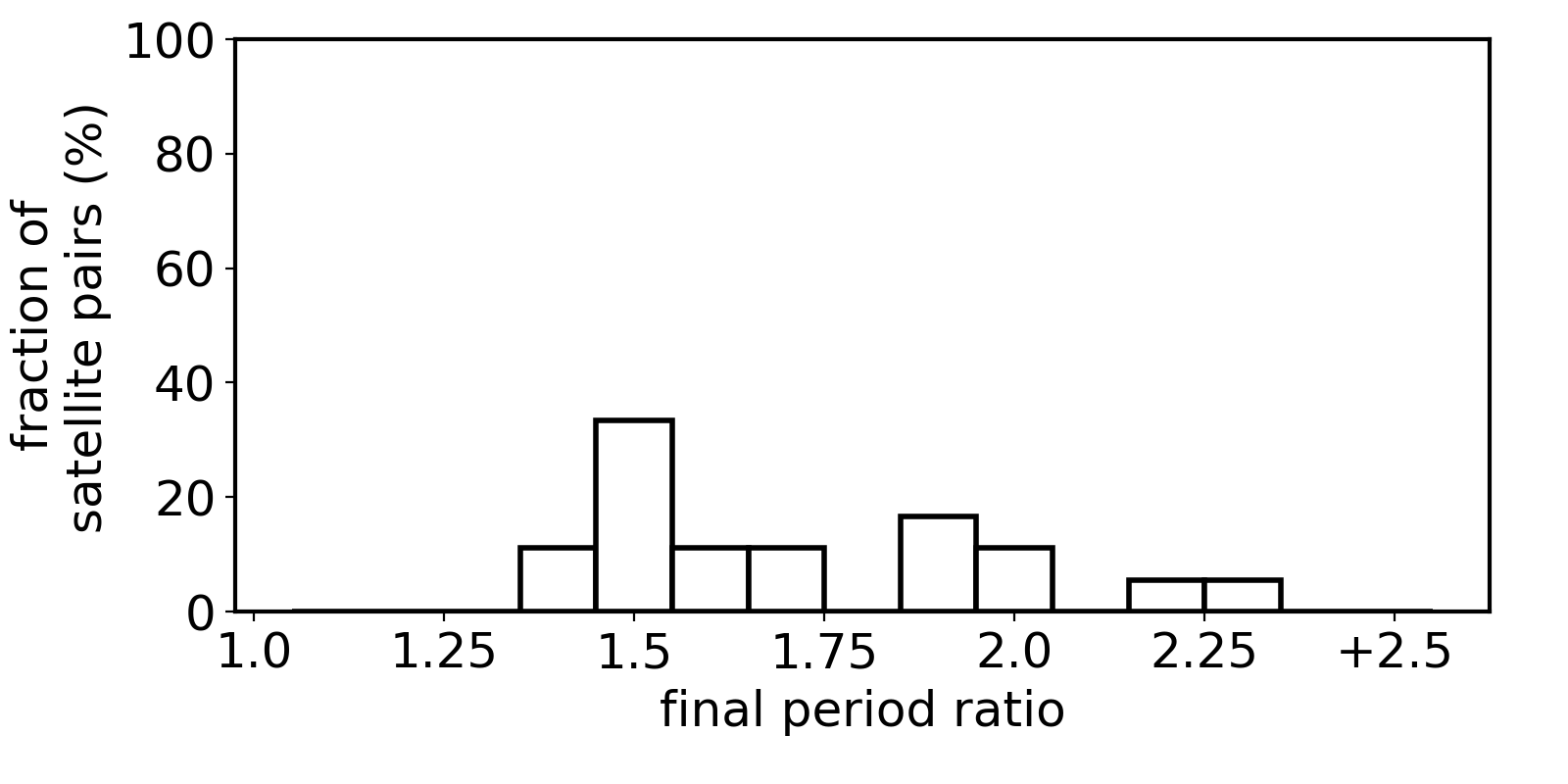} }
\centering
\subfigure[$\dot{\rm M}_{\rm p0}=1.5\times 10^{-9}~{\rm M_J}$/yr and 30 satellitesimals]{\includegraphics[width=0.8\columnwidth]{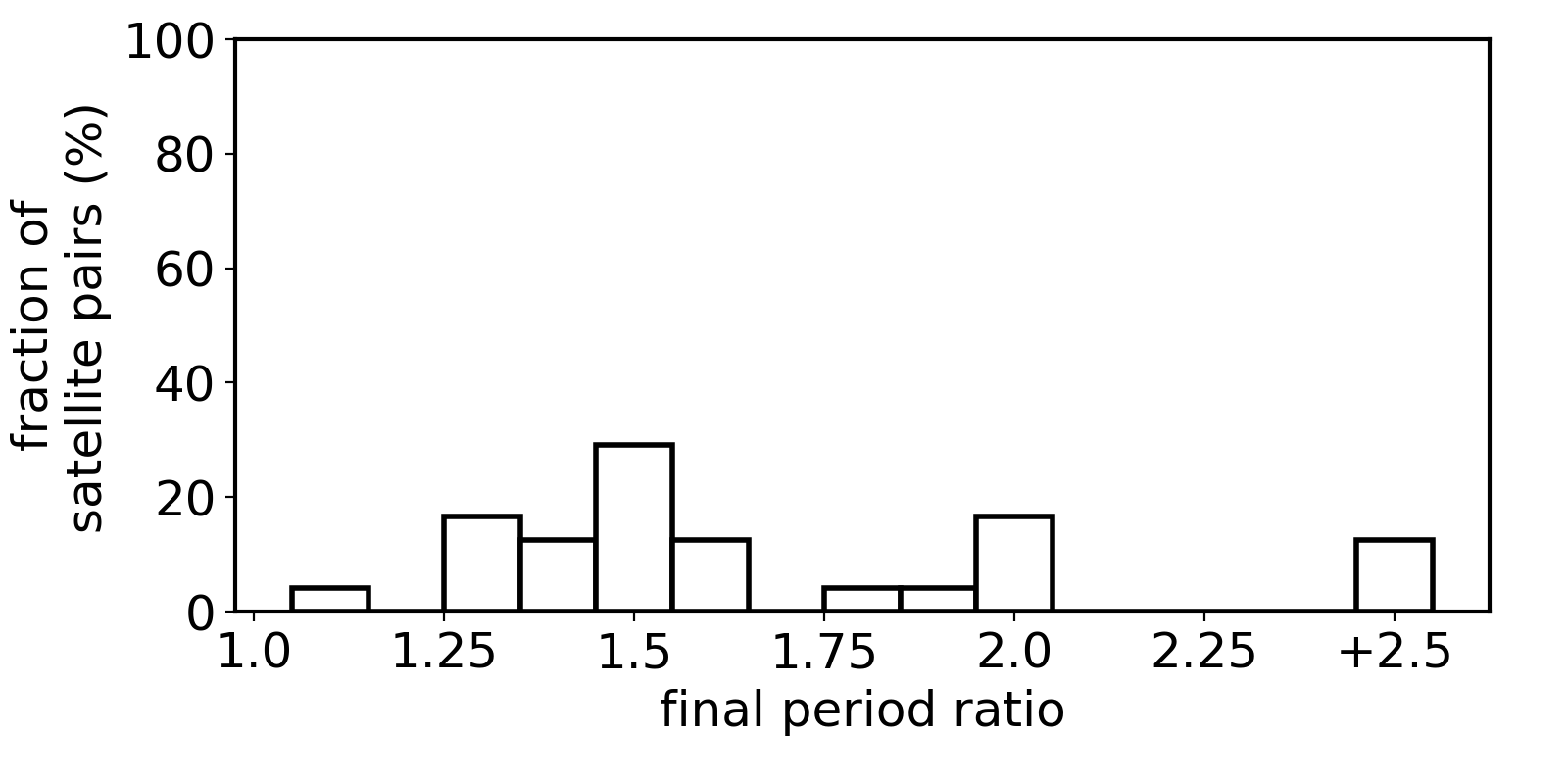} }
\quad \quad \quad
\subfigure[$\dot{\rm M}_{\rm p0}=1.5\times 10^{-9}~{\rm M_J}$/yr and 50 satellitesimals]{\includegraphics[width=0.8\columnwidth]{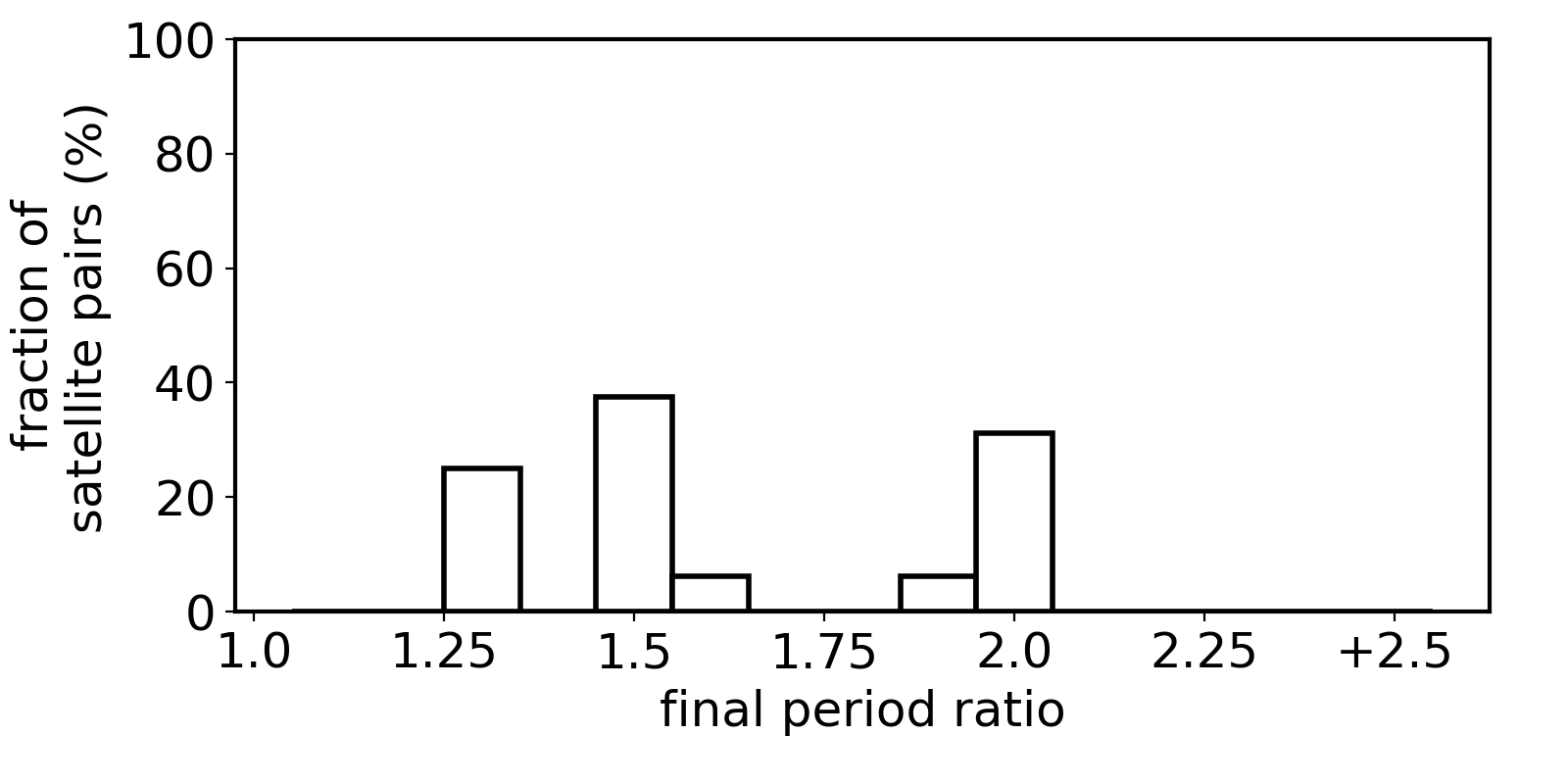} }
\centering
\subfigure[$\dot{\rm M}_{\rm p0}=10^{-9}~{\rm M_J}$/yr and 30 satellitesimals]{\includegraphics[width=0.8\columnwidth]{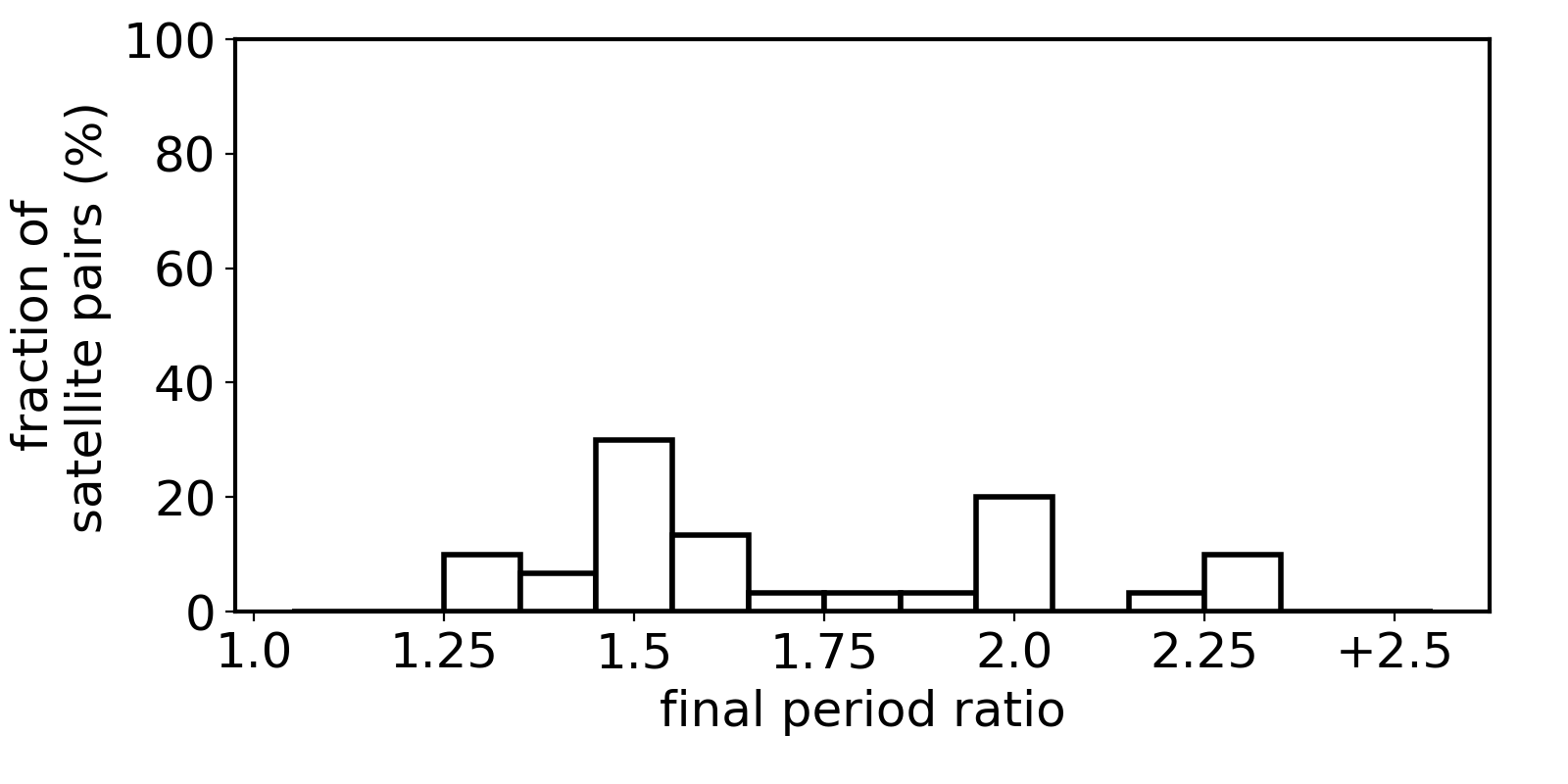} }
\quad \quad \quad
\subfigure[$\dot{\rm M}_{\rm p0}=10^{-9}~{\rm M_J}$/yr and 50 satellitesimals]{\includegraphics[width=0.8\columnwidth]{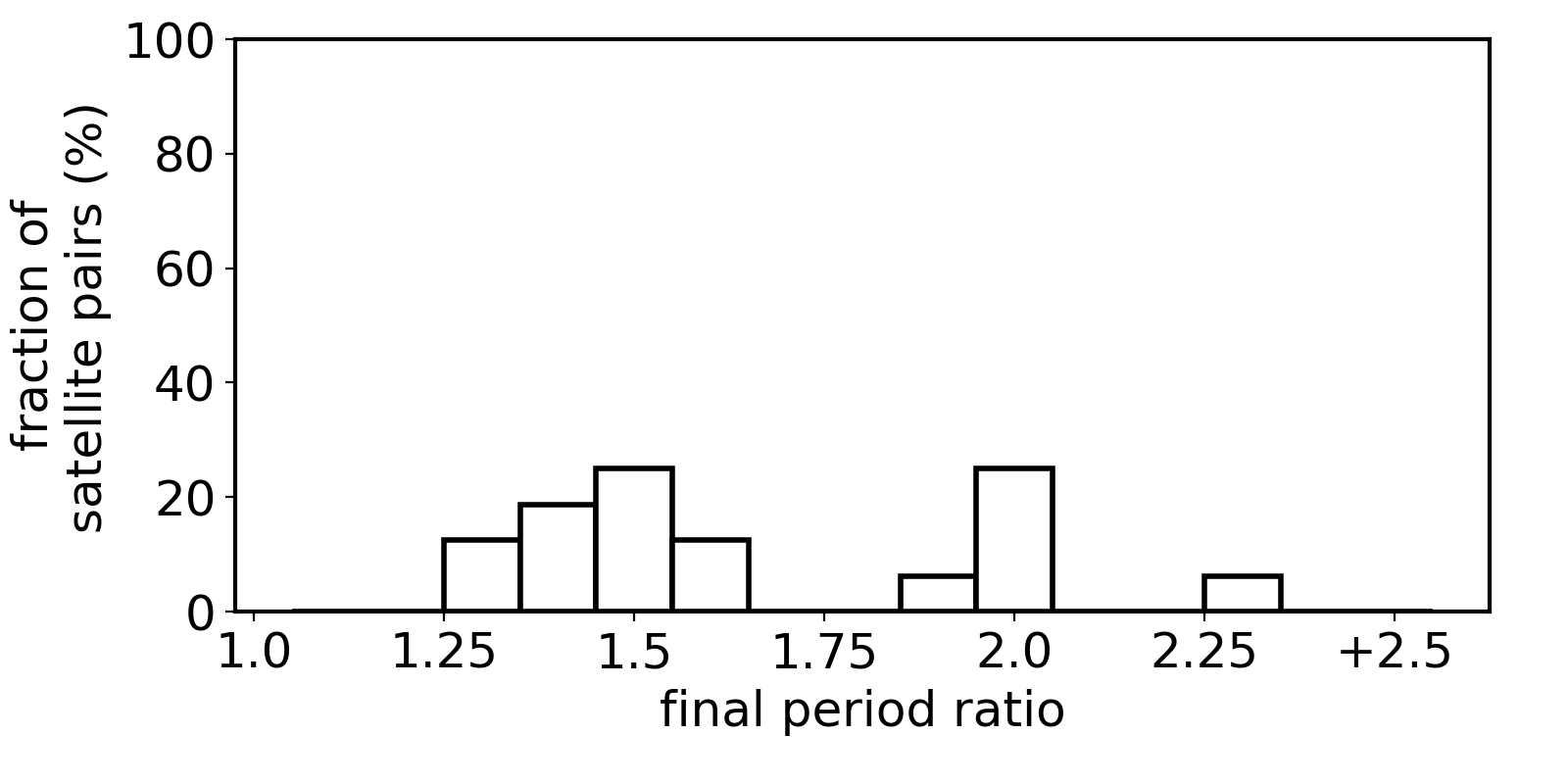} }
\caption{Period ratio distribution of adjacent satellite-pairs. Each panel shows the results of 10 simulations. The left and right side panels show the results of simulations starting with 30 and 50 satellitesimals, respectively. From top-to-bottom the panels show the results of simulations  with different pebble fluxes:  a) and b) $\dot{\rm M}_{\rm p0}=5\times 10^{-9}~{\rm M_J}$/yr, c) and d) $\dot{\rm M}_{\rm p0}=3\times  10^{-9}~{\rm M_J}$/yr, e) and f) $\dot{\rm M}_{\rm p0}=1.5\times  10^{-9}~{\rm M_J}$/yr, and g) and h) $\dot{\rm M}_{\rm p0}=10^{-9}~{\rm M_J}$/yr.\label{fig:period}}
\end{figure*}

Figure \ref{fig:number} shows that when the pebble flux decreases, from top to bottom, the final masses of the satellites also decrease. Final masses of satellites in simulations starting with 30 and 50 satellitesimals, and same pebble fluxes, are not dramatically different from each other. However, both sets of simulations show that an increase in the pebble flux by a factor of $\gtrsim$2 (from  $\dot{\rm M}_{\rm p0}=1.5 \times 10^{-9}~{\rm M_J}$/yr to  $\dot{\rm M}_{\rm p0}=5\times 10^{-9}~{\rm M_J}$/yr) is enough to change the final structure of our satellite systems from systems where satellites have masses fairly similar to those of the Galilean satellites to systems where satellites have systematically larger masses. Our overall best match to masses Galilean satellite system comes from simulations with  $\dot{\rm M}_{\rm p0}=1.5\times 10^{-9}~{\rm M_J}$/yr. The time-integrated pebble flux in this latter case corresponds to $\sim10^{-3}{\rm M_J}$. Lower pebble fluxes ($\dot{\rm M}_{\rm p0}= 10^{-9}~{\rm M_J}$/yr) produce too low-mass satellites, also inconsistent with the real system. Our simulations that best reproduce the mass of the Galilean satellites typically produce between 3 and 5 satellites. The efficiency of pebbles accretion in these systems is about 9\%, which is similar to efficiencies calculated by \cite{Ro20} ($\sim$10\%).

Figure \ref{fig:period} shows the final period ratio distribution of adjacent satellites in our simulations. Again, the left and right panels show the results of simulations starting with 30 and 50 satellitesimals, respectively. The two top panels in this figure a) and b) -- which correspond to the highest pebble flux -- show that systems with more massive satellites tend to have dynamically less compact systems (e.g. a larger fraction of pairs with period ratio $\geq2$). Our sets of simulations that better match the masses of the Galilean satellites (Figure \ref{fig:period}; panels e) and f) ) typically produce $\sim$10-20\% of satellite pairs locked in 2:1 MMR. The majority of satellite pairs in our systems is locked in more compact resonant configurations. 

\begin{figure}
\includegraphics[width=\columnwidth]{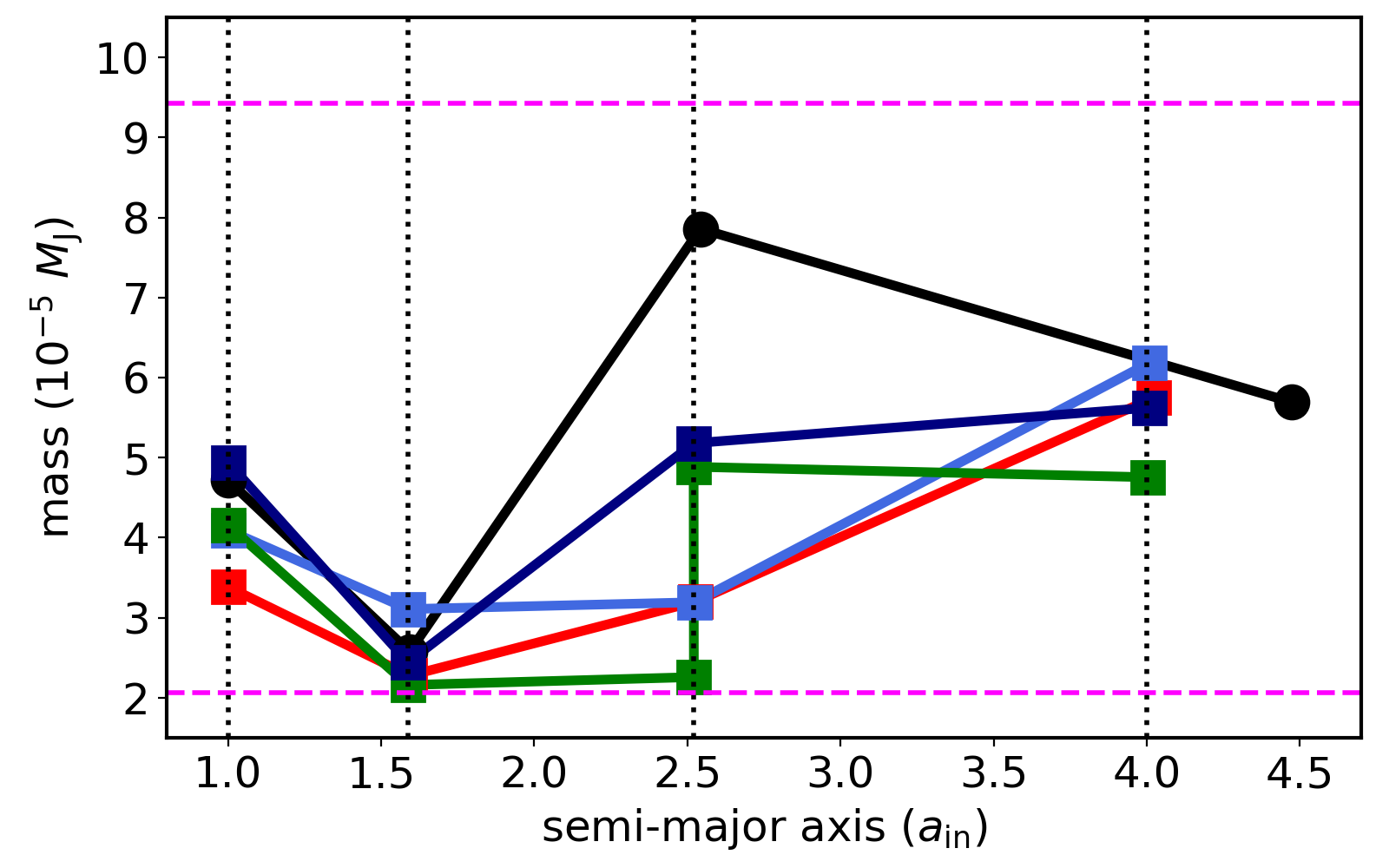}
\caption{Final masses and semi-major axes of satellites produced in our best-case simulations. In all cases, only 4 satellites are formed. For presentation purposes, we have re-scaled the position of the satellites  by a factor of order of unity to make the position of the innermost satellite in our simulations to coincide with the distance of Io to Jupiter ($a_{\rm in}$). The real Galilean system is given by the black line with a circle and the vertical dotted lines give the location of the 8:4:2:1 resonant chain and the pink vertical lines give the limits on mass of our third constraint (0.8${\rm M_E}$ and 1.2${\rm M_G}$). Our final systems show no radial mass ranking. The system represented by the green solid line shows a co-orbital satellite with the third innermost satellite.\label{fig:triangular}}
\end{figure}

Figure~\ref{fig:triangular} shows the final semi-major axis versus the mass of satellites in selected systems that satisfy all constraints defined in Section \ref{sec:constraints}. As in Figure~\ref{fig:number}, each satellite is shown by a colorful point (symbol) -- where lines connect satellites in a same system. The system shown in red represents our nominal analogue that will be further discussed later in the paper. The real Galilean system is shown in black. The three innermost vertical dotted lines mark the respective positions of the three innermost Galilean satellites. The outermost vertical dotted line shows the location of Callisto if it was also locked in a 2:1 mean motion resonance with Ganymede. The horizontal pink lines show the limiting masses defined in constraint iii) of Section \ref{sec:constraints}. These systems match fairly well the masses and resonant configurations of the Galilean satellite system. However, the outermost satellite in our simulations is always locked in a 2:1 MMR with the second outermost one. As discussed before, Callisto is not locked in resonance with Ganymede. We do not consider this to be a critical issue for our model because Callisto may have left the resonance chain via divergent migration due to tidal-interaction of satellites with Jupiter \citep{Fu16,downey2020inclination}.

Figure~\ref{fig:triangular}  also shows that in our best analogues,  the mass of the innermost satellite is typically larger than that of the second innermost one, which is also true in the Galilean system. However, the masses of our Ganymede-analogues are typically lower than that of the real satellite. Finally, one of the important result of our model is that the radial-mass distribution of our satellites do not present any radial mass-ranking (e.g. satellites' mass increases with their distance to Jupiter)  or triangular-mass distribution (e.g. the innermost and outermost satellites are the less massive). These are typical issues found in previous models \cite[e.g.][]{Mo03a,Mo03b,Cr12}.

\subsection{The Dynamical Architecture of our Systems}

\begin{figure}
\subfigure[]{\includegraphics[width=0.95\columnwidth]{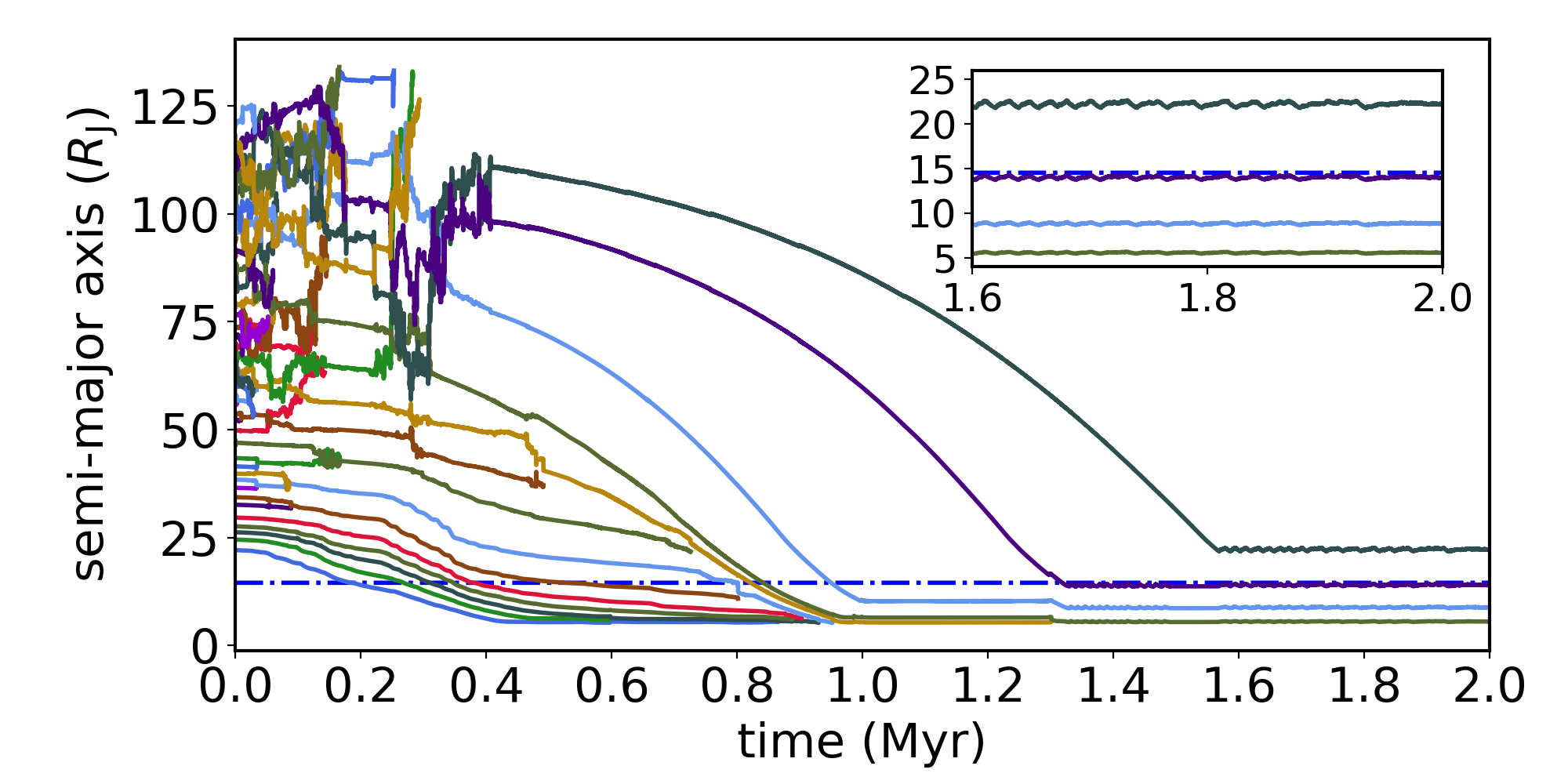}}
\subfigure[]{\includegraphics[width=0.95\columnwidth]{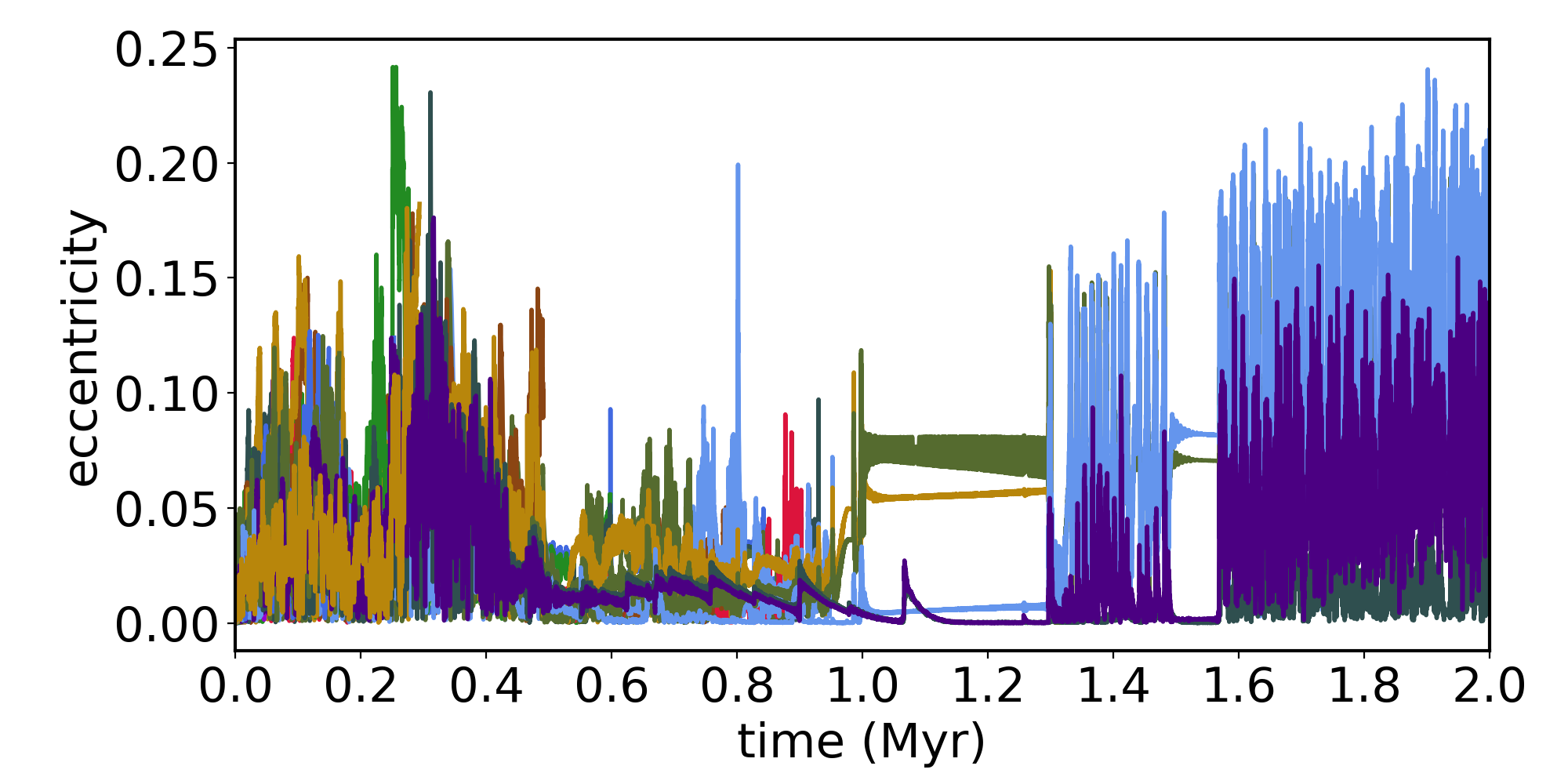}}
\subfigure[]{\includegraphics[width=0.95\columnwidth]{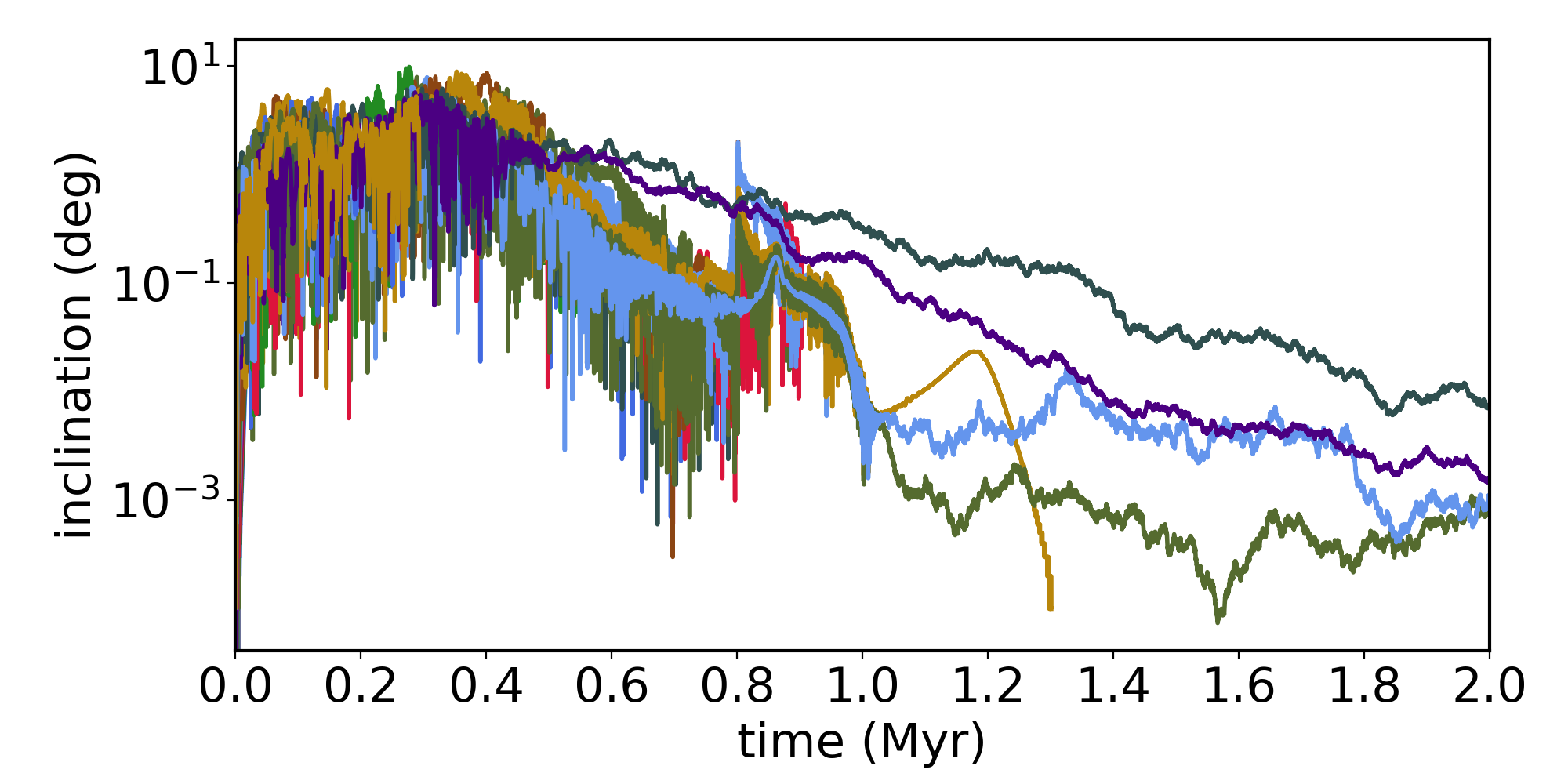}}
\subfigure[]{\includegraphics[width=0.95\columnwidth]{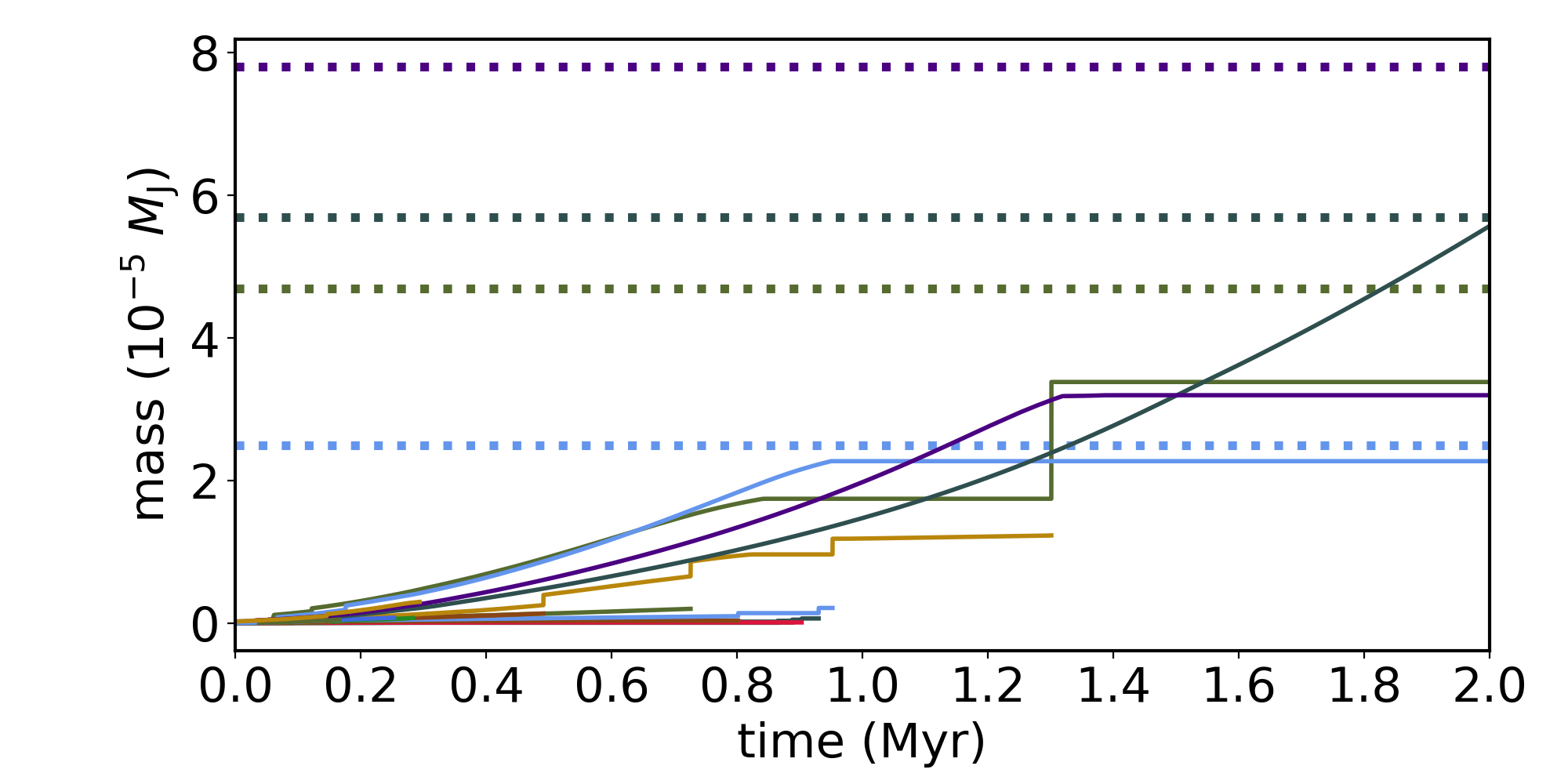}}
\caption{{\bf From to to bottom:} Evolution of semi-major axes, orbital eccentricities, orbital inclinations, and masses of satellites in a simulation with $\dot{\rm M}_{\rm p0}=1.5\times 10^{-9}~{\rm M_J}$/yr and starting with 50 satellitesimals. The dot-dashed line in the top panel shows the snowline location, and horizontal lines in the bottom panel show the masses of the real Galilean satellites. These horizontal line matches the color of the analogues to indicate the mass they should have to be a perfect match. All satellite pairs evolve in a 2:1 MMR.\label{fig:system3}}
\end{figure}

In this section, we present the dynamical evolution of some of our best Galilean system analogues. Figure~\ref{fig:system3} shows a simulation  with $\dot{\rm M}_{\rm p0}= 1.5\times 10^{-9}~{\rm M_J}$/yr, and starting with 50 satellitesimals. From top to bottom, the panels show the temporal evolution of semi-major axis, orbital eccentricity, orbital inclination, and masses of satellitesimals/satellites. The colorful horizontal dotted lines in the bottom panel match the color of one of the analogues, and they show the mass that each analogue should have to exactly match the mass of its corresponding Galilean satellite. Satellitesimals first grow by pebble accretion and start to migrate inwards. The innermost satellitesimal reaches the inner edge and stop migrating. As additional satellitesimals converge to the inner edge of the disk, orbital eccentricities and inclinations start to increase, which leads to scattering events and collisions. Finally, the satellitesimals converge into a resonant configuration anchored at the disk inner edge. In this case, all adjacent satellite pairs are evolving in a 2:1 MMR.  Their final orbits typically exhibit very low orbital inclinations ($\sim$0.01~deg). However, the final orbital eccentricities of our satellites are modestly high ($\sim$0.05-0.2). Table  \ref{tab:intro} shows that the orbital eccentricities of the Galilean satellites are much lower than those of analogues in Figure \ref{fig:system3}. More importantly, this is a trend present in all our best analogues. In the next section, we attempt to address this issue by slightly changing our disk model.

\subsection{Simulations with a more realistic cooling CPD} \label{subsec:nominal}

In all our previous simulations, we have neglected the evolution of the disk temperature. However, as the disk evolves, it looses masses and gets colder and thin \citep[e.g.][]{Sz16}.  In this section we replace Eq. \ref{eq:Tprofile}, representing our original CPD temperature profile, by the following new temperature profile \citep{Sa10,Ro17}
\begin{equation}
T_{\rm new}=225\left(\frac{r}{10~{\rm R}_{\rm J}}\right)^{-3/4}e^{-\frac{t}{4\tau_d}}~K \label{eq:Tprofilenew},
\end{equation}
where $t$ represents the time, and $\tau_d$=1~Myr.

In our simulations where the disk temperature does not evolve in time, the CPD aspect ratio at 15~R$_{\rm J}$ (Ganymede's distance to Jupiter) is $\sim0.06-0.07$. In our new disk setup, the disk aspect ratio at 15~R$_{\rm J}$ also at the end of the gas disk phase is $\sim0.04$. We have conducted 40 new simulations, considering that the CPD temperature decays in time as Eq.~\ref{eq:Tprofilenew}.

\begin{figure}
\subfigure[]{\includegraphics[width=0.95\columnwidth]{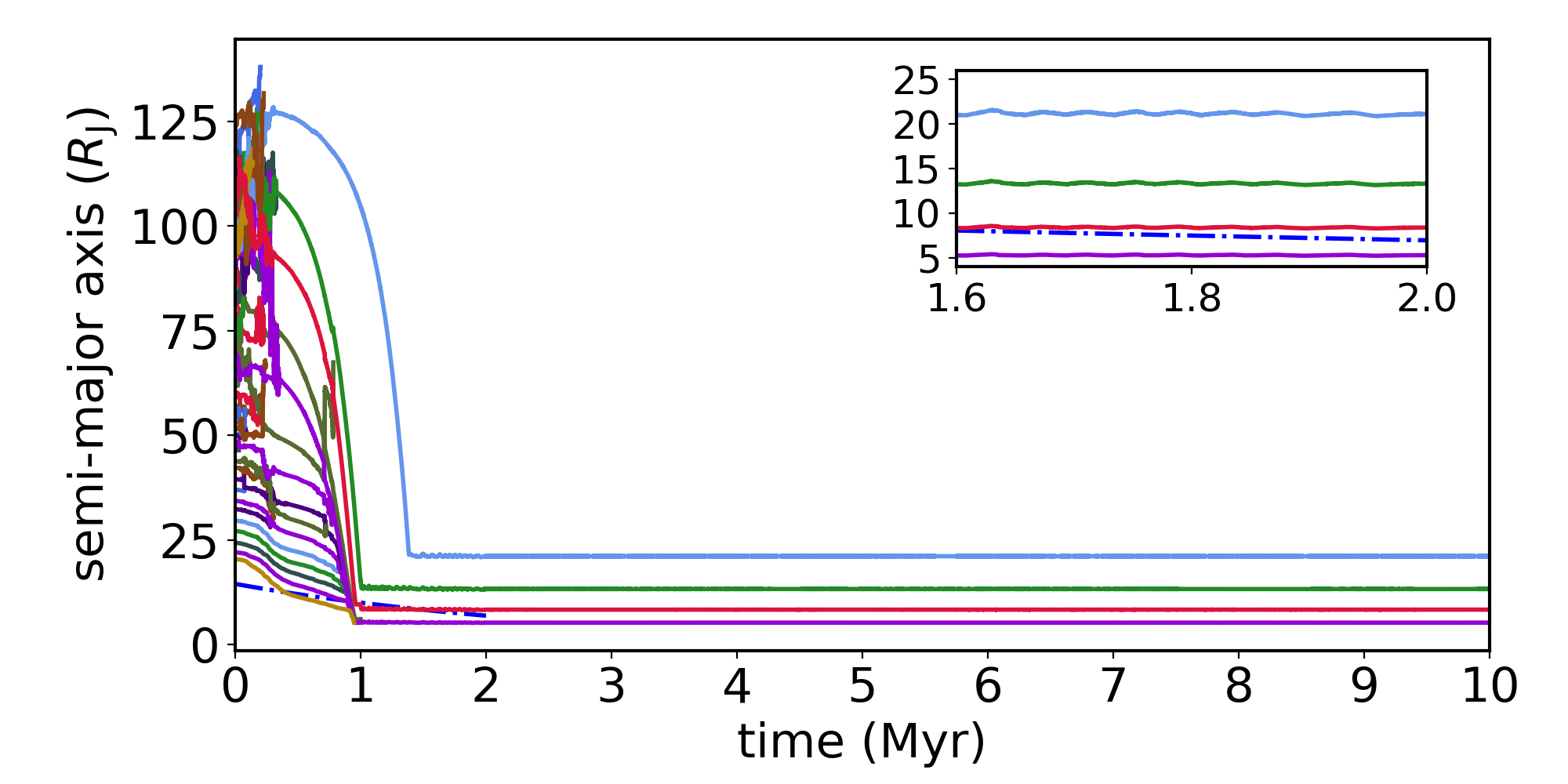}}
\subfigure[]{\includegraphics[width=0.95\columnwidth]{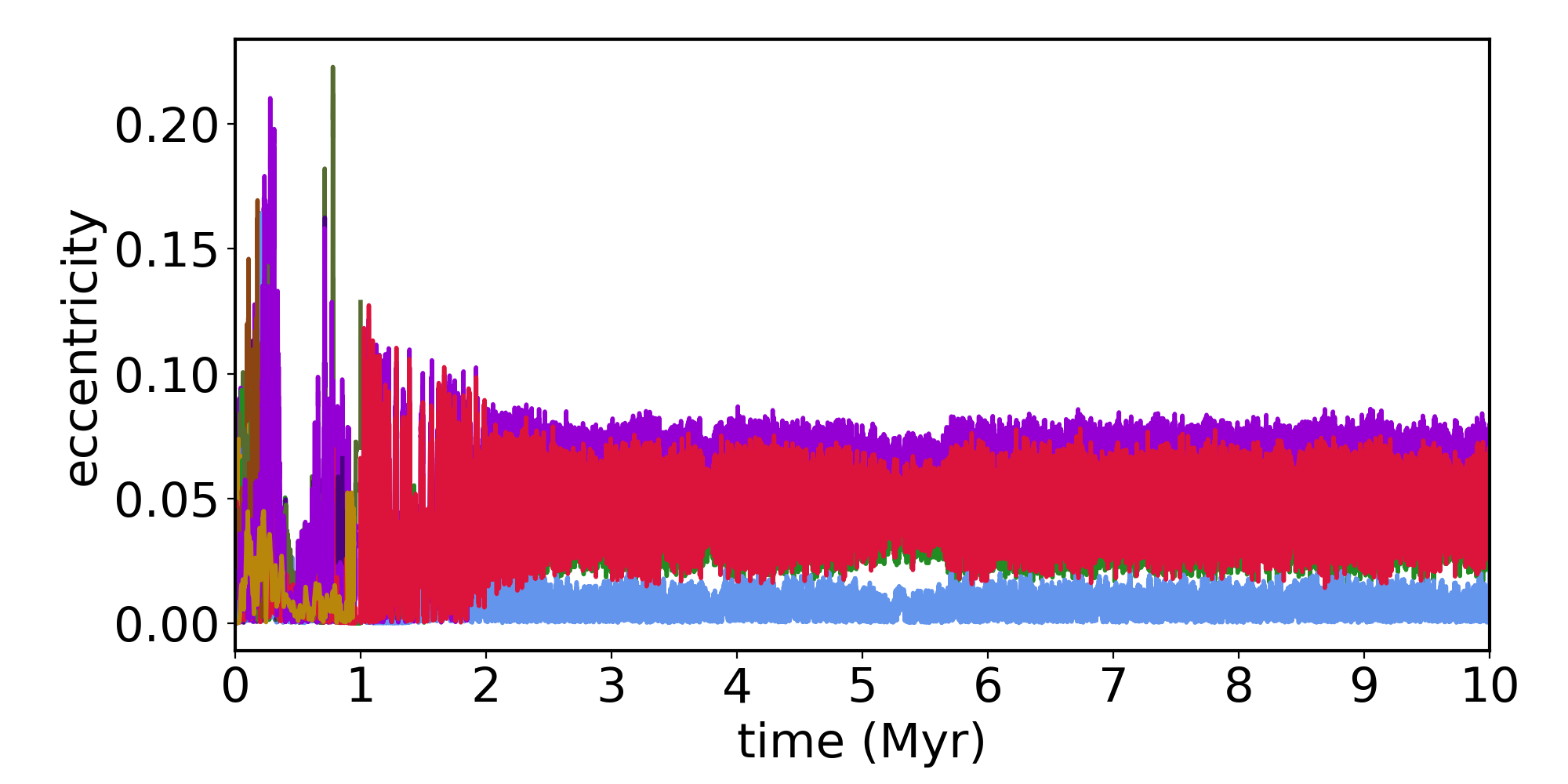}}
\subfigure[]{\includegraphics[width=0.95\columnwidth]{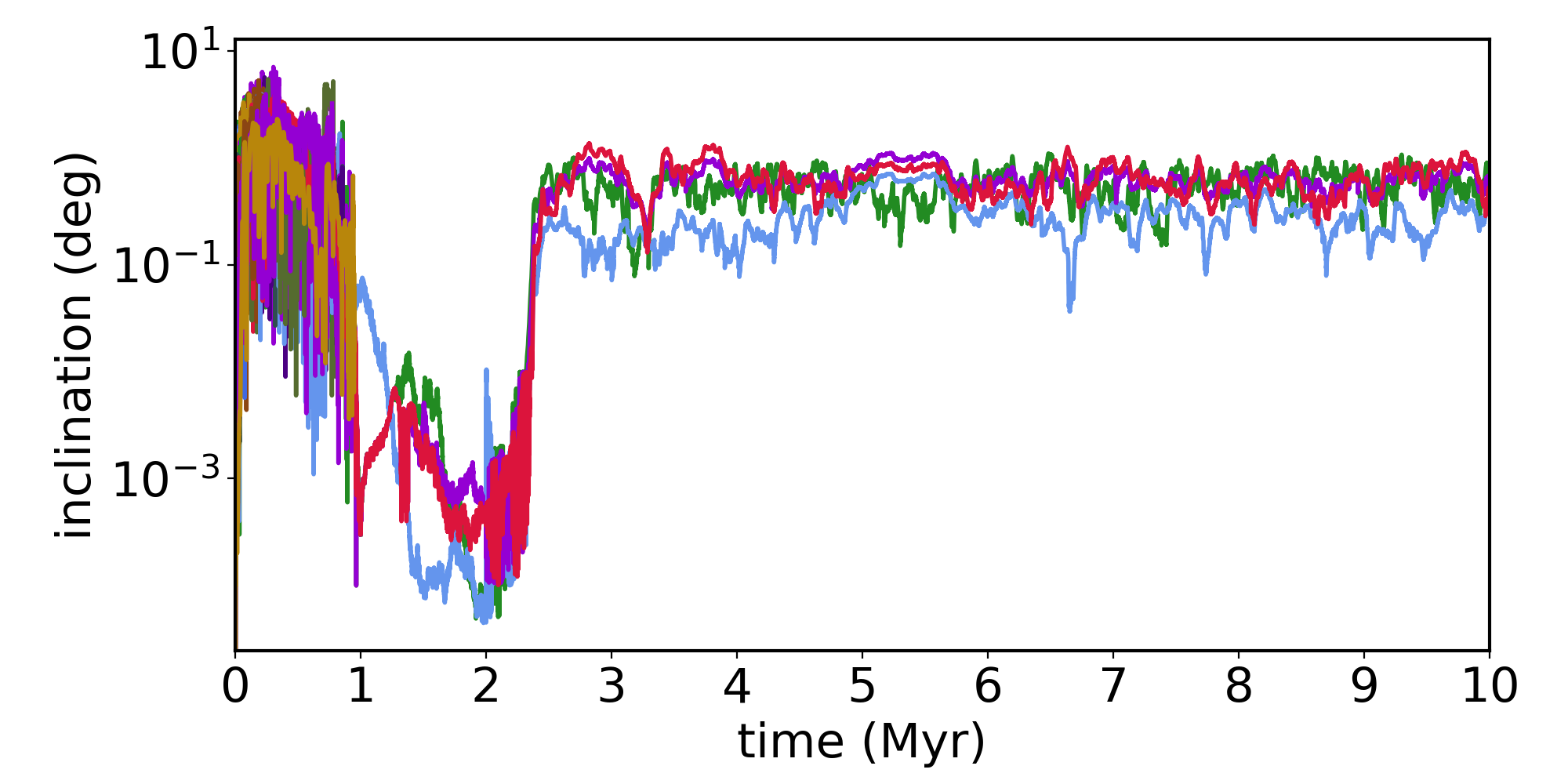}}
\subfigure[]{\includegraphics[width=0.95\columnwidth]{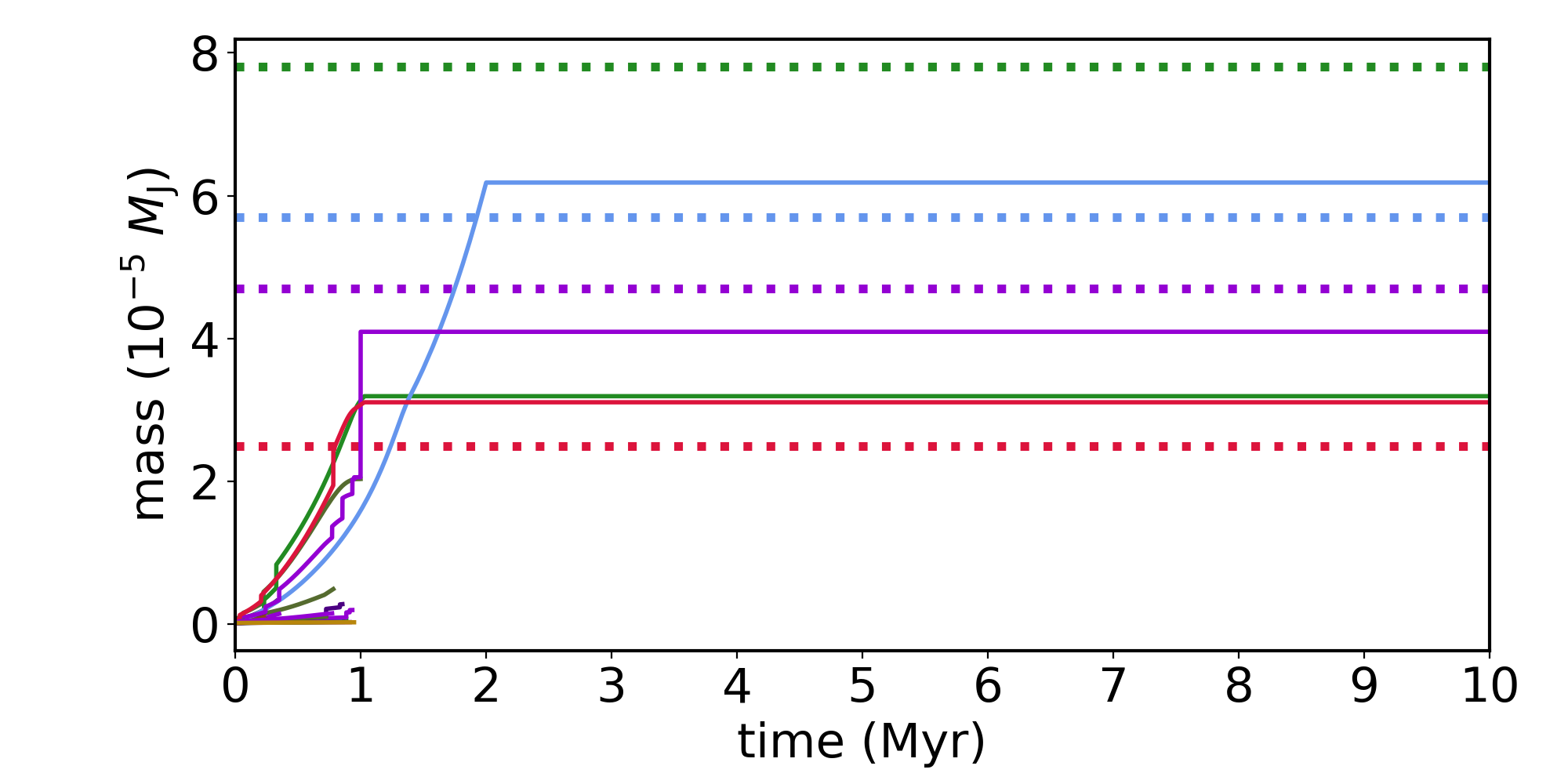}}
\caption{{\bf From top to bottom:} Temporal evolution of semi-major  axis,  orbital  eccentricity,  inclination,  and  mass of satellitesimals in a simulation starting with 50 satellitesimals. As the disk cools down, the snowline moves inwards as shown by the blue dotted line in the top panel. The horizontal lines in the bottom panel show the masses of the real Galilean satellites. These horizontal line matches the color of the analogues to indicate the mass they should have to be a perfect match. All satellite adjacent pairs are involved in a 2:1 MMR, forming a resonant chain. The gas disk dissipates at 2~Myr and the system remains dynamically stable up to 10~Myr. \label{fig:nom2}}
\end{figure}

Figure~\ref{fig:nom2} shows the dynamical evolution of one of our good analogues (see Section \ref{sec:constraints}). The gas in the CPD dissipates at 2~Myr and we follow the long-term dynamical evolution of these satellites up to 10~Myr. Figure~\ref{fig:nom2} shows, from top to bottom, the evolution of semi-major axis, eccentricity, inclination, and masses of the satellitesimals. This simulation starts with 50 satellitesimals. The evolution of satellitesimals in this case is remarkably similar to that of simulations where the disk temperature is kept fixed in time. The combined pebble accretion efficiency of the system is about 13\%, which is also similar to our other simulations with the exponential decay of the disk temperature. At the end of the gas disk phase, at 2~Myr, the satellites form a resonant chain anchored at the disk inner edge with all-three satellite pairs evolving in a 2:1 MMR. The final orbital eccentricities of satellites are lower than those observed in Figure \ref{fig:system3}, as expected. However, the orbital eccentricities of our analogues are still too high compared to those of the Galilean system.

\begin{figure}
\includegraphics[width=\columnwidth]{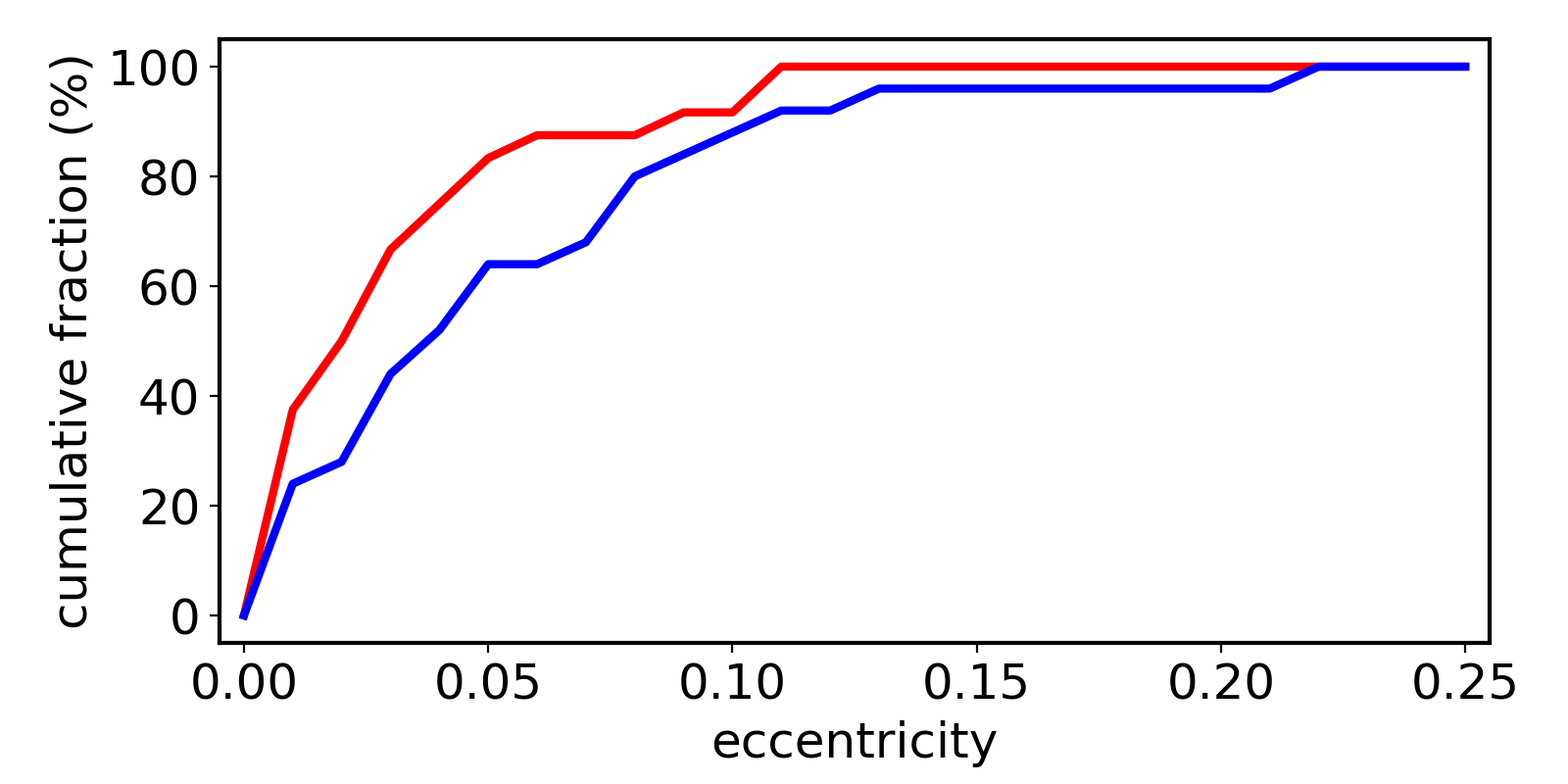}
\caption{Eccentricity distribution of our final satellites in simulations with $\dot{\rm M}_{\rm p0}=1.5\times 10^{-9}~{\rm M_J}$/yr and 50 satellitesimals. The blue curve shows the case where the disk temperature is kept constant during the gas disk phase. The red line shows the case where the disk temperature decays exponentially in time (see Eq. \ref{eq:Tprofilenew}).   \label{fig:cumulative}}
\end{figure}

We have found that the orbital eccentricities of satellites in simulations where the disk cools down in time are overall lower than those where the disk temperature is kept constant. This is not a surprising result because the equilibrium eccentricity of capture in resonance  scales as $e_{\rm eq}\sim h_g$~\citep{goldreichschlichting14,deckbatygin15,pichierietal18}. This is in agreement with the results of our simulations (see Figure~\ref{fig:4sat} and \ref{fig:system3}). However, the observed eccentricities of our final satellites are still too high compared to those of the real satellites today (Figure \ref{fig:cumulative}). One could conjecture that this is because the disk aspect ratio of our new disk is still not low enough. However, in order to have a disk aspect ratio of $\leq$0.01-0.001 at $<$15~$R_{\rm J}$ -- which could more easily lead to final satellites with eccentricities similar to those of Galilean satellites -- it would require a CPD with gas temperature of about $\leq$5~K at  15~$R_{\rm J}$, which is  unrealistic \citep[e.g.][]{Sz16}. So we suggest that in order to damp the orbital eccentricities of our analogues down to the level of the Galilean satellites, a different mechanism should be invoked. We will conduct new simulations and discuss a possible solution to this issue in Section \ref{sec:faketides}.

\subsection{The Water-Ice mass fraction of our satellites}

\begin{figure*}
\includegraphics[width=1.8\columnwidth]{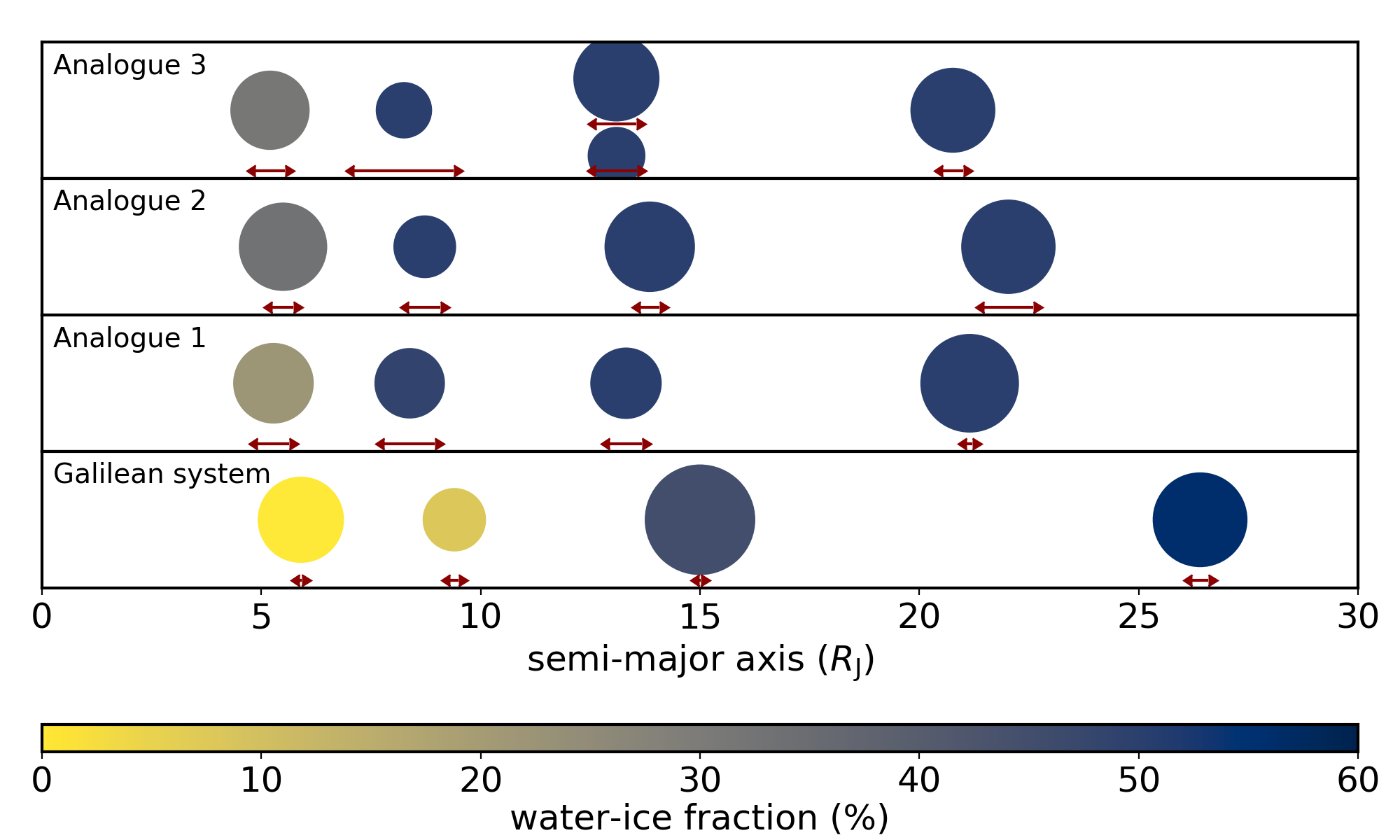}
\caption{Galilean system analogues at the end of the gas disk phase (2~Myr). Each line shows a satellite system produced in our simulations. Individual satellites are represented by color-coded dots. Dot's size  scales linearly with the satellite's mass and color represents its water-ice fraction. The horizontal axis shows satellites' orbital semi-major axis. Orbital eccentricities  are represented by the horizontal red bars showing the variation in heliocentric distance over semi-major axis ($R_{\rm J}$). The three systems presented here are also shown in Figure~\ref{fig:triangular}.\label{fig:compilation}}
\end{figure*}

In this section, we analyze the water mass fraction of the satellites in our best Galilean system analogues. Figure~\ref{fig:compilation} shows three of our best analogues. Each line shows one satellite system and the horizontal axis shows its distance from Jupiter. The color-coded dots show individual satellites, where the color represents their water-ice fraction and dot sizes scale linearly with mass. The real Galilean system is shown at the bottom, for reference. The orbital eccentricity of each satellite is represented by the horizontal red bars showing the variation in the planetocentric distance over the semi-major axis ($R_{\rm J}$). In all these systems, adjacent satellite pairs evolve in a 2:1 MMR. 

It is clear in Figure \ref{fig:compilation} that the water-ice fraction of the innermost satellites in our simulations are significantly higher than those of Io and Europa. However, in all our systems the two outermost satellites have water-ice rich compositions similar to those of Ganymede and Callisto. This is a trend observed in all our simulations. In this work, we assume that pebbles inside the snowline are 0.1~cm size silicate particles and those outside are larger, 1~cm size icy-particles. Therefore, pebbles outside the snowline are far more efficiently accreted by the satellitesimals than those inside it. Consequently, satellitesimals beyond the snowline grow faster than those inside it and tend to starve the innermost satellitesimals by reducing the flux of pebbles that they receive. As distant more massive objects migrate inwards they collide with lower-mass satellites growing by the accretion of silicate pebbles. Typically, the innermost satellite in our simulations is a satellitesimal that -- on its way to the inner edge of the disk -- has collided with several satellitesimals growing inside the snowline. This tends to affect its final water-mass-fraction as observed in Figure \ref{fig:compilation}.  However, in none of our simulations, our innermost analogue is as dry as Io or Europa. We have performed a limited number of simulations where we increase the sizes of the pebbles inside the snowline by a factor of 3, to attempt to accelerate the growth of the innermost satellites, but none of these simulations produce good analogues satisfying the i)-iv) conditions of Section \ref{sec:constraints}.

As discussed before, it is not clear if Io and Europa were born with water-ice poor compositions and then lost (most/all) their water-content after formation \citep{Bi20} or formed with their current compositions. So if Io and Europa formed by pebble accretion and with water-poor compositions they must have formed very early, probably much earlier than Ganymede and Callisto,  and at a time where satellitesimals existed (or were able to efficiently grow via pebble accretion) only well inside the disk snowline.

On the other hand, if Io and Europa lost their water via hydrodynamic escape \citep{Bi20} the masses of the innermost satellites in our simulations should be reduced by a factor of 1.2-1.7 to account for the water-ice component loss. We do not simulate this effect here, but if this is the case our innermost satellite would still have mass satisfying constraint iii) in most of our analogues. This is not true for the second innermost satellites in our analogues (Figure \ref{fig:compilation}). But, as some of our simulations do produce Europa analogues  twice as massive as the current Europa (Figure \ref{fig:compilation}) our lack of a good match in our analogues' sample  is probably a consequence of the stochasticity of these simulations and small number statistics. 

Finally, the top system in Figure \ref{fig:compilation} shows that the third innermost analogue shares its orbit with a co-orbital satellite.  Note that in our system analogues, Ganymede analogues are typically not as massive as the real Ganymede (but masses typically agree within a factor of $\sim$2).  A potential future collision of these co-orbital satellites could result in a satellite with final mass even closer to that of Ganymede.

\section{Mimicking the long-term dynamical evolution of our Galilean system analogues}\label{sec:faketides}

In our simulations, when the gas in the CPD dissipates, at 2~Myr, orbital eccentricity damping and gas-driven migration cease. Our simulations are numerically integrated at most for additional 8~Myr in a gas-free environment where satellites and Jupiter gravitationally interact as point-mass particles. However, in reality, the long-term dynamical evolution of the Galilean satellites is also modulated by tidal interactions with Jupiter and other satellites \citep{Pe02,Fu16,lari2020long}. Tidal effects tend to increase the angular momentum of the satellites -- that migrate outwards because the Galilean satellites are outside the centrifugal radius of Jupiter -- and decrease their orbital eccentricities \citep{Gr73}. The resonant configuration of the Galilean satellites is expected to enhance the effects of planet-satellite tidal dissipation \citep{lari2020long}.

In the theory of dynamical tides, if each satellite has a different effective tidal quality factor $Q$ due to resonance locking between moons and internal oscillation modes of Jupiter (see Eq.~\ref{fuller} and \ref{k2Q}), tidal migration can be divergent~\citep{Fu16}. To estimate the potential effects of the resonance locking on the long-term dynamical evolution of our system analogues, we conduct some additional numerical simulations where we use a simple prescription to crudely mimic the effects of tides on our satellite systems. Our main goal here is to test if the three innermost satellites remain in resonance when subject to tidal dissipation forces and to infer the level of dynamical excitation of our final systems.

\begin{figure}
\includegraphics[width=\columnwidth]{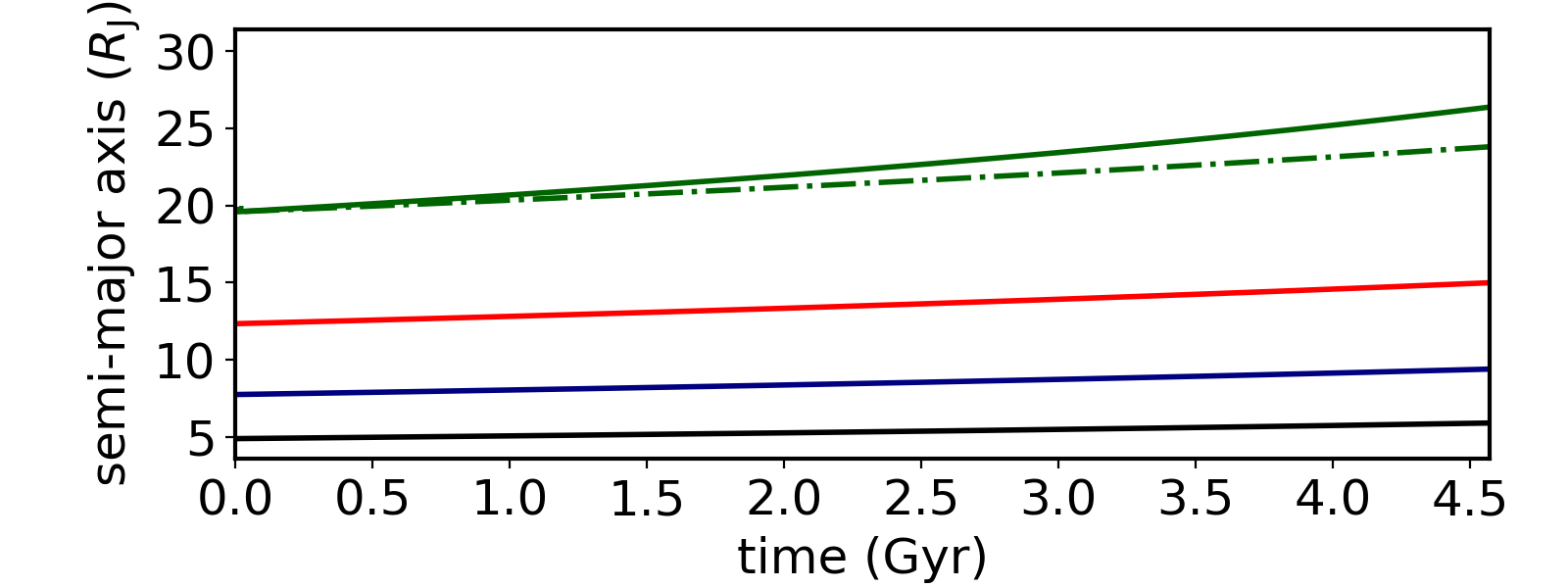}
\caption{Temporal evolution of the semimajor axis of the Galilean satellites over the Solar System age as estimated via resonance locking theory (Fuller et al 2016). The values of $t_\alpha$ for Io, Europa, Ganymede, and Callisto are 44, 101, 217, and 318~Gyr, respectively. The dot-dashed line corresponds to the evolution of the 2:1 MMR with Ganymede.\label{fig:edampSMA}}
\end{figure}

The migration timescale of a satellite via dynamical tides is given by \citep{Fu16,nimmo2018,downey2020inclination}:
\begin{equation}
\frac{1}{t_a}=\frac{|\dot{a}|}{a}=\frac{2}{3}\frac{1}{t_\alpha}\left(\frac{\Omega}{n}-1\right), \label{fuller} 
\end{equation}
with
\begin{equation}
\frac{k_2}{Q}=\frac{1}{3nq_s}\left(\frac{a}{\rm R_J}\right)^5\frac{1}{t_a}, \label{k2Q}
\end{equation}
where $\Omega$ and $t_\alpha$ are Jupiter's rotation frequency and internal oscillation modes timescale and $k_2$ is the Love number. Astrometric observations \citep{valery2009} suggest that three innermost Galilean satellites migrate outwards with timescales of 20~Gyr, which translates to $t_\alpha$ equal to 44, 101, and 217~Gyr, from the Io to Ganymede \citep{downey2020inclination}. There is no estimate of the current tidal migration timescale of Callisto via astrometric observations~\citep{downey2020inclination}. So we use equation~\ref{fuller} to estimate Callisto's current tidal migration timescale assuming that: i) Callisto is initially locked in a 2:1 MMR with Ganymede; ii) Ganymede is initially at 13~$R_{\rm J}$ (consistent with our Analogue 1); iii) Callisto migrates to its current position over the age of the Solar System; iv) the three innermost satellites remain locked in the Laplace resonance.  We obtained for Callisto a $t_a=12$~Gyr ($t_\alpha=318$~Gyr). Figure~\ref{fig:edampSMA} shows the estimated position of the satellites as a function of time, given by equation~\ref{fuller}. The solid lines show the semi-major axis evolution of each satellite. The green dot-dashed line shows the position of the 2:1~MMR with Ganymede (red solid line). As one can see, at the end of the simulation at 4.57~Gyr, Callisto migrates outwards faster than the 2:1~MMR with Ganymede moves outwards, leaving the resonance configuration. The three innermost satellites, on the other hand, remain locked in the Laplace resonance as we will show later. With the migration timescales of all satellites in hand, we can now perform  N-body numerical simulations to probe the long-term dynamical evolution of our analogues.

To perform our simulations mimicking the effects of dynamical tides, we use the final orbital configuration of the system at the end of the gas disk phase (Analogue~1) and we consider the subsequent evolution of the system in a gas-free scenario. Due to the high computational cost of these simulations, we do not numerically integrate our system for the entire age of the Solar System, but only for 20~Myr (which requires 3 weeks of CPU-time in a regular desktop).  We have assumed  for the three innermost satellites $t_a=20$~Gyr, as suggested by observations~\citep{valery2009}, and $t_a=12$~Gyr for Callisto as discussed before. When performing our N-body simulations we also apply eccentricity damping to the satellites, which is assumed to correlate with the tidal migration timescale via the factor $S$ given as :
\begin{equation}
S=\frac{1/t_e}{1/t_a}=\frac{|\dot{e}/e|}{|\dot{a}/a|}   
\end{equation}
here we test $S=10^5$ and $S=10^6$. These values are based on the ratio between semimajor axis and eccentricity variation timescales for Galilean satellites in classical tidal theory \citep{goldreich1966q,zhang2004,lainey2017new}.

\begin{figure}
\subfigure[]{\includegraphics[width=\columnwidth]{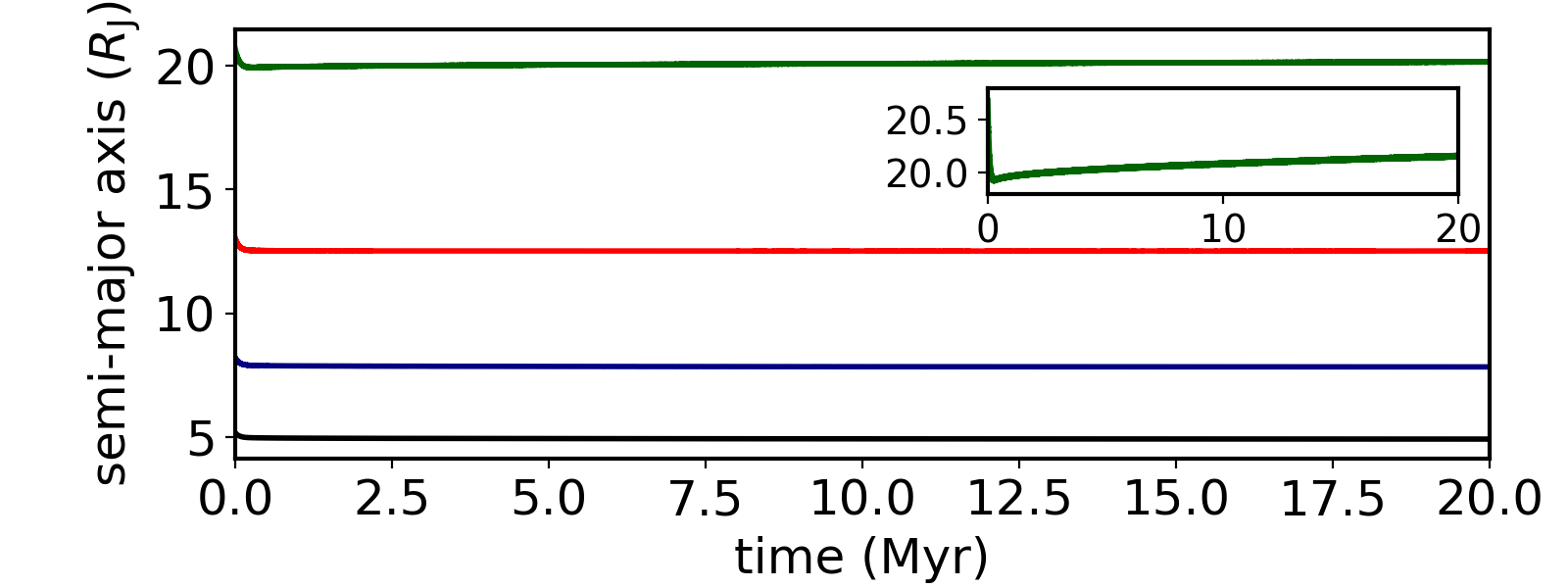}}
\subfigure[]{\includegraphics[width=\columnwidth]{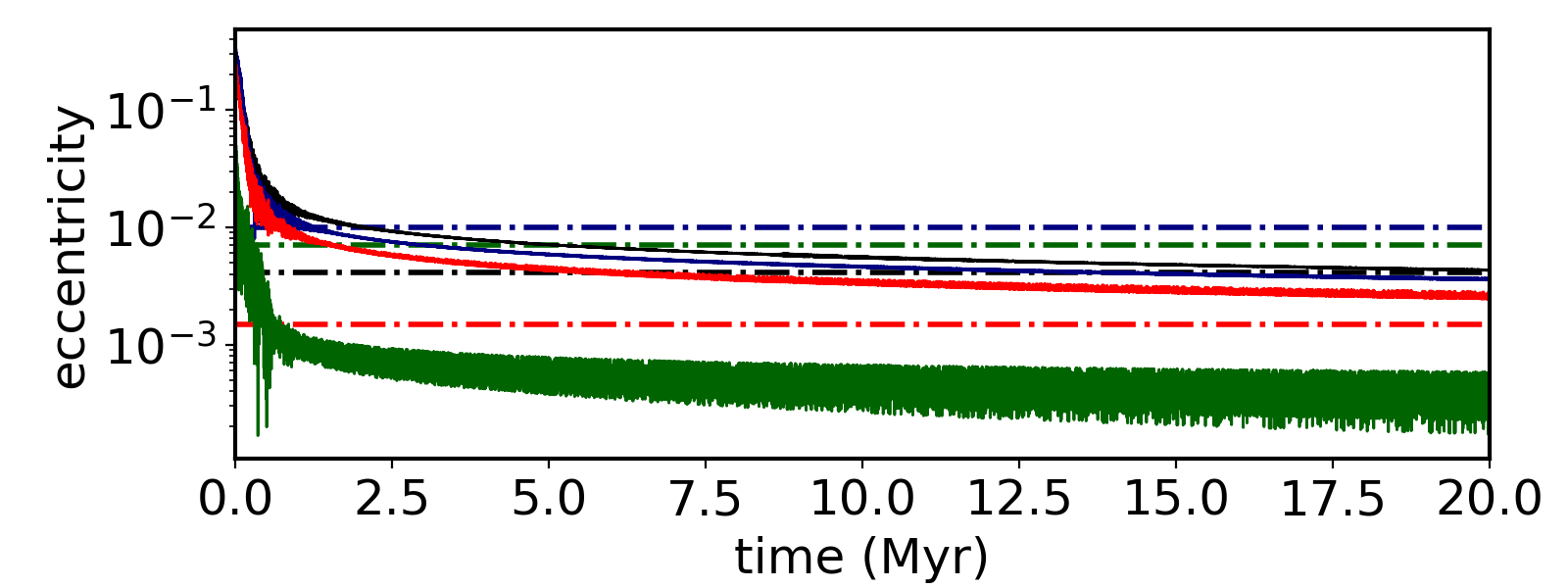}}
\caption{a) Semi-major axis and b) eccentricity of the satellites in Analogue 1 for $S=10^5$. Each solid line shows the evolution of one satellite. Dot-dashed horizontal lines show the orbital eccentricities of the Galilean satellites for reference.\label{fig:edampF}}
\end{figure}
Figure~\ref{fig:edampF} shows the semi-major axes and orbital eccentricities' evolution of satellites in Analogue~1 during 20~Myr. Our analogue satellites migrate outwards very slowly and the system remains dynamically stable. The bottom panel of Figure~\ref{fig:edampF} shows that the orbital eccentricities of our analogues are damped to values consistent with those of the real Galilean satellites in a relatively short timescale. The Figure~\ref{fig:edampF} corresponds to $S=10^5$. Our simulation where $S=10^6$ resulted in an even lower level of orbital excitation at 20~Myr. 

\begin{figure}
\subfigure[$\phi=2\lambda_2-\lambda_1-\varpi_1-180^{\circ}$ (red line) and $\phi=2\lambda_2-\lambda_1-\varpi_2$ (blue line)]{\includegraphics[width=\columnwidth]{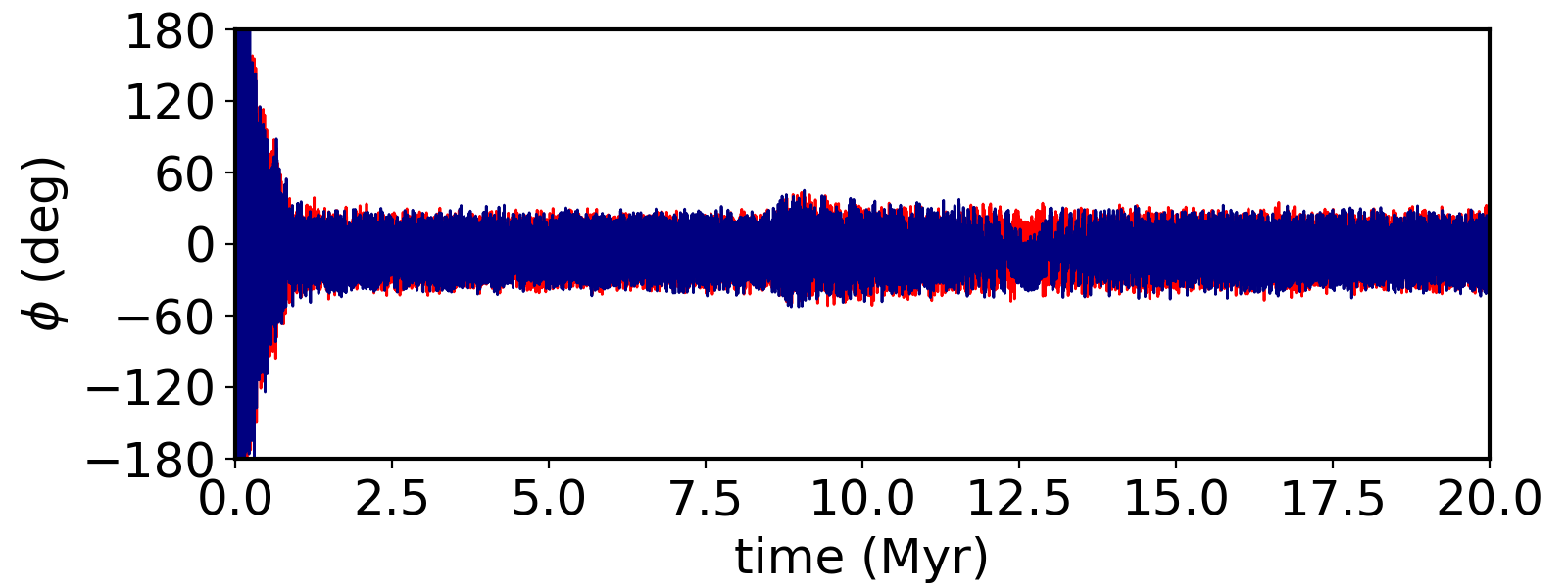}}
\subfigure[$\phi=2\lambda_3-\lambda_2-\varpi_2-180^{\circ}$~(red~line)~and~$\phi=2\lambda_3-\lambda_2-\varpi_3$~(blue~line)]{\includegraphics[width=\columnwidth]{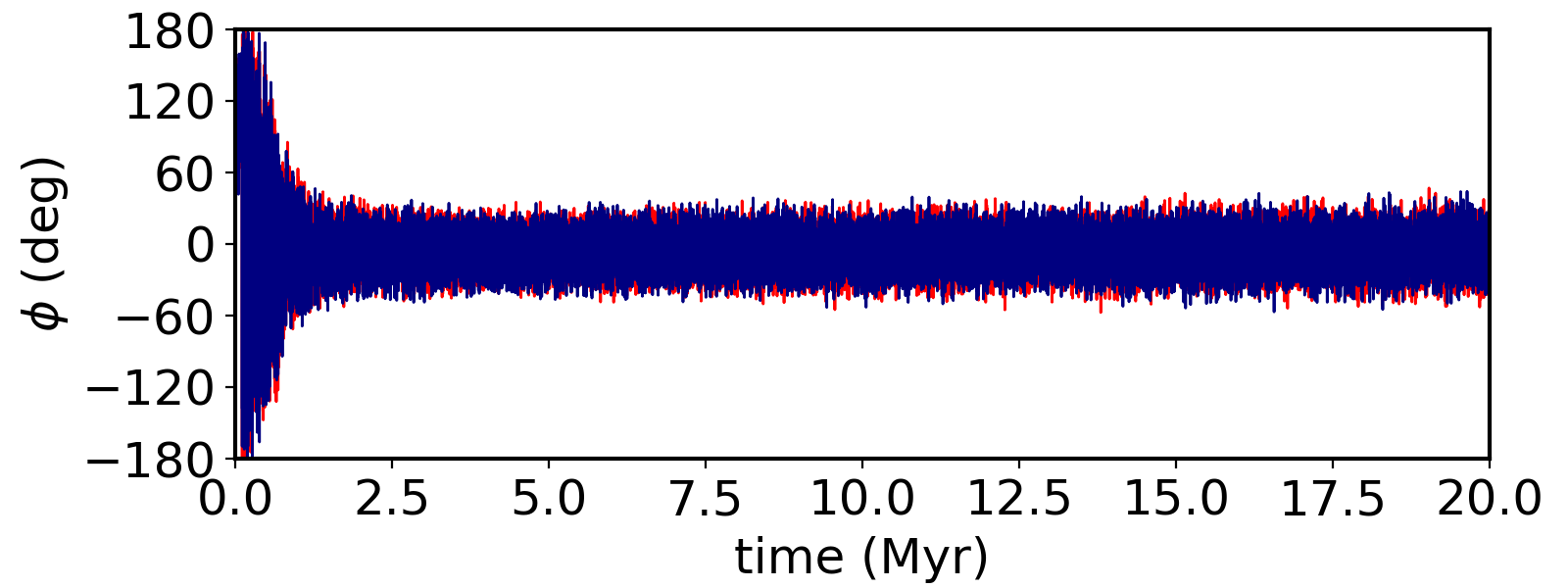}}
\subfigure[$\phi=2\lambda_4-\lambda_3-\varpi_3-180^{\circ}$ (red line) and $\phi=2\lambda_4-\lambda_3-\varpi_4$ (blue line)]{\includegraphics[width=\columnwidth]{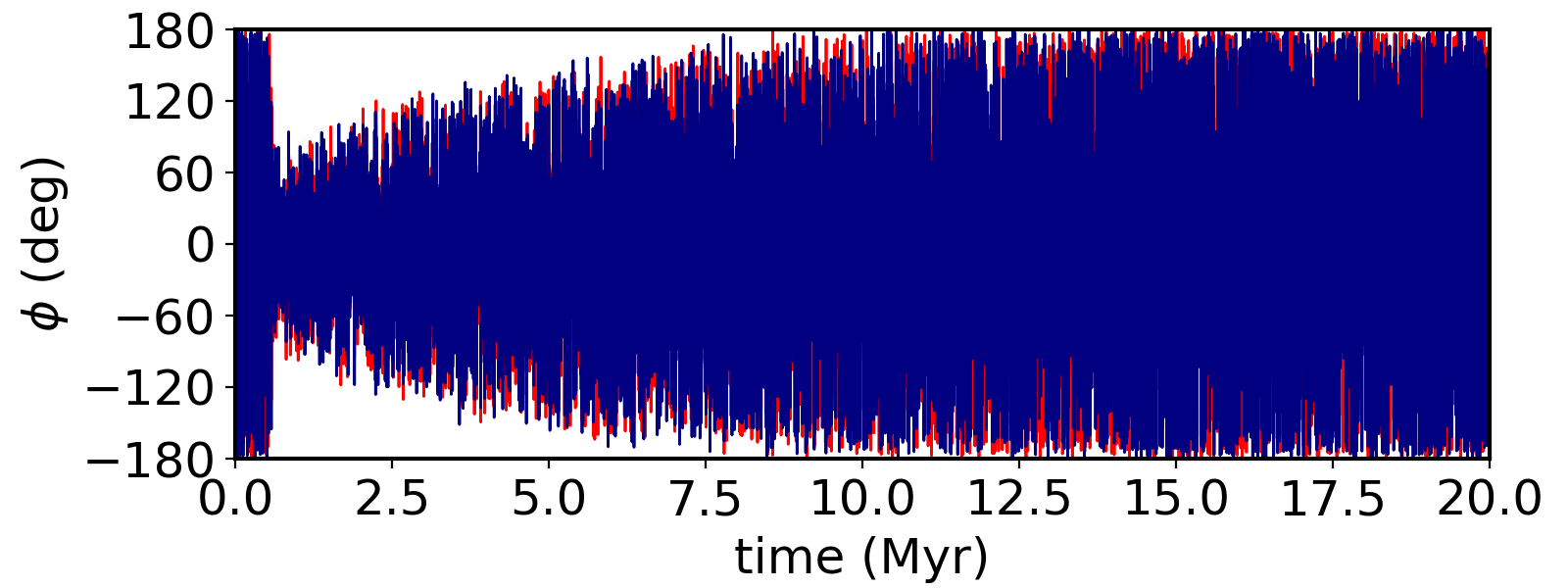}}
\subfigure[$\phi=2\lambda_3-3\lambda_2+\lambda_1$]{\includegraphics[width=\columnwidth]{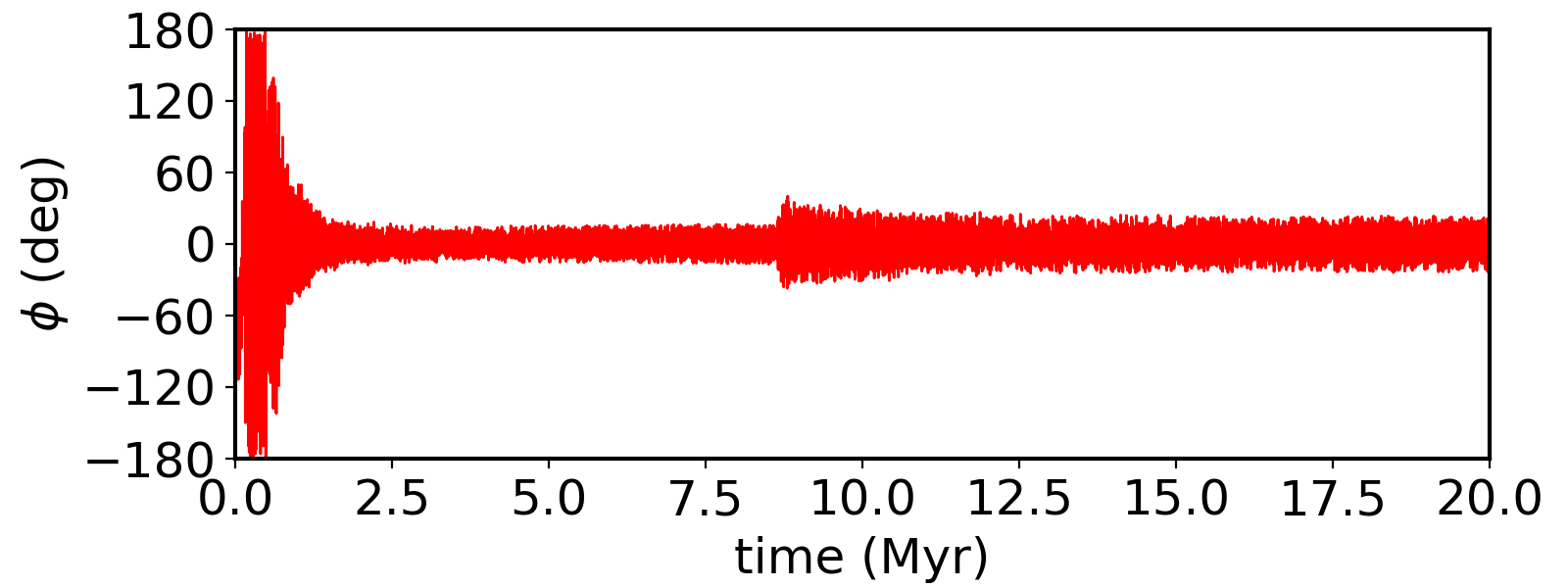}}
\caption{From top to bottom, the first three panels show the evolution of resonant angles associated with the 2:1 MMR of different satellite pairs. The bottom panels shows the resonant angle associated with the Laplacian resonance characterized by the three innermost satellites. The labels 1, 2, 3, and 4 corresponds to the innermost, second innermost, third innermost, and outermost satellite in our simulation, respectively. All these resonant angles librate at the beginning of the simulation, which corresponds to the the end of the gas disk phase. The orbital eccentricity of the innermost satellite is damped in a timescale $t_{e}=t_{a}/S$, where $S=10^5$ (see Figure \ref{fig:edampF}), to mimic the effects of tidal dissipation. \label{fig:resangF}}
\end{figure}

Finally, in Figure \ref{fig:resangF}, we analyze the behavior of the resonant angles characterizing our resonant chain in the same simulation of Figure~\ref{fig:edampF}. From top to bottom,  the first three panels show the resonant angles associated with the 2:1 MMR. These plots show that the innermost and second innermost satellite pairs remain locked in a 2:1 MMR when we mimic the tidal dissipation effects on eccentricity. This is indicated by the reduction -- relative to the beginning of the simulation -- in the libration amplitudes of the resonant angles ($\phi$) which librate around zero with small amplitude after the simulation timespan. However, in the case of the outermost satellite pair, the amplitudes of libration of the associated resonance angles gradually increase until they start to circulate, dissolving the resonant configuration, without breaking the resonant configuration of the other pairs. The bottom panel shows that the amplitude of the resonant angle associated with the Laplacian resonant slightly increases at $\sim 8$~Myr -- the moment when the outermost satellite leaves the resonance with the second outermost satellite -- but the three innermost satellites remain locked in this configuration. Therefore, our results also suggest that the Galilean satellite system is a primordial resonant chain, where Callisto was once in resonance with Ganymede but left this configuration via divergent migration due to dynamical tides \citep{Fu16,downey2020inclination,lari2020long,Ha20,Durante2020,Idini2021}. Of course, a complete validation of this result may require self-consistent simulations modeling tidal planet-satellite dissipation effects but this is beyond the scope of this paper.

\section{Discussion}\label{sec:discussion}

We have assumed a fully formed Jupiter from the beginning of our simulations. We believe this is a fair approximation because Jupiter's CPD and the Galilean satellites mostly likely started to form during the final phase of Jupiter's growth. \cite{Sz16} showed that the characteristics of a circumplanetary disk are mainly determined by the temperature at the planet's location -- which is driven by the gas accretion rate onto the planet -- and have a weak dependency on the planet mass. In our model, the total flux of gas from the CSD to the CPD is about $\sim0.1~{\rm M_J}$. So, even if one assumes that all  gas entering the CPD is accreted by Jupiter, it should have no less than 90\% of its current mass at the beginning of our simulations. Thus, we do not expect this small difference in the planet's mass than we have considered here to change the quality of our results. One potential impact could be seen in the migration timescale of the satellites. However, the migration timescale depends also on CPD's model, as surface density and aspect ratio. As we have performed simulations considering different dissipation modes for the gas disk we consider our main findings to be fairly robust against this issue.

\begin{figure}
\includegraphics[width=\columnwidth]{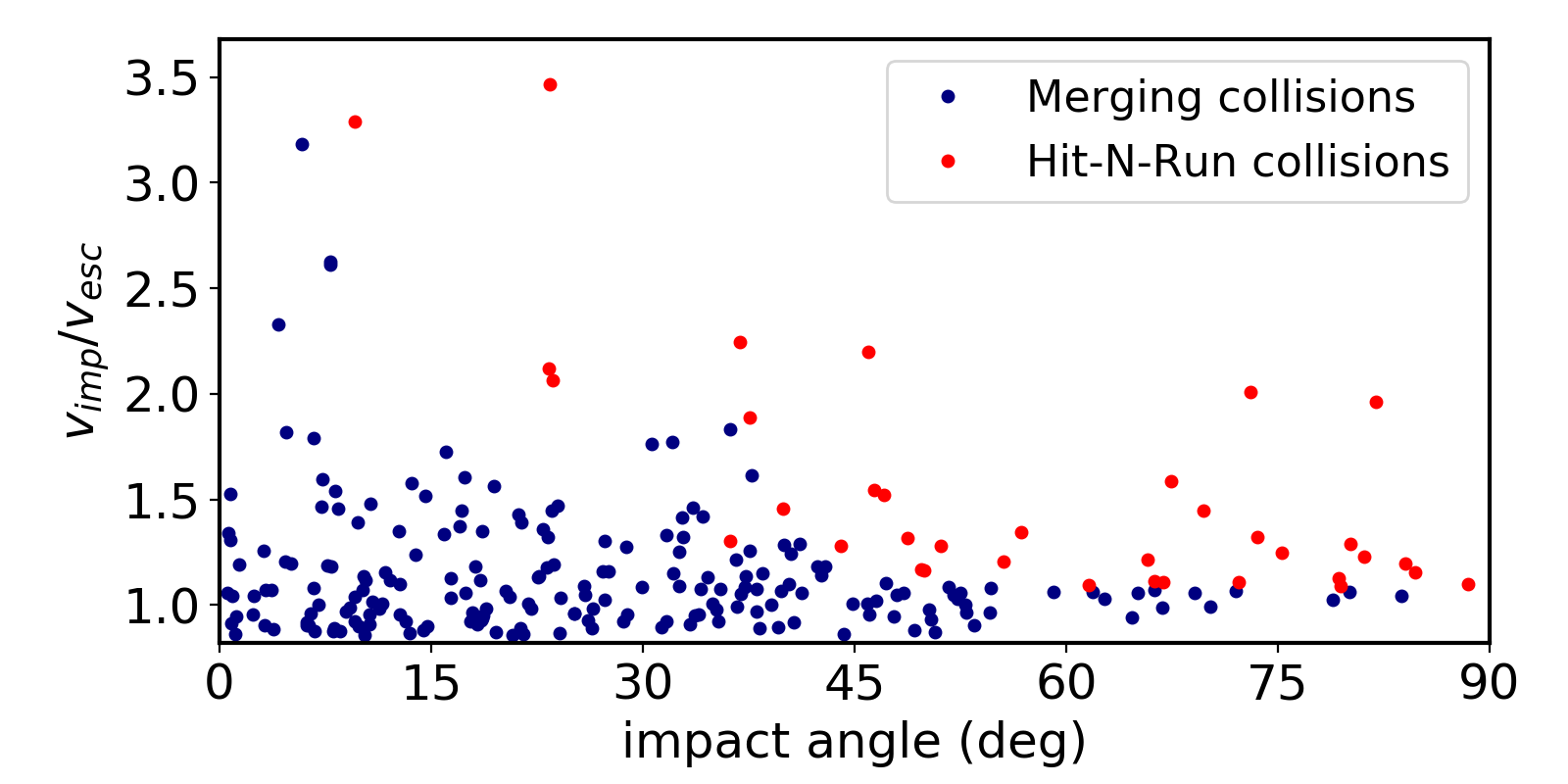}
\caption{ Normalized impact velocity (${\rm v_{imp}/v_{esc}}$) as a function of the impact angle in simulations that produced Galilean satellite analogues. The color-coding shows the predicted outcome following  Kokubo \& Genda (2010) and Genda et al. (2012). Blue (red) dots correspond to the impacts that fall in the merging (hit-and-run) regime. 85\% of the impacts in these selected simulation qualify as perfect merging events. \label{fig:impact}}
\end{figure}

Our treatment of collisions is also simplistic. All of our collisions are modeled as perfect merging events that conserve mass and linear momentum. To evaluate whether or not our assumption is adequate, we have analyzed the expected outcome of collisions in our simulations following \cite{2010ApJ...714L..21K} and \cite{2012ApJ...744..137G}. We have found that about 85\% of our collisions qualify as perfect merging events, and only 15\% of our simulations fall into the hit-and-run regime (see Figure~\ref{fig:impact}). Impact velocities in our simulations are very low because collisions happen during the gas disk phase when satellites have low orbital eccentricities and inclinations due to gas-tidal damping and drag. These results show that our treatment of impacts is fairly appropriate to study the formation of the Galilean satellites.

For simplicity, our simulations started with satellitesimals distributed initially between 20 and 120~${\rm R_J}$. We have argued that these objects were most likely captured from the CSD \citep{Zh12} rather than having been born in-situ. However, we do not model the capture of planetesimals from the CSD in the CPD in this work. The CPD's location where planetesimals from the CSD are preferentially captured depends on planetesimals sizes, orbits, gas surface density, and ablation degree when they enter the CPD \citep{Fu13,Ro20}. Simulations from \cite{Ro20} suggest the total time necessary to Jupiter capture the total mass in satellitesimals assumed at the beginning of our simulations would be only $\sim 10^3-10^4$~years.

Our best Galilean satellite analogues have typically four satellites as the real Galilean system. However, Jupiter hosts a complex system of moons. Jupiter hosts four satellites with sizes of tens of kilometers inside  Io's orbit and a large set of regular satellites of a few kilometers in size outside Callisto's orbit. Here we speculate that these two populations of objects represent fragments, late captured, or leftover satellitesimals that were too small to efficiently grow by pebble accretion. This hypothesis is under investigation.

Our model is honestly simplified and we do not plead to definitively explain the formation of the Galilean satellites. However, it is astonishing that using the same physical processes invoked in models to explain the formation of planets in the Solar System and around other stars \cite[e.g.][]{levisonetal15b,Iz19}, our simulations can still form satellites systems that resemble fairly well the Galilean moons.  Our results also show the importance of future studies to provide firmer constraints on the original composition of Io and Europa. Were they born with water-rich compositions or not?

\section{Conclusions}\label{sec:conclusions}

We have performed a suite of  N-body numerical simulations to study the formation of the Galilean satellites in a gas-starved disk scenario \citep{Ca02,Ca06,Ca09,Sa10} invoking pebble accretion and migration. Our simulations start considering an initial population of $\sim$200~km-size satellitesimals in the circumplanetary disk that are envisioned to be planetesimals captured from the sun's natal disk. We have performed simulations testing different initial number of satellitesimals in the CPD and pebble fluxes. Our pebbles fluxes are consistent with those estimated via the ablation of planetesimals in the circumplanetary disk \citep{Ro20}. Our best Galilean system analogues were produced in simulations where the time-integrated pebble flux is about $10^{-3}~{\rm M_J}$. In these simulations, satellitesimals grow via pebble accretion and migrate to the disk inner edge where they stop migrating at the inner edge trap. When they approach this location, dynamical instabilities and orbital crossing promote further growth via impacts. Our simulations typically produced between 3 and 5 final satellites anchored at the disk inner edge, forming a resonant chain. Their masses match relatively well those of the Galilean satellites. All our system analogues show 3 pairs of satellites locked in a 2:1 mean motion resonance. Thus, we propose that Callisto -- the outermost Galilean satellite -- was originally locked in resonance with Ganymede but left this primordial configuration via divergent migration due to tidal dissipative effects \citep{Fu16,downey2020inclination}.  We also proposed that the orbital eccentricities of the Galilean satellites were much higher in the past and were damped to their current values via tidal dissipation without destroying the resonant configuration of the innermost satellites \citep{Fu16,downey2020inclination,lari2020long,Ha20,Durante2020,Idini2021}. Finally, we proposed that the Galilean system represents a primordial resonant chain that did not become unstable after the circumplanetary gas disk dispersal. Thus the formation path of the Galilean system is probably similar to that of systems of close-in super-Earths around stars, as the Kepler-223 \citep{mills2016resonant}, TRAPPIST-1 \citep{gillon2017seven}, and TOI-178 \citep{Leleu21} systems~\citep{Iz17,Iz19}.  

Our simulations do not reproduce the current low water-ice fractions of Io and Europa, and require efficient water loss via hydrodynamic escape \citep{Bi20} to occur to match their current bulk compositions. If efficient water loss via hydrodynamic took place it is expected that Europa should have developed a higher deuterium-to-hydrogen ratio compared with Ganymede and Callisto \citep{Bi20}. This prediction may be tested in the future via in-situ measurements by the Europa Clipper spacecraft or infrared spectroscopic observations \citep{Bi20}. In all our simulations, the two outermost satellites have water-ice rich compositions similar to Ganymede and Callisto. Our results suggest that if Io and Europa were born water-ice depleted, they should have formed much earlier than Ganymede and Callisto and well inside the CPD's snowline.  Additional constraints on Io and Europa are now crucial to constrain Galilean satellites' formation models. 

\section*{Acknowledgements}
We thank the referee, Konstantin Batygin, for carefully reading our paper and providing valuable comments and suggestions. G.M. thanks FAPESP for financial support via grant 2018/23568-6.
S.M.G.W. thanks FAPESP (2016/24561-0), CNPq (309714/2016-8; 313043/2020-5) and Capes for the financial support.
A.I thanks financial support from FAPESP (2016/12686-2;  2016/19556-7) and CNPq (313998/2018-3) during the initial preparation of this work. A.~I. also thanks financial support via NASA grant 80NSSC18K0828 during the final preparation and submission of this work.
Research carried out using the computational resources of the Center for Mathematical Sciences Applied to Industry (CeMEAI) funded by FAPESP (grant 2013/07375-0).

\section*{Data Availability}
The data underlying this article will be shared on reasonable request to the corresponding author.


\bibliographystyle{mnras}
\bibliography{galileanmoons} 








\bsp	
\label{lastpage}
\end{document}